\documentclass[english,aps,prb,notitlepage,twocolumn,superscriptaddress,floatfix]{revtex4-2} 
\usepackage{CJK}
\usepackage{graphicx}
\usepackage{dcolumn}
\usepackage{bm}
\usepackage{mathtools}
\usepackage{amssymb,amsmath,amsfonts,dsfont}
\usepackage{hyperref}
\usepackage{color}

\newcommand{\tr}{\operatorname{Tr}}
\newcommand{\emb}{\text{emb}}
\newcommand{\eff}{\text{eff}}

\def\bra#1{\mathinner{\langle{#1}|}}
\def\ket#1{\mathinner{|{#1}\rangle}}
\def\braa#1{\mathinner{\langle\!\langle{#1}|}}
\def\kett#1{\mathinner{|{#1}\rangle\!\rangle}}

\newcommand{\braakett}[2]{\langle\!\langle #1|#2\rangle\!\rangle}
\newcommand{\proj}[1]{\ket{#1}\!\!\bra{#1}}
\newcommand{\mP}{\mathcal{P}}
\newcommand{\hc}{\text{H.c.}}

\makeatletter
\let\inserttitle\@title
\makeatother

\begin{document}
\begin{CJK*}{UTF8}{mj}

\title{Tensor network influence functionals in the continuous-time limit: Connections to quantum embedding, bath discretization, and higher-order time propagation}

\author{Gunhee Park (박건희)}
\affiliation{Division of Engineering and Applied Science, California Institute of Technology, Pasadena, California 91125, USA}
\author{Nathan Ng}
\affiliation{Department of Chemistry, Columbia University, New York, New York 10027, USA}
\author{David R.\ Reichman}
\affiliation{Department of Chemistry, Columbia University, New York, New York 10027, USA}
\author{Garnet Kin-Lic Chan}
\affiliation{Division of Chemistry and Chemical Engineering, California Institute of Technology, Pasadena, California 91125, USA}

\begin{abstract}
    We describe two developments of tensor network influence functionals (in particular, influence functional matrix product states (IF-MPS)) for quantum impurity dynamics within the fermionic setting of the Anderson impurity model. The first provides the correct extension of the IF-MPS to continuous time by introducing a related mathematical object, the boundary influence functional MPS. The second connects the dynamics described by a compressed IF-MPS to that of a quantum embedding method with a time-dependent effective bath undergoing nonunitary dynamics. 
    Using these concepts, we implement higher-order time propagators for the quench dynamics of the Anderson impurity model within the boundary IF-MPS formalism. The calculations illustrate the ability of the current formulation to efficiently remove the time step error in standard discrete-time IF-MPS implementations as well as to interface with state vector propagation techniques.
    They also show the advantages of IF-MPS dynamics, with its associated highly compact effective bath dynamics, over state vector propagation with a static bath discretization.
\end{abstract}

\maketitle
\end{CJK*}

\section{Introduction}

Quantum impurity models, such as the Anderson impurity model (AIM), consist of an interacting impurity coupled to a (possibly noninteracting) bath. They provide simple settings in which to study quantum many-body physics, including the Kondo effect in magnetic impurities~\cite{Hewson1993}, and serve as a starting point of computational embedding frameworks, such as dynamical mean-field theory~\cite{Georges1996} or density matrix embedding theory~\cite{Knizia2012, doi:10.1021/acs.jctc.6b00316}, to solve lattice problems. The challenge of describing non-equilibrium dynamics in such models has spurred the development of many computational techniques~\cite{PAECKEL2019167998, PhysRevB.79.235336, PhysRevB.90.235131, Ganahl2015, Kohn_2022,Muehlbacher2008, Cohen2011, Gull2011, Gull2010, Cohen2013, Cohen2014a, Cohen2014b}.

One way to describe quantum impurity dynamics is to use the equation of motion of the impurity reduced density operator obtained by tracing out the bath. The influence of the bath is clearly expressed in a path integral language via the Feynman-Vernon influence functional (IF) \cite{FEYNMAN2000547}, which reweights the paths in the path integral. Although the influence functional for a noninteracting bath with linear coupling can be expressed in a compact mathematical form, its effect on the impurity dynamics must still be numerically approximated in practice~\cite{Tanimura1989,10.1063/1.2938087, Dong2014,PhysRevLett.110.086403,PhysRevB.89.165105,PhysRevB.92.125145, PhysRevLett.122.186803, Chen_2019,Makri1992, Makri1995a, PhysRevB.77.195316, PhysRevB.82.205323,Makri1992, Makri1995a}.

Recently, tensor network methods have been explored within the IF language to overcome some limitations of other numerical IF techniques, and in particular, to capture longer-time memory effects \cite{Banuls2009,PhysRevA.91.032306,Strathearn2018,Cygorek2022,Gribben2022,10.1063/5.0047260, PhysRevLett.123.240602, SONNER2021168677,PhysRevB.107.195101,PhysRevB.107.125103,PhysRevB.107.L201115, PhysRevB.108.205110, PhysRevB.109.045140}. By treating the IF as a temporal wavefunction expressed as a temporal matrix product state (MPS), one can exploit low entanglement in time, leading to a compact representation of certain temporal correlations. Such tensor network-based IF methods have been successfully demonstrated in several contexts, including the spin-boson model \cite{Strathearn2018,Cygorek2022,Gribben2022,10.1063/5.0047260, PhysRevLett.123.240602}, hard-core bosons \cite{10.1063/5.0047260}, one-dimensional spin chains \cite{Banuls2009,PhysRevA.91.032306,SONNER2021168677}, and more recently, in the context of the AIM \cite{PhysRevB.107.195101,PhysRevB.107.125103,PhysRevB.107.L201115, PhysRevB.108.205110, PhysRevB.109.045140, PhysRevB.109.165113}.

The current paper is concerned with two developments of the tensor network IF approach. Although the ideas generalize to IFs of any linearly coupled noninteracting bath, for concreteness we work specifically with the AIM, which expands on the previous developments based on the standard IF-MPS framework~\cite{PhysRevB.107.195101,PhysRevB.107.125103,PhysRevB.107.L201115, PhysRevB.108.205110}. The first development removes the time step error in the standard IF treatment obtained from a Trotterized representation of the path integral.  Formulating the problem in the continuous-time limit yields a continuous MPS version of the IF-MPS. We show that the standard IF-MPS in fact does not have a useful continuous-time limit, and instead one must consider a closely related object, the boundary IF-MPS, to define this limit. We provide an explicit construction of the continuous-time boundary IF-MPS for the AIM.

The second development is concerned with the relationship between the (boundary) IF-MPS  and discrete bath dynamics. Standard state vector methods for the dynamics of the AIM use a discrete bath with a fixed set of bath energies and couplings~\cite{PAECKEL2019167998, PhysRevB.79.235336, PhysRevB.90.235131, Ganahl2015, Kohn_2022}. We show that impurity dynamics defined by an IF-MPS of fixed bond dimension is equivalent to propagating in a space of effective bath orbitals in Liouville space, where the bath energies and couplings dynamically vary with time. This naturally connects tensor network influence functionals to dynamical quantum embedding theories, such as real-time density matrix embedding, which also utilize a dynamical set of bath energies and couplings~\cite{doi:10.1063/1.5012766}. An important difference, however, is that the IF-MPS defines a nonunitary dynamics in the bath. 

We implement the boundary IF-MPS propagation for the AIM in both the discrete-time and continuous-time formulations. Directly comparing to standard static bath discretizations, we show that the IF-MPS method converges extremely rapidly with respect to the effective bath size, and shows none of the discretization artifacts of standard bath discretizations. Further, the combination of the above two insights suggests that the boundary IF-MPS dynamics in the continuum limit can be efficiently implemented using standard state vector time-propagation techniques. We use this to implement a high-order Runge-Kutta time propagator for boundary IF-MPS dynamics and demonstrate the high-order error with time step. In contrast, we show that higher than first-order Trotter methods in the standard discrete-time IF-MPS still suffer from first-order time step errors, due to the IF-MPS bond truncation. By Trotterizing the correct continuous-time dynamics, we derive a corrected version of the Trotterized error with the correct time-order scaling, significantly improving on the second-order Trotter formulation currently used in tensor network IF approaches.

The paper is organized as follows. In Sec.~\ref{sec:sec2}, we first recapitulate the formulation of the IF-MPS for noninteracting fermionic baths, describing in detail the formalism we use here in terms of number-conserving Slater determinants. Through this picture, we establish the connection between the IF-MPS dynamics and Liouville state vector propagation of an impurity coupled to a set of effective bath orbitals, that is, the dynamics of a quantum embedding of the impurity.  In Sec.~\ref{sec:sec3}, we analyze the continuous-time limit of the IF-MPS and show that the correct object to consider is the boundary IF-MPS and we provide its continuous-time limit. Using the state vector formalism, we rewrite the continuous-time propagation in terms of a differential equation of motion for the Liouville state vector in the quantum embedding space.  In Sec.~\ref{sec:sec4}, we provide numerical results using the boundary IF-MPS for the single impurity Anderson model and compare the discretization errors associated with a static set of bath orbitals with the dynamic set of bath orbitals defined by the IF-MPS. We further analyze the time step errors from both the discrete-time and continuous-time formulations. In Sec.~\ref{sec:sec5}, we conclude with discussions of some implications of our results.

\section{Influence Functional theory in discrete time}\label{sec:sec2}

\subsection{Influence functional for single impurity Anderson model}

We consider a single impurity Anderson model,
\begin{equation}
    \begin{gathered}
    \hat{H} = \hat{H}_S + \hat{H}_{SB}+\hat{H}_{B},\\
    \hat{H}_S = U \hat{n}_{\uparrow} \hat{n}_{\downarrow} + \sum_{\sigma} \varepsilon_{\sigma} \hat{n}_{\sigma}, \\
    \hat{H}_{SB} = \sum_{i, \sigma} \left( t_i \hat{c}^{\dag}_{i, \sigma} \hat{d}_{\sigma} + \hc \right),\\
    \hat{H}_{B} = \sum_{i, \sigma} E_{i} \hat{c}^{\dag}_{i, \sigma} \hat{c}_{i,\sigma},
    \end{gathered}
\end{equation}
where $\hat{d}^{\dag}_{\sigma}$ ($\hat{c}^{\dag}_{i, \sigma}$) creates a fermion of spin $\sigma \in \{\uparrow,\downarrow\}$ in the impurity (bath) orbitals and $\hat{n}_{\sigma} = \hat{d}^{\dag}_{\sigma} \hat{d}_{\sigma}$ is the number density operator of the impurity orbital of spin $\sigma$. $\hat{H}_S$ and $\hat{H}_B$ are an impurity-only and bath-only Hamiltonian, respectively, and $\hat{H}_{SB}$ is an impurity-bath coupling Hamiltonian. Note that $\hat{H}_{SB}$ and $\hat{H}_B$ are of noninteracting (quadratic) form. Hereafter,  we assume a discrete and finite set of bath orbitals and also assume that the impurity is initially decoupled from the bath, $\hat{\rho}(0) = \hat{\rho}_S(0) \otimes \hat{\rho}_B$. We further assume a Gaussian initial bath, in particular, the thermal state $\hat{\rho}_B \propto e^{-\beta \hat{H}_B}$ with inverse temperature $\beta$. This assumption is appropriate for the problem of quench dynamics from decoupled finite-temperature baths with $\hat{H}_{SB}=0$.~\footnote{Within the influence functional formalism, this assumption can be relaxed to general Gaussian system-bath initial states, including a quantum quench from the ground state of the model at $U=0$ to finite $U$. } For quadratic operators, we omit the hats when expressing their matrix elements in a single-particle basis.

While the time evolution of the density operator can be described by the von Neumann equation,
\begin{equation}
    i \frac{d}{dt} {\hat{\rho}} = [\hat{H},\hat{\rho}],
\end{equation}
we instead will adopt the super-fermion representation of Liouville space \cite{Schmutz1978,10.1063/1.3548065, HARBOLA2008191,Prosen_2008,PhysRevLett.110.086403}. This uses a super-Fock space with twice the number of orbitals as the original Hilbert space, obtained by applying a particle-hole transformation to the bra of the density operator, and assuming the resulting operator acts on a vacuum to produce a state. We denote states in the super-Fock space using a double bra/ket notation. For example, the initial density operator is written as $\kett{\rho(0)}=\kett{\rho_S(0)}\otimes\kett{\rho_B}$. The von Neumann equation now becomes a Hamiltonian time evolution with the Liouville operator $\hat{L}$,
\begin{equation}
    i\frac{d}{dt}\kett{\rho} = \hat{L}\kett{\rho}.
\end{equation}
The Liouville operator for the Anderson impurity model takes the form (using tildes on the operators from the particle-hole transformed Fock space),
\begin{equation}
    \begin{gathered}
    \hat{L} = \hat{L}_S + \hat{L}_{SB}+\hat{L}_{B},\\
    \hat{L}_S = U \hat{n}_{\uparrow} \hat{n}_{\downarrow} -U (1-\hat{\Tilde{n}}_{\uparrow})(1-\hat{\Tilde{n}}_{\downarrow}) \\
    + \sum_{\sigma}  \varepsilon_{\sigma} (\hat{n}_{\sigma} + \hat{\Tilde{n}}_{\sigma}),  \\
    \hat{L}_{SB} = \sum_{i, \sigma} \left( t_i \hat{c}^{\dag}_{i, \sigma} \hat{d}_{\sigma} +t_i \hat{\Tilde{c}}^{\dag}_{i, \sigma} \hat{\Tilde{d}}_{\sigma} + \hc \right),\\
    \hat{L}_{B} = \sum_{i, \sigma} E_{i} (\hat{c}^{\dag}_{i, \sigma} \hat{c}_{i,\sigma} + \hat{\Tilde{c}}^{\dag}_{i,\sigma} \hat{\Tilde{c}}_{i,\sigma} ),
    \end{gathered}
\end{equation}
where we have omitted all constant terms. We will collectively refer to the tilde and non-tilde creation operators as $\hat{a}^{\dag}_{i,\sigma}$,
\begin{equation}
    \hat{a}^{\dag}_{i,\sigma} =
    \begin{cases}
        \hat{c}^{\dag}_{i,\sigma} & 1 \le i \le N_b \\
        \hat{\Tilde{c}}^{\dag}_{i-N_b,\sigma} & N_b + 1 \le i \le 2N_b
    \end{cases},
\end{equation}
for $N_b$ bath orbitals.

The initial bath state in the super-Fock space $\kett{\rho_B}$ is given by a Slater determinant
\begin{equation}
    \kett{\rho_B} = \prod_{i,\sigma} \left( f_{+}(E_{i,\sigma},\beta) \hat{c}^{\dag}_{i,\sigma} + 
f_{-}(E_{i,\sigma},\beta) \hat{\Tilde{c}}^{\dag}_{i,\sigma}  \right)\kett{0}
\label{eq:thermal}
\end{equation}
where $\kett{0}$ is a vacuum state in the super-Fock space, $f_{+}(E,\beta)=(1+e^{\beta E})^{-1}$, and $f_{-}(E,\beta)=1-f_{+}(E,\beta)$. We are often interested in the reduced density operator of the impurity, $\hat{\rho}_S = \tr_B(\hat{\rho})$. In the super-Fock space, the trace over the bath is equivalent to taking the overlap with a trace vector that can be expressed as a Slater determinant,
\begin{equation}\label{eq:trace_vec}
    \kett{\tr_B} = \prod_{i,\sigma} \left( \hat{c}^{\dag}_{i,\sigma} + \hat{\Tilde{c}}^{\dag}_{i,\sigma} \right) \kett{0},
\end{equation}
and the amplitude of a configuration $s$ in $\rho_S$ can be expressed as
\begin{equation}
    \braakett{s}{\rho_S} = (\braa{s}\otimes \braa{\tr_B})\kett{\rho}.
\end{equation}

\begin{figure}[t]
    \centering
    \includegraphics[width=0.6\columnwidth]{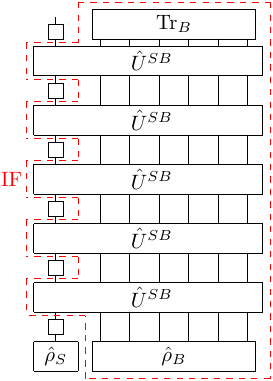}
    \caption{A schematic diagram for the real-time evolution of the Anderson impurity model in Liouville space after $N_t=5$ time steps. The initial density operator is given by $\hat{\rho}(0) = \hat{\rho}_S \otimes \hat{\rho}_B$, described by a vectorized state in a Liouville space. The time evolution of the density operator is described by time-evolution operators following a second-order Trotter decomposition, alternating between $\hat{U}^{S}$ (squares) and $\hat{U}^{SB}$ (rectangles). After the time evolution, the bath degrees of freedom are traced out, which is equivalent to applying the trace tensor $\text{Tr}_B$ to the bath. The influence functional (IF) tensor corresponds to the tensor after the contraction of the bath modes in all $\hat{U}^{SB}$, $\hat{\rho}_B$, and $\text{Tr}_B$ tensors, grouped with the red-dashed lines.}
    \label{fig:Liou}
\end{figure}

The discretized time evolution of the density operator with timestep $\Delta t$ can be expressed via a second-order Trotter decomposition
\begin{align}\label{eq:Liou_evo}
    \kett{\rho_S(t_N)} & = \tr_B \left[ \left(e^{-\frac{i}{2}\hat{L}_{S} \Delta t}e^{-i\hat{L}_{SB}\Delta t} e^{-\frac{i}{2}\hat{L}_{S} \Delta t}  \right)^{N_t} \kett{\rho(0)}\right] \nonumber \\
    & = \tr_B \Bigl[ \hat{U}^{S} \hat{U}^{SB} \hat{U}^{S} \cdots \hat{U}^{SB}\hat{U}^{S} \kett{\rho(0)} \Bigr]
\end{align}
where $t_{N_t} = N_t \Delta t$ and $\hat{U}^{S}$ ($\hat{U}^{SB}$) is the time-evolution operator for $\hat{L}_S$ ($\hat{L}_{SB}$ and $\hat{L}_B$). The tensor network diagram for this time evolution is shown in Fig.~\ref{fig:Liou}, where we see that the  $\hat{U}^{S}$ tensor is applied only within the impurity $S$, whereas the tensor for $\hat{U}^{SB}$ is applied to both $S$ and $B$.

The influence functional (IF) tensor is defined as the tensor arising from the contraction of all the bath degrees of freedom in $\hat{U}_{SB}$, the initial bath density operator, and the trace vector (Fig.~\ref{fig:Liou}). This IF tensor 
is indexed by the state of the impurity orbitals before and after each $\hat{U}_{SB}$, for $N_t$ time steps (denoted $s^{i}_{m}$ and $s^{f}_{m}$, respectively, for the $m$-th time step) and so is indexed by the configuration of $2N_t$ impurity orbitals. In addition, we mention that the IF tensors for different spins are constructed separately thanks to the absence of spin-mixing terms in $\hat{L}_{SB}$, and hence, the spin indices are omitted for brevity.

Due to the one-dimensional structure of the temporal axis, the IF can be rewritten in terms of $N_t$ tensors that are directly obtained from $\hat{U}^{SB}$ by inserting the bath configurations between each time step. Denoting the impurity configurations, $\bm{s}=(s^{i}_{1},s^{f}_{1}, \cdots, s^{i}_{N_t}, s^{f}_{N_t})$, and the corresponding IF tensor element as $I(\bm{s})$, the matrix product state (MPS) representation of the IF can be written as
\begin{equation}\label{eq:IF_mps}
    I(\bm{s}) = l^T \cdot  A_{N_t}^{s^{i}_{N_t},s^{f}_{N_t}} \cdots A_{1}^{s^{i}_{1},s^{f}_{1}} \cdot r,
\end{equation}
where the matrix elements for $A_m$ are given by
\begin{equation}\label{eq:mps_elem}
    \left( A_m^{s_m^i,s_m^f} \right)_{b_{m},b_{m-1}} = \braa{s_m^f,b_{m}} \hat{U}^{SB} \kett{s_m^i,b_{m-1}},
\end{equation}
and $b_m$ is the bath configuration after $m$ applications of $\hat{U}^{SB}$, $r_{b_0} = \braakett{b_0}{\rho_B(0)}$, and $l_{b_{N_t}}=\braakett{\tr_B}{b_{N_t}}$.

In the above MPS representation, the bond dimension is given by the dimension of the super-Fock bath space, which, in many cases, is too large to deal with directly. Previous studies \cite{PhysRevB.107.125103, PhysRevB.107.L201115} have made use of the fermionic Gaussian properties of the IF (arising from the linear coupling and quadratic bath) to find a compressed form of the MPS representation \cite{PhysRevB.92.075132, PhysRevB.103.125161}. In this paper, we will use a slightly different language to formulate the MPS compression in terms of finding Schmidt vectors in the bath. This algorithm (described in Sec.~\ref{sec:mps_compression}) is closely related to that in Ref.~\cite{PhysRevB.92.075132} and improves on the computational scaling in Ref.~\cite{PhysRevB.103.125161}.

\subsection{From influence functionals to state vector propagation} \label{sec:IF_to_statevec}

\begin{figure*}[t]
    \centering
    \includegraphics[width=0.8\textwidth]{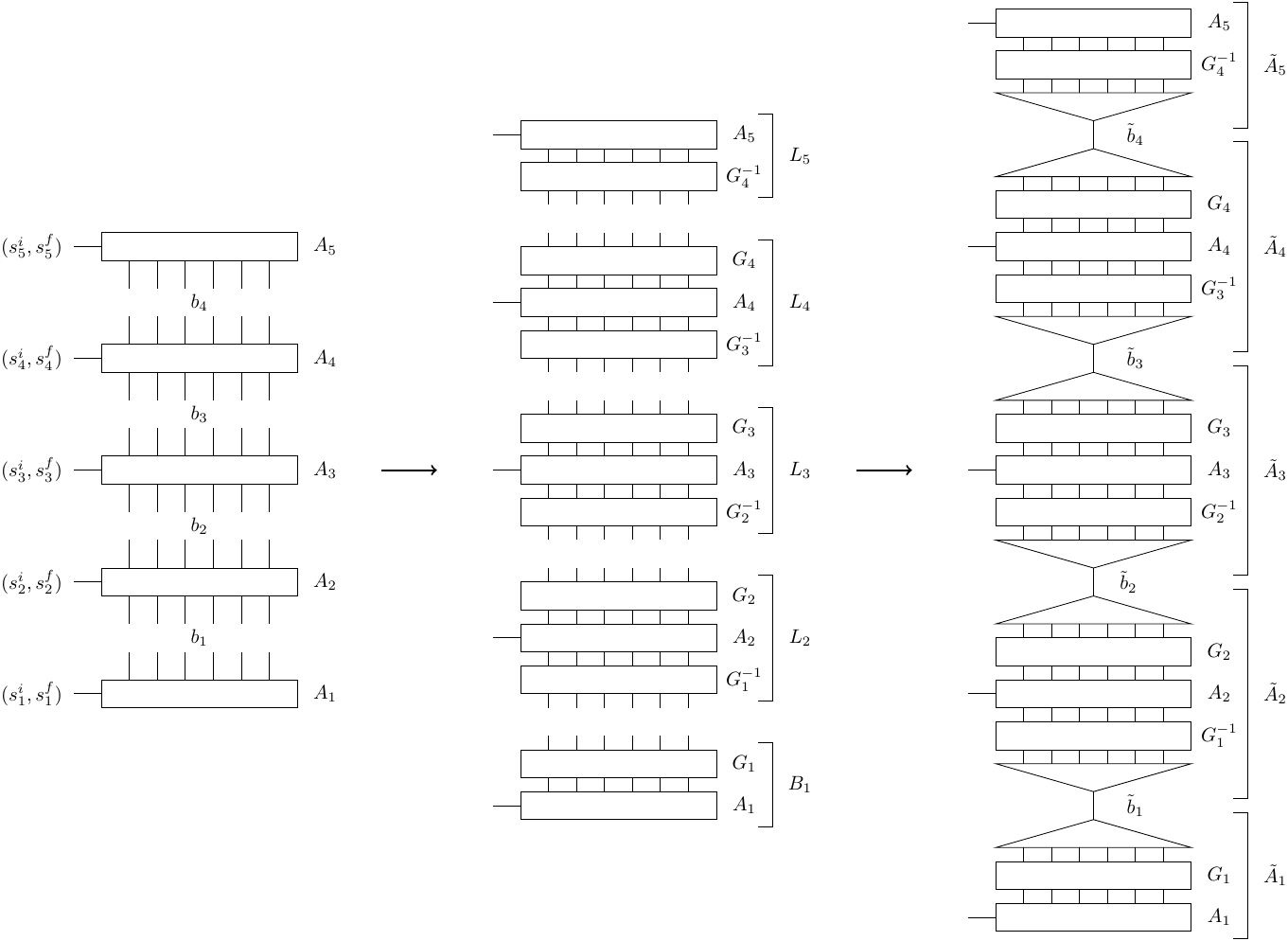}
    \caption{General matrix product state (MPS) compression scheme for an influence functional MPS (IF-MPS). The bond dimension of the initial uncompressed MPS is given by the dimension of the bath Liouville space, indexed by bath configurations, $b_m$ (left). The MPS needs to be converted into canonical form for an optimal truncation. The gauge matrices $G_m$ are inserted in the bath Liouville space to convert the MPS into the canonical form (middle). Gauge matrices transform the original matrix elements of the MPS into left-normalized matrices, denoted as $L_m=G_{m} A_m G_{m-1}^{-1}$. Afterward, projectors are inserted into the bath subspace that contains the bath states with the largest singular values (right). The projected bath configurations are denoted as $\Tilde{b}_m$.}
    \label{fig:mps_compression}
\end{figure*}

In this paper, we will often switch between two equivalent pictures: dynamics encoded by a compressed IF-MPS, and a state vector propagation corresponding to a quantum embedding. 
To understand this mapping, we first describe the conventional MPS compression scheme (i.e., without using any Gaussian properties of the bath) and explain how it can be interpreted as a type of projected bath dynamics. (We recall that the IF-MPS can, in general, be expressed through Eqs.~\ref{eq:IF_mps} and \ref{eq:mps_elem}, for arbitrary system-bath quantum dynamics, i.e., even for an interacting bath).

In conventional MPS compression, the MPS is first transformed to a canonical form to enable an optimal truncation of the bond dimension. The canonical form is defined using the gauge degrees of freedom in the MPS~\cite{Schollwoeck2011},
\begin{gather}
    A_m^{s^i_m,s^f_m} \rightarrow L_m^{s^i_m,s^f_m}=G_{m} A_m^{s^i_m,s^f_m} G_{m-1}^{-1}, \\
    \sum_{s^i_m,s^f_m} L_m^{s^i_m,s^f_m \: \dag } L_m^{s^i_m,s^f_m} = I, \label{eq:left_norm}
\end{gather}
where the matrices $L_m^{s^i_m,s^f_m}$ satisfying  Eq.~\ref{eq:left_norm}, are called left-normalized matrices. The gauge matrices, $G_m$, are inserted in the bond space, or the Liouville bath space of the IF-MPS (Fig.~\ref{fig:mps_compression}). The MPS that is composed of the left-normalized matrices is called left-canonical. The left-canonical MPS is then
\begin{equation}
    I(\bm{s}) =   L_{N_t}^{s^{i}_{N_t},s^{f}_{N_t}} \cdots L_2^{s^{i}_{2},s^{f}_{2}} B_1^{s^{i}_{1},s^{f}_{1} },
\end{equation}
where $B_1^{s_1^i,s_1^f} = G_1 A_1^{s_1^i,s_1^f}$. After choosing this gauge, the MPS is compressed by iteratively applying truncated singular value decompositions (SVD) from right to left. 

We can group together the effect of the gauging and compression together with the system-bath evolution to define new matrices of the IF-MPS, 
$\Tilde{A}_m$, 
\begin{gather}   \left(\Tilde{A}_m^{s^i_m,s^f_m}\right)_{\Tilde{b}_{m},\Tilde{b}_{m-1}} = \nonumber \\
    \braa{s_m^f,\Tilde{b}_{m}} \mP_m \hat{G}_m \hat{U}^{SB} \hat{G}^{-1}_{m-1} \mP_{m-1} \kett{s_m^i,\Tilde{b}_{m-1}}, \label{eq:eff_timeevo}
\end{gather}
where $\mP_m$ denotes the projectors 
onto the bath states associated with the largest singular values (or equivalently, the largest eigenvalues of the bath density matrix) after the gauge transformation (Fig.~\ref{fig:mps_compression}), 
and $\Tilde{b}_m$ denotes the projected bath configurations.

$\Tilde{A}_m$ can be viewed as defining a (nonunitary) evolution in the Liouville space of the system and the bath.
Alternatively, the MPS gauging and compression procedure can be seen as a pure projected bath dynamics in the Liouville space $\hat{G}_m^{-1} \mP_m \hat{G}_m$, inserted between the system-bath evolution $\hat{U}^{SB}$. The coarse graining (i.e., projection) of the bath degrees of freedom is referred to as a quantum embedding, and consequently, the truncated IF-MPS dynamics is a quantum embedding scheme with a dynamically evolved bath, similar to (real-time) density matrix embedding theory (DMET)~\cite{Knizia2012,doi:10.1021/acs.jctc.6b00316,doi:10.1063/1.5012766,10.1063/5.0146973}, but with the important difference that the dynamics of the projected bath is nonunitary, as $\hat{G}_m$ is nonunitary.

As the time step $\Delta t\to 0$, we are led to the continuous-time limit of the IF-MPS. We consider the subtleties of continuous-time construction in Sec.~\ref{sec:boundaryif}. 
However, the above shows that 
we can also view time evolution as being performed on an embedded 
state vector (``wavefunction'') in the projected Liouville space (embedding space) $\{ | s_m^i \tilde{b}_m \rangle \rangle \}$. 
We can then formulate the 
continuous-time dynamics in terms of the equations of motion for the embedded wavefunction and the bath projectors (or equivalently, the bath density matrix). 
The dynamics of such a bath density matrix has previously been considered in the MPS language in Ref.~\cite{PhysRevA.91.032306} for a 1D spin chain, which derived the dissipative contribution of the system-bath coupling to the density matrix dynamics.

We will be interested in the above constructions for the case of a fermionic Gaussian bath where we can replace the discussion of many-body states and density matrices with orbitals and one-particle reduced density matrices (1-RDM). We now turn to the formulation of the IF-MPS operations in terms of these quantities.

\subsection{Schmidt decomposition of Slater determinants}\label{sec:SD_schmidt}

In this section, we review the Schmidt decomposition of Slater determinants \cite{Peschel2012, doi:10.1021/acs.jctc.6b00316}. A Slater determinant is given by
\begin{equation}\label{eq:SD_exp}
\ket{\psi}=\prod_{p=1}^{N_{\text{occ}}} \hat{c}_{p}^{\dag}\ket{0}, \quad \hat{c}^{\dag}_p = \sum_i C_{ip} \hat{a}^{\dag}_i,
\end{equation}
where $\hat{c}^{\dag}_p$ is a creation operator of the $N_{\text{occ}}$ occupied orbitals, $\hat{a}^{\dag}_{i}$ is a creation operator of orthonormal orbitals in the basis of $n$ sites, and $C_{ip}$ is the orbital coefficient matrix. Given a bipartite Hilbert space, $\mathcal{H}=\mathcal{H}_A \otimes \mathcal{H}_B$, where the first $n_A$ orbitals belong to subsystem A and the other $n_B = n - n_A$ orbitals belong to subsystem B, the Schmidt decomposition of the Slater determinant can be obtained by diagonalizing the one-particle reduced density matrices (1-RDM), $\Gamma_{ij} = \bra{\psi} \hat{a}^{\dag}_j \hat{a}_i \ket{\psi}$, of the subsystems. Assuming $n_A < n_B$ and $N_{\text{occ}}>n_A$, the Schmidt decomposition can be written as
\begin{equation}\label{eq:SD_Schmidt}
    \ket{\psi} = \prod_{k=1}^{n_A} \left( \sqrt{\nu_k} \hat{c}^{\dag}_{A,k} +
    \sqrt{1-\nu_k} \hat{c}^{\dag}_{B,k} \right) \prod_{l=n_A+1}^{N_{\text{occ}}} \hat{c}^{\dag}_{B,l} \ket{0},
\end{equation}
where $\nu_k$ ($1-\nu_k$) denote the eigenvalues of the 1-RDM of $A$ ($B$) with values between 0 and 1, $\hat{c}^{\dag}_{A,k}$ ($\hat{c}^{\dag}_{B,k}$) create the corresponding eigenmodes, and $\hat{c}^{\dag}_{B,l}$ create eigenmodes with eigenvalue 1 of the 1-RDM of $B$.

Based on the above, the orbitals in $B$ can be classified into three different categories: (1) \textit{entangled} orbitals, $\hat{c}^{\dag}_{B,k}$, $1 \leq k \leq n_A$, which are entangled with $A$, (2) \textit{core} orbitals, $\hat{c}^{\dag}_{B,l}$, $n_A +1 \leq l \leq N_{\text{occ}}$, which are fully occupied in $B$ and not entangled with $A$, (3) \textit{virtual} orbitals, which are unoccupied and so do not appear in Eq.~\ref{eq:SD_Schmidt} and also are not entangled with $A$. 

Note that the Slater determinant with all $\nu_k = \frac{1}{2}$ corresponds to a maximally entangled fermionic state $\ket{\phi}$, where the reduced density operator of subsystem $A$ is proportional to the identity. It is possible to write the 1-RDM of $A$ from $\ket{\psi}$ as that of a maximally entangled fermionic state $\ket{\phi}$ after a gauge transformation within the subsystem $A$, i.e., $\hat{G}_{A} \ket{\phi}$, where
\begin{equation}
    \hat{G}_A = \text{exp}\left(\sum_k \log g_k \hat{c}_{A,k}^{\dag} \hat{c}_{A,k}\right), \quad g_k = \sqrt{\frac{\nu_k}{1-\nu_k}}, 
    \label{eq:gauge_maxentangled}
\end{equation}
up to a normalization constant factor, assuming $0<\nu_k<1$ for all $k$. $\hat{G}_A$ satisfies the following:
\begin{equation}
    \hat{G}_A \hat{c}^{\dag}_{A,k} \hat{G}_A^{-1} = g_k \hat{c}^{\dag}_{A,k}, \: \hat{G}_A \hat{c}_{A,k} \hat{G}_A^{-1} = g_k^{-1} \hat{c}_{A,k}.
\end{equation}
This gauge transformation is related to the gauge transformation introduced in Sec.~\ref{sec:IF_to_statevec} because the state $\ket{\phi}$ is ``left-normalized" with respect to the subsystem $A$. We will use this gauge transformation to convert  Slater determinants into left-normalized forms in the next section.

The Schmidt decomposition can be truncated by treating the entangled orbitals with $\nu_k \approx 1$ ($\nu_k \approx 0$) as core (virtual) orbitals, retaining the orbitals with larger $\sqrt{\nu_k(1-\nu_k)}$. In other words, a projection operator on the entangled orbital space that keeps only $\nu_k$ close to $\frac{1}{2}$ can be applied to truncate the Schmidt decomposition. This truncation scheme is sometimes called a ``mode'' truncation~\cite{PhysRevB.105.L081101,PhysRevB.92.075132, PhysRevB.100.245121} and has been utilized in the context of tensor network truncations of fermionic Gaussian states.

After a truncation to $n_{\text{ent}}$ entangled orbitals, the Slater determinant can be written as
\begin{multline}
    \ket{\psi} = \prod_{k=1}^{n_{\text{ent}}} \left( \sqrt{\nu_k} \hat{c}^{\dag}_{A,k} +
    \sqrt{1-\nu_k} \hat{c}^{\dag}_{B,k} \right) \\ \times \prod_{l=n_{\text{ent}}+1}^{n_{\text{ent}}+n_{A,c}} \hat{c}^{\dag}_{A,l}  \prod_{l=n_{\text{ent}}+1}^{n_{\text{ent}}+n_{B,c}} \hat{c}^{\dag}_{B,l} \ket{0},
\end{multline}
where $n_{A,c}$ ($n_{B,c}$) is the number of core orbitals in $A$ ($B$).

\subsection{Influence functional tensors as Slater determinants}

\label{sec:ifsd}

We now describe how the bipartitions of the exact influence functional tensor can be expressed as Slater determinants, which are obtained by propagating a finite number of steps forward in time from $\kett{\rho_B(0)}$ or backward in time from $\braa{\tr_B}$. Thanks to the noninteracting nature of the bath, both the IF and its partitions correspond to fermionic Gaussian states, specifically, Bardeen-Cooper-Schrieffer (BCS) states~\cite{PhysRevB.107.195101,PhysRevB.107.L201115, PhysRevB.107.125103}. Despite this fact, we will prefer to work with Slater determinants and convert the BCS states to Slater determinants through a particle-hole transformation. This is because when building the matrix elements of the IF-MPS in Eq.~\ref{eq:eff_timeevo}, the basis of Slater determinants will allow us to use number symmetry, which significantly reduces the prefactors in the numerical computations.

\begin{figure}
    \centering
    \includegraphics[width=0.85\columnwidth]{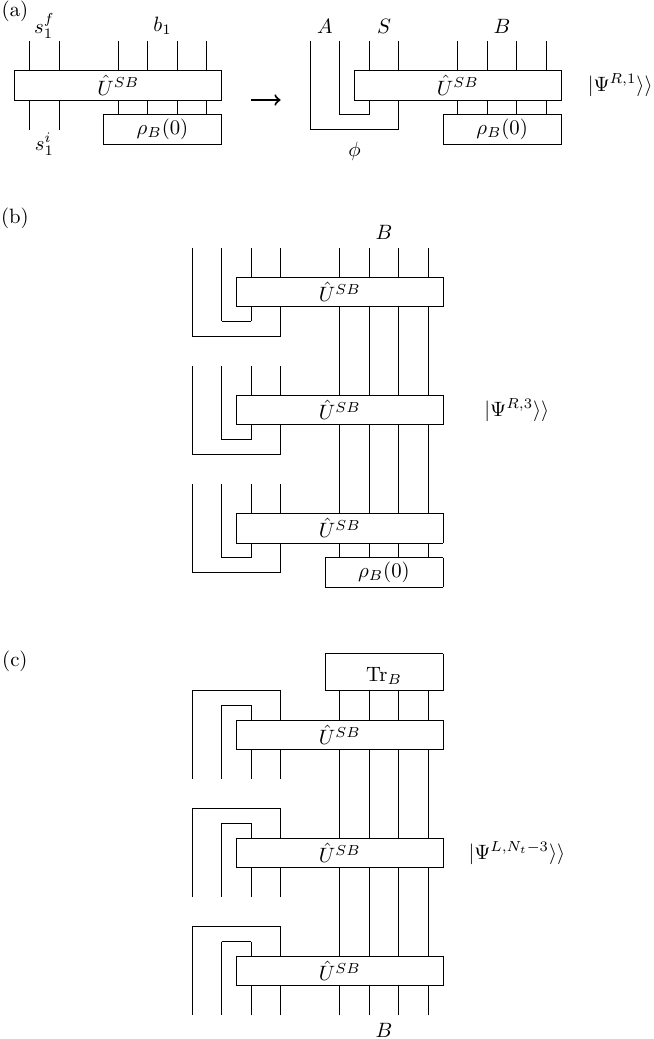}
    \caption{(a) A right IF state after one time step $\kett{\Psi^{R,1}}$. It represents a partial contraction of the bath degrees of freedom between the time-evolution operator $\hat{U}^{SB}$ and the initial bath density operator $\rho_B(0)$. It can be expressed as a Slater determinant by introducing a maximally entangled fermionic state, $\phi$, which has impurity and auxiliary fermionic orbitals denoted by $S$ and $A$, respectively. Coupling the maximally entangled state to the time-evolution operator on the input impurity orbitals yields the Slater determinant state. (b) A right IF state after the $m$-th time step $\kett{\Psi^{R,m}}$, ($m=3$ in the figure) can also be represented as a Slater determinant by introducing maximally entangled fermionic states at each time step. (c) A left IF state $\kett{\Psi^{L,m}}$ ($m=N_t-3$ in the figure) can be constructed by contracting the time-evolution operator from the top downwards, and can also be represented as a Slater determinant by inserting maximally entangled fermionic states starting from the top.}
    \label{fig:IF_SD_explicit}
\end{figure}

We define the right and left bipartitions of IF (right and left IF for short) $\prod_m \hat{U}_m^{SB} \cdot \kett{\rho_B(0)}$ and $\braa{\tr_B} \cdot \prod_m \hat{U}_m^{SB}$, respectively, where the products follow the order in the MPS representation and here, $\cdot$ indicates the partial contraction of the intermediate bath configurations.
We start with the first right IF, $\hat{U}^{SB} \cdot \kett{\rho_B(0)} $ for simplicity. Its tensor elements are determined from $\braa{s_1^f, b_1} \hat{U}^{SB} (\kett{s_1^i} \otimes \kett{\rho_B(0)})$, and its external degrees of freedom are $s_1^i$, $s_1^f$, and $b_1$ (see Fig.~\ref{fig:IF_SD_explicit}(a)). Its occupation numbers satisfy the relationship
\begin{equation}
    n(s_1^f) + n({b_1}) - n({s_1^i}) = n({\rho_B(0)}),
\end{equation}
which is not consistent with a number-conserving state.

After applying the particle-hole transformation to the input configuration, $s_1^i \rightarrow \Bar{s}_1^i$, (for example, $00 \rightarrow 11$, $01 \rightarrow 10$, $10 \rightarrow 01$, and $11 \rightarrow 00$), we obtain
\begin{equation}
    n(s_1^f) + n({b_1}) + n({\Bar{s}_1^i}) = n({\rho_B(0)}) + 2,
\end{equation}
where $n(\Bar{s}^i_1) = 2 - n(s^i_1)$. Therefore, after the transformation, the right partition IF is a Slater determinant with $n({\rho_B(0)}) + 2$ occupied orbitals.

Formally, the particle-hole transformations can be expressed in terms of fermionic tensor network contractions by inserting maximally entangled fermionic states,
\begin{equation}
    \kett{\phi} = \prod_s \frac{1}{\sqrt{2}} (\hat{\alpha}_s^{\dag}+\hat{d}_s^{\dag})\kett{0},
\end{equation}
where the index $s$ denotes the input impurity orbitals from the super-Fock Liouville space and $\hat{\alpha}_s^{\dag}$ indicates a creation operator of auxiliary fermionic orbitals. Note that the occupation numbers of the auxiliary fermionic orbitals have particle-hole transformed occupation numbers compared to the original input impurity orbitals. The right IF state representation of $\hat{U}^{SB} \cdot \kett{\rho_B(0)} $, which we will denote  $\kett{\Psi^R} \equiv \kett{\Psi^{R,1}}$, is then written as follows (Fig.~\ref{fig:IF_SD_explicit}a),
\begin{equation}
    \kett{\Psi^{R,1}} = \left( \hat{I}^{A} \otimes \hat{U}^{SB} \right) \kett{\phi} \otimes \kett{\rho_B(0)},
\end{equation}
where $\hat{I}^A$ is the identity operator on the auxiliary fermionic orbitals. Because the initial state $\kett{\phi} \otimes \kett{\rho_B(0)}$ is given by a Slater determinant and $\hat{I}^A \otimes \hat{U}^{SB}$ is number conserving, we see again that the right IF state is also a Slater determinant. 
The right IF state for $\prod_m \hat{U}_m^{SB} \cdot \kett{\rho_B(0)}$ can be similarly expressed by inserting maximally entangled fermionic states at each time step and coupling one of the modes to $\hat{U}^{SB}$, as shown schematically in Fig.~\ref{fig:IF_SD_explicit}b.

\begin{figure*}[bt]
    \centering
    \includegraphics[width=\textwidth]{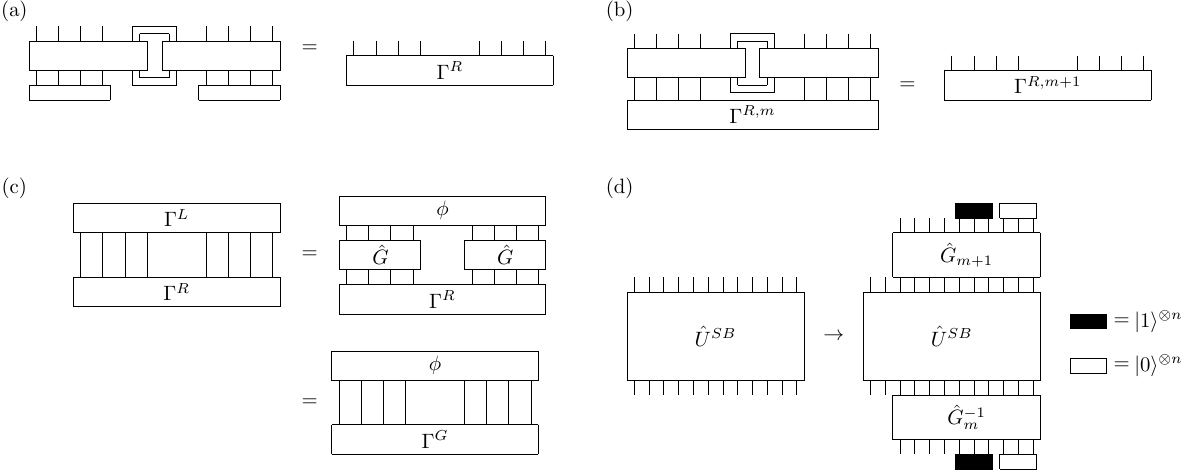}
    \caption{(a) The bath 1-RDM of the right IF state, $\Gamma^{R}$, is obtained by tracing out the impurity orbitals, i.e., by contracting the state with its complex conjugate, within the impurity space. (b) The bath 1-RDM of the right IF state after the $m$-th time step, $\Gamma^{R,m}$ can be updated to $\Gamma^{B,m+1}$ by applying the time-evolution operator and tracing out the impurity orbitals. (c) A gauge transformation, $\hat{G}$, is extracted from the bath 1-RDM of the left IF state, $\Gamma^{L}$, by converting it to a maximally entangled fermionic state, $\phi$. The gauge-transformed right IF state is obtained by applying the gauge transformation, $\hat{G}$, to the bath, and its bath 1-RDM is transformed to $\Gamma^{G}$. (d) The impurity-bath time-evolution operator, $\hat{U}_{SB}$, is approximated by an effective time-evolution operator by projecting the bath to the effective bath orbitals after the gauge transformation, $\mP_{m+1} \hat{G}_{m+1}\hat{U}^{SB}\hat{G}_m^{-1} \mP_m$. The filled and unfilled rectangles represent the core and virtual orbital spaces and these spaces are projected into fully occupied and unoccupied states, respectively.}
    \label{fig:Gamma_fig}
\end{figure*}

To construct the compression of the IF-MPS later, we require the bath 1-RDMs. Given the right IF state at the $m$-th time step $\kett{\Psi^{R,m}}$, we define the bath 1-RDM of the right IF state (Fig.~\ref{fig:Gamma_fig}a) as
\begin{equation}
    \Gamma^{R,m}_{ij} = \frac{\braa{\Psi^{R,m}} \hat{a}^{\dag}_j \hat{a}_i \kett{\Psi^{R,m}}}{\braakett{\Psi^{R,m}}{\Psi^{R,m}}},
\end{equation}
where the indices $i$ and $j$ refer to the super-Fock bath orbitals. The 1-RDM at the next time step can be computed from the evolution of $\kett{\phi}\otimes \kett{\Psi^{R,m}}$ under $\hat{I}^A \otimes \hat{U}^{SB}$. The bath 1-RDM after the evolution can be written as
\begin{equation}\label{eq:gammaB_update}
    \Gamma^{R,m+1}_{ij} = \left[U^{SB} \Gamma^{R,m} U^{SB\dag} \right]_{ij} + \left[U^{ASB} \Gamma^{\phi} U^{ASB\dag} \right]_{ij},
\end{equation}
where $\Gamma^{\phi}$ denotes the 1-RDM of $\kett{\phi}$ and $U^{ASB}$ denotes a single particle-basis representation of $\hat{I}^A\otimes\hat{U}^{SB}$. The diagrammatic representation of the evolution of the bath 1-RDM is drawn in Fig.~\ref{fig:Gamma_fig}b. 

We can construct the left IF state in an analogous way, but where the state propagates in the inverse (negative) time direction using $\hat{U}^{SB \dag}$ (Fig.~\ref{fig:Gamma_fig}c). Denoting the left IF state and its bath 1-RDM, $\kett{\Psi^L}$ and $\Gamma^L$, respectively, the bath 1-RDMs at successive time steps (from top downwards) are related by 
\begin{equation}\label{eq:gammaL_update}
    \Gamma^{L,m-1}_{ij} = \left[U^{SB\dag} \Gamma^{L,m} U^{SB} \right]_{ij} + \left[U^{ASB\dag} \Gamma^{\phi} U^{ASB} \right]_{ij}.
\end{equation}
Note that the initial state is given by the trace vector $\kett{\Psi^{L,N_t}} = \kett{\tr_B}$, which is also given by a Slater determinant as in Eq.~\ref{eq:trace_vec}. With both the left and right IF state, the IF-MPS can be written as $\braa{\Psi^{L,m}}\cdot\kett{\Psi^{R,m}}$ for any $m$.

\subsection{Influence functional matrix product state compression}
\label{sec:mps_compression}

We now revisit the MPS compression of the IF-MPS, described in Sec.~\ref{sec:IF_to_statevec}, but now utilizing the noninteracting nature of the bath, which allows all the steps to be expressed at the level of orbitals and single-particle quantities.

In Sec~\ref{sec:IF_to_statevec}, the optimal compression of the MPS required the IF-MPS to be in a canonical form. This was achieved by transforming matrices into left-normalized matrices by inserting gauge matrices into the MPS. As discussed in Sec.~\ref{sec:ifsd} the partitions of the IF-MPS for a noninteracting bath are Slater determinants, and the gauge transformation (Eq.~\ref{eq:gauge_maxentangled})  to convert Slater determinants to maximally entangled fermionic pairs, which are left-normalized, was introduced in Sec.~\ref{sec:SD_schmidt}.

We therefore have all the ingredients to convert the IF-MPS to canonical form. We start with the left IF state $\kett{\Psi^L}$ and determine the gauge transformation $\hat{G}$ from the eigenvalues and eigenvectors of the bath 1-RDM $\Gamma^{L}_{ij}$. We denote the eigenvalues of $\Gamma^{L}_{ij}$ as $\nu_k$, where $k$ indexes the eigenvalues, and the rotation matrix as $R^{L}_{ik}$, whose columns are the eigenvectors of $\Gamma^L$ in a single-particle basis. With this gauge transformation, we can represent the IF-MPS as, $\braa{\Psi^L} \cdot \kett{\Psi^R} = \braa{\phi}\cdot \hat{G} \cdot \kett{\Psi^R}$. The gauge matrix in the single-particle basis, $G$, can be written as
\begin{equation}
    G_{ki} = g_k R^{L*}_{ik},
\end{equation}
where $g_k = \sqrt{\nu_k/(1-\nu_k)}$ (Eq.~\ref{eq:gauge_maxentangled}). Note that $g_k$ diverges when $\nu_k \to 1$, so in practice, we regularize $\nu_k$ with a small threshold $\varepsilon$ so that $\nu_k = \varepsilon$ when $\nu_k < \varepsilon$ and $\nu_k = 1 - \varepsilon$ when $1-\nu_k < \varepsilon$. A similar regularization scheme for the 1-RDM has been used in multiconfiguration time-dependent Hartree theory \cite{10.1063/1.463007,MEYER199073} and real-time density matrix embedding theory \cite{doi:10.1063/1.5012766,10.1063/5.0146973}.

The gauge matrices need to be absorbed into the right IF state to further canonicalize the IF-MPS. 
We call the state, $\kett{\Psi^G} = \hat{G} \cdot \kett{\Psi^R}$, a gauge-transformed right IF state and its bath 1-RDM, $\Gamma^G$. The gauge transformation is a nonunitary transformation, so we always normalize the state when computing $\Gamma^G$.
\begin{equation}\label{eq:gamma_tilde}
    \begin{gathered}
        {\Gamma}^G_{ij} = \frac{\braa{\Psi^G}\hat{a}^{\dag}_j \hat{a}_i \kett{\Psi^G}}{\braakett{\Psi^G}{\Psi^G}}  \\
    =\left[G\sqrt{\Gamma^{R}}
(\sqrt{\Gamma^R}G^{\dag}G\sqrt{\Gamma^R}+I-\Gamma^R)^{-1}\sqrt{\Gamma^{R}}G^{\dag}\right]_{ij}
    \end{gathered}
\end{equation}
where $\Gamma^R$ is the bath 1-RDM of the $\kett{\Psi^R}$. A detailed derivation of this expression is included in the Supplemental Material (SM).

Subsequently, we decompose the gauge-transformed right IF state into a right-normalized maximally entangled fermionic state and another gauge transformation, and this gauge transformation contains the singular values of the IF-MPS at this bipartition. Taking the largest singular values in the truncated SVD in the conventional MPS compression is equivalent to taking the eigenvectors of $\Gamma^G$ with eigenvalues closest to $\frac{1}{2}$ (or large $\nu_k^G(1-\nu_k^G)$) as done in the mode truncation approach introduced in Sec.~\ref{sec:SD_schmidt}. We will call the bath orbitals from the selected eigenvectors of $\Gamma^G$ the effective bath orbitals.

Hence, the procedure for IF-MPS compression at the orbital level is as follows: (1) Diagonalize the bath 1-RDM of the gauge-transformed right IF state $\Gamma^G$ and obtain its eigenvalues $\nu_k^G$, and rotation matrix $R^G$. (2) Take the eigenvectors with the $N_{\eff}$ largest $\nu_k^G(1-\nu_k^G)$, where $N_{\eff}$ is a number of the effective bath orbitals we select, defining the truncated rotation matrix $R^\text{eff}$.  (3) Construct the projection operator $\mP$ from $R^\text{eff}$, i.e., the eigenvectors corresponding to the effective bath orbitals. The configurations from the other bath orbitals in the projected subspace are fixed to be either fully occupied (core orbitals, $\nu_k^G \approx 1$) or unoccupied (virtual orbitals, $\nu_k^G \approx 0$).

For the Liouville time evolution of the Anderson impurity model with the initial Gaussian thermal bath, it is possible to prove that if $\nu_k^G$ is an eigenvalue of ${\Gamma}^G$, so is $1-\nu_k^G$ (see the SM). In this case, we obtain the same number of core and virtual orbitals and choose the effective bath orbitals symmetrically by occupancy.

The tensor elements of the IF-MPS can then be computed from Eq.~\ref{eq:eff_timeevo}, after applying the gauge transformation and the projectors $\mP_m$ for each $m$-th time step, 
\begin{equation}\label{eq:el_timeevo_gauge}
    \braa{s_m^f,\Tilde{b}_{m}} \mathcal{P}_m \hat{G}_{m} \hat{U}^{SB} \hat{G}_{m-1}^{-1} \mathcal{P}_{m-1} \kett{s_m^i,\Tilde{b}_{m-1}},
\end{equation} 
with the configurations of the effective bath orbitals $\Tilde{b}_m$. This tensor element defines an effective time-evolution operator for the embedded wavefunction defined on the impurity and effective bath orbitals. The diagrammatic representation for the effective time-evolution operator is illustrated in Fig.~\ref{fig:Gamma_fig}d. 

The above tensor elements can be efficiently computed from determinant formulas. Since the configurations in the core and virtual orbitals are fixed, we can compute the determinant of block matrices, keeping the core and virtual orbital block matrices fixed. Therefore, the computational complexity to compute determinants for all configurations is $\mathcal{O}(N_{b}^{3} + 2^{2N_{\eff}}N_{\eff}^{3})$, where the first term corresponds to the computation of the determinant and inverse of the core and virtual orbital block matrices and the second term corresponds to the computation of the determinant of the effective bath orbital block matrices. The cost for computing the full set of MPS tensor elements at time step $N_t$ is  $\mathcal{O}(N_{b}^{3}N_t + 2^{2N_{\eff}}N_{\eff}^3N_t)$, which is \textit{linear} in the number of time steps $N_t$.

\section{Continuous-Time Formulation of IF-MPS}\label{sec:sec3}

\subsection{Boundary influence functional tensor}
\label{sec:boundaryif}

Defining the continuous-time limit of the influence functional tensor network is in principle one way to eliminate the time step error from the standard second-order Trotter decomposition, and in numerical applications allows for the introduction of a wide variety of higher-order differential integrators.
However, as shown in \cite{SONNER2021168677}, the IF-MPS shows a nonphysical entanglement entropy scaling 
in the limit of $\Delta t \to 0$, as the entanglement entropy always scales to zero. This suggests that the continuous-time limit requires a more careful treatment. 

In particular, the formalism of continuous matrix product states (cMPS) \cite{PhysRevLett.104.190405,PhysRevLett.118.220402} describes a quantum wavefunction of continuous variables that (in general) supports a finite entanglement entropy as the discretization approaches the continuum limit.  
In this section, we show that the usual IF-MPS does not support a standard cMPS representation in the continuous-time limit,
and instead a closely related object, the \textit{boundary} influence functional MPS should be used.
The boundary influence functional MPS is implicitly used in transverse contraction ~\cite{Pirvu_2010, Banuls2009} (e.g., Ref.~\cite{PhysRevA.91.032306} states that this becomes a continuous MPS in the continuum limit, without providing an explicit construction) and has previously been used in influence functional calculations with interacting baths~\cite{10.1063/5.0047260}.

Consider the continuous-time limit $\Delta t \to 0$, where $\hat{U}^{SB}$ can be expressed as
\begin{equation}\label{eq:U^SB}
    \hat{U}^{SB} = \hat{I} - i(\hat{L}_{SB}+\hat{L}_{B}) \Delta t,
\end{equation}
and the corresponding matrix elements of the IF-MPS $A^{s_i,s_f}$ are,
\begin{equation}\label{eq:A_IFMPS}
    \begin{gathered}
        A^{0,0} = I - i\hat{L}_B \Delta t, \\
        A^{0,1} = A^{1,0} = -i \hat{L}_{SB} \Delta t, \\
        A^{1,1} = I - i\hat{L}_B \Delta t.
    \end{gathered}
\end{equation}
These do not have the same form as the tensor entries in a cMPS~\cite{PhysRevLett.104.190405,PhysRevLett.118.220402}, which take the following form:
\begin{equation}\label{eq:contMPS}
    \begin{gathered}
        A^{0} = I + \epsilon Q, \\
        A^{1} = \sqrt{\epsilon} R, \\
        A^{2} = \frac{1}{2} \epsilon R^2,
    \end{gathered}   
\end{equation}
where $Q$ and $R$ are arbitrary matrices within the virtual bond space of the MPS, $\epsilon$ is the infinitesimal interval corresponding here to $\Delta t$ on the temporal axis, and the upper indices 0, 1, and 2 label the number of excitations in the physical bond.

There are two main differences between Eq. ~\ref{eq:A_IFMPS} and Eq. ~\ref{eq:contMPS}. First, the IF has an additional ``$I$" term in $A^{1,1}$, that is not in $A^2$ in Eq.~\ref{eq:contMPS}. Second, the $A^1$ terms are proportional to $\Delta t$, not $\sqrt{\Delta t}$. It is clear to see the effect of these two differences in the formulation of the differential equation of motion for the 1-RDM $\Gamma^R$. In the infinitesimal limit of Eq.~\ref{eq:gammaB_update}, after expanding $U^{SB}$ the first term becomes \begin{equation}\label{eq:gammaB_update_dt}
    \left[ U^{SB}\Gamma^{R,m} U^{SB \dag} \right]_B = \Gamma^{R,m} - i[L_B,\Gamma^{R,m}]\Delta t.
\end{equation}
For the second term,
\begin{equation}\label{eq:gamma_second}
    \left[U^{ASB} \Gamma^{\phi} U^{ASB\dag} \right]_B = - \frac{1}{2} L_{SB} L_{SB}^{\dag} \Delta t^2,
\end{equation}
which vanishes at $\Delta t \rightarrow 0$. Therefore, the differential equation of motion for $\Gamma^R$ becomes
\begin{equation}\label{eq:gammaB_eom0}
    \frac{d\Gamma^R}{dt} = -i [{L}_B,\Gamma^R].
\end{equation}
This corresponds to unitary dynamics in the super-Fock bath orbital space, which preserves the spectrum of $\Gamma^R$. The initial $\Gamma^R$ is given by the pure state $\kett{\rho_B}$, so its single-particle spectrum consists of only 0 and 1. Thus, the entanglement entropy of the IF is zero. Note that this result is the manifestation of the fact that the term in Eq.~\ref{eq:gamma_second} is proportional to $(\Delta t)^2$, instead of $(\Delta t)^1$, which implies that the bath dynamics does not have effective ``dissipation" terms.

To obtain a more proper continuum limit, we split $\hat{U}^{SB}$ into two parts. For clarity, we write $\hat{I}$ in Eq.~\ref{eq:U^SB} as $\hat{I}_S \otimes \hat{I}_B$. Similarly, $\hat{L}_B$ can also be written in product form, $\hat{I}_S \otimes \hat{L}_B$, and $\hat{L}_{SB}$ is the only term that acts on both the impurity and bath. We can split $\hat{L}_{SB}$ using its singular value decomposition,
\begin{align}
    \hat{L}_{SB} &= \sum_{s,i} t_{si} \hat{d}_s^{\dag} \hat{a}_i + \hc \nonumber \\
    &= \sum_{a=1}^{n_S} \hat{O}_{a}^S \otimes \hat{O}_{a}^B + \hc,
\end{align}
\begin{equation}
    \begin{gathered}
        t = USV = (US^{1/2})(S^{1/2}V) = t^{S} t^{B},  \\
    \hat{O}_{a}^S = \sum_{s} t^{S}_{sa}  \hat{d}_s^{\dag}, \quad \hat{O}_{a}^B = \sum_{i} t^{B}_{ai} \hat{a}_i,
    \end{gathered}
\end{equation}
where $n_S$ is the number of singular values. This leads us to write $\hat{U}^{SB}$ as follows:
\begin{equation}
    \begin{gathered}
        \hat{U}^{SB}   = \: \hat{I}_S \otimes(\hat{I}_B -i \hat{L}_B \Delta t)  \\
       +  \sum_{a=1}^{n_S} (\hat{O}_a^S \sqrt{\Delta t}) \otimes (-i\hat{O}_a^B \sqrt{\Delta t}) + \hc \\
       = \: \sum_{a=0}^{2n_S} \: \hat{W}^{S}_a \otimes \hat{W}^{B}_a  
    \end{gathered}
\end{equation}
\begin{equation}
    \begin{gathered}
        \hat{W}^S_a = 
        \begin{cases}
            \hat{I}_S & a=0 \\
            \hat{O}^S_a \sqrt{\Delta t} & 1 \le a \le n_S \\
            \hat{O}^{S \dag}_{a-n_S} \sqrt{\Delta t} & n_S+1 \le a \le 2n_S,
        \end{cases}
        \\
        \hat{W}^B_a = 
        \begin{cases}
            \hat{I}_B - i\hat{L}_B \Delta t & a=0 \\
            -i\hat{O}^B_a \sqrt{\Delta t} & 1 \le a \le n_S \\
            -i\hat{O}^{B \dag}_{a-n_S} \sqrt{\Delta t} & n_S+1 \le a \le 2n_S.
        \end{cases}
    \end{gathered}
\end{equation}

\begin{figure}
    \centering
    \includegraphics[width=\columnwidth]{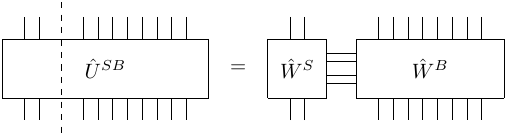}
    \caption{The time-evolution operator, $\hat{U}^{SB}$ can be split into two tensors, $\hat{W}^{S}$ and $\hat{W}^B$, which can be expressed as $\hat{U}^{SB}=\sum_a \hat{W}^S_a \otimes \hat{W}^B_a$. In the single impurity case with two impurity orbitals in the Liouville space, the index $a$ has four components.}
    \label{fig:U_split}
\end{figure}

Therefore, instead of defining the elements of the MPS using $\hat{U}^{SB}$, we use $\hat{W}^B_a$. 
These are the tensor elements of the boundary IF-MPS. Its diagrammatic representation is drawn in Fig.~\ref{fig:U_split}. We can introduce a set of maximally entangled orbitals, $\hat{f}_a^{\dag}$, on the auxiliary indices $a$ thereby expressing $\hat{W}^B$ in a number-conserving format, 
\begin{equation}
    \hat{W}^B = \hat{I}_B - i\hat{L}_B \Delta t -i \sqrt{\Delta t} \sum_a \left( \hat{f}_a^{\dag} \hat{O}_a^B + \hc \right).
\end{equation}
The matrix elements of the boundary IF-MPS are then given by
\begin{equation}
    \begin{gathered}
        W^{B,0} = {I}_B - i{L}_B \Delta t, \\
        W^{B,1} = -i \sqrt{\Delta t} \sum_a \left( f_a^{\dag} O_a^B + \hc \right), \\
        W^{B,2} = \Delta t \sum_a |\hat{O}_a^B|^2, 
    \end{gathered}
\end{equation}
which satisfies the form of Eq.~\ref{eq:contMPS} \footnote{The factor of 1/2 difference from $W^{B,2}$ comes from the fact that Eq.~\ref{eq:contMPS} describes double excitations of indistinguishable bosonic particles whereas $W^{B,2}$ takes into account different auxiliary fermions.}. Therefore, the boundary IF-MPS has a well-defined continuum limit. It is also possible to construct $\hat{W}^B$ from $\hat{U}^{SB}$ for general $\Delta t$ outside the continuous-time limit, which is described in the SM. Henceforth, we will implicitly assume that the IF-MPS refers to the boundary IF-MPS except where the distinction is important.

\subsection{Gauge dynamics in the continuous-time IF-MPS}
\label{sec:eom}

The boundary IF-MPS matrix elements define a new effective time-evolution operator in the Liouville space of the system and bath with a well-defined continuous-time limit with nonvanishing entanglement entropy. This makes it possible to express and compress the MPS in the continuous-time limit.

As described in Sec.~\ref{sec:ifsd}, in the case of a noninteracting bath, the object that defines the conversion to canonical form and the compression is the 1-RDM of the right/left IF state, $\Gamma^R$ and $\Gamma^L$. Using 
$\hat{W}^B$, we can construct the corresponding iterative procedure to update the 1-RDM of the right/left IF state. In the continuous-time limit, $\Delta t \rightarrow 0$, this leads to an equation of motion for the 1-RDM of the right/left IF state.

The equation of motion for the 1-RDM of the right IF state $\Gamma^R$ is given by
\begin{equation}\label{eq:gamma_eom}
    \frac{d \Gamma^R}{dt} = t^{B \dag}t^B -i[{L}_B,\Gamma^R] - \{t^{B \dag}t^B,\Gamma^R\},
\end{equation}
where $\{A,B\}=AB+BA$ is an anticommutator. A detailed derivation of this equation of motion is in the SM. It clearly shows that $t^B$ acts as a dissipation term. Similarly, the equation of motion for the 1-RDM of the left IF state $\Gamma^{L}$ is given by
\begin{equation}\label{eq:gammatr_eom}
    \frac{d \Gamma^{L}}{dt} = t^{B \dag}t^B +i[{L}_B,\Gamma^{L}] - \{t^{B \dag}t^B,\Gamma^{L}\}.
\end{equation}

From $\Gamma^L$, we can find the gauge transformation $\hat{G}$. This is then absorbed into the right IF state to canonicalize the IF-MPS. For ${\Gamma}^G$, the 1-RDM of the gauge-transformed right IF state, we first define the gauge-transformed time-evolution operator
\begin{equation}\label{eq:WB_gauge}
    \begin{gathered}
        \hat{W}^G = \hat{G} \hat{W}^B \hat{G}^{-1} + \frac{d\hat{G}}{dt} \hat{G}^{-1} \Delta t =\\
        \hat{I}_B -i \hat{L}_{GB} \Delta t -i \sqrt{\Delta t} \sum_a \left( \hat{f}_a^{\dag} \hat{F}^{1}_a + \hat{F}^{2 \dag}_a \hat{f}_a \right), \\
        \hat{{L}}_{GB} = \hat{G} \hat{L}_B \hat{G}^{-1} + i \frac{d\hat{G}}{dt}\hat{G}^{-1}, \\
        \hat{F}_a^1 = \hat{G} \hat{O}_a^B \hat{G}^{-1} = \sum_i \kappa^{1}_{ai} \hat{a}_i, \\
        \kappa^{1} = t^B G^{-1}, \\
        \hat{F}_a^{2 \dag} = \hat{G} \hat{O}_a^{B \dag} \hat{G}^{-1} = \sum_i \kappa^{2}_{ia} \hat{a}^{\dag}_i, \\
        \kappa^{2} = G t^{B \dag}.
        \end{gathered}
\end{equation}
Note that in the definition of $\hat{L}_{GB}$, we have taken $\Delta t \to 0$, and there is a term that takes into account the time dependence of the gauge transformation. The time derivative of the gauge transformation can be written as,
\begin{equation}
    \begin{gathered}
        \left[\frac{d\hat{G}}{dt}\hat{G}^{-1}\right]_{kl} = \dot{g}_k g^{-1}_k \delta_{kl} -  \sum_i g_k R^{L *}_{ik} \dot{R}^{L}_{il} g^{-1}_l , \\
        \dot{g}_k g^{-1}_k  = \frac{1}{2\nu_k(1-\nu_k)} \dot{\nu}_k , \\
        \dot{\nu}_k  = - \left[R^{L \dag} \dot{\Gamma}^{L} R^{L} \right]_{kk} = -\sum_a |\sum_i t^{B}_{ai}R^{L}_{ik}|^2(1-2\nu_k), \\
        \left[R^{L\dag}\dot{R}^{L}\right]_{kl} = - \frac{\left[R^{L \dag} \dot{\Gamma}^{L} R^{L} \right]_{kl}}{\nu_l - \nu_k} \: (k \neq l) = \\
        \frac{1-\nu_k-\nu_l}{\nu_k-\nu_l}\left[R^{L \dag}t^{B\dag}t^{B}R^{L}\right]_{kl} +i\left[R^{L \dag}{L}_B R^{L}\right]_{kl},
    \end{gathered}
\end{equation}
where we set the $\dot{R}^{L}$ terms to be zero when $k=l$ and $|\nu_k - \nu_l| < \varepsilon$. 
Note that $R^L$ is only defined up to degenerate eigenvectors of $\Gamma^{L}$; therefore, to fix this redundancy, we propagate $R^{L}$ with the regularized $\dot{R}^{L}$ from a reference time, which we set to be the final time of the $\Gamma^{L}$ propagation. 

With this gauge-transformed time-evolution operator, the equation of motion for ${\Gamma}^G$ can be written as follows:
\begin{align}\label{eq:gammatilde_eom}
    \frac{d{\Gamma}^G}{dt} &= \kappa^2 \kappa^{2 \dag} -i ({L}_{GB} {\Gamma}^G - {\Gamma}^G {L}_{GB}^{\dag})-\{\kappa^2 \kappa^{2 \dag},{\Gamma}^G\} \nonumber \\
    &+ {\Gamma}^G \left[ \kappa^2 \kappa^{2 \dag} - \kappa^{1 \dag} \kappa^{1} +i({L}_{GB} - {L}_{GB}^{\dag})  \right]{\Gamma}^G.
\end{align}
After obtaining $\Gamma^G$ in this continuous-time picture, the effective bath orbitals and their rotation matrix $R^{\eff}$ (see Sec.~\ref{sec:ifsd}) can be defined from the eigenvectors of $\Gamma^G$. $R^{\eff}$ is then used to construct the time-dependent projection operator $\mathcal{P}$, which defines the compression of the IF-MPS in the continuous-time setting. 

\subsection{Embedding Liouville operator for the embedded wavefunction}
\label{sec:ctwavefunction}

Once we solve the equation of motion for ${\Gamma}^G(t)$, its truncated spectrum defines a set of effective bath orbitals at all (continuous) times $t$. By projecting the original Liouville dynamics into the time-dependent embedding space (of system and bath orbitals) and retaining terms first order in $\Delta t$, 
we can extract the generator for the embedded wavefunction, which we call here the embedding Liouville operator $\hat{L}^{\emb}$. This yields a continuous-time state vector propagation governed by the boundary IF-MPS.

First, by expanding Eq.~\ref{eq:el_timeevo_gauge}, we have $\mP \hat{{L}}_{G}\mP$ where $\mP$ is the projection operator into the effective bath orbitals and $\hat{{L}}_{G} = \hat{G}\hat{L}_{SB}\hat{G}^{-1}+\hat{{L}}_{GB}$. In addition, there is also a term for the time dependence of the effective bath orbitals in the projection operator. Using the rotation matrix of the effective bath orbitals $R^{\eff}$, the additional Liouville operator from this time dependence can be written as
\begin{equation}\label{eq:hatX}
    \hat{X} = -i \sum_{m,n} \left[R^{\eff \dag} \dot{R}^{\eff} \right]_{mn}\hat{a}_{m}^{\dag} \hat{a}_{n},
\end{equation}
where $m$ and $n$ index the effective bath orbital basis. Finally, the impurity-only Liouville operator can be simply added. The embedding Liouville operator becomes
\begin{equation}\label{eq:embedding_liouville}
    \hat{L}^{\emb} = \hat{L}_S + \mP\hat{{L}}_{G}\mP + \hat{X} = \hat{L}_S + \hat{L}^{\emb}_{SB}.
\end{equation}
We can now write down the equation of motion for the embedded wavefunction $\kett{\Psi^{\emb}}$ with which we can carry out state vector propagation,
\begin{equation}\label{eq:eom_emb}
    i \frac{d}{dt} \kett{\Psi^{\emb}} = \hat{L}^{\emb} \kett{\Psi^{\emb}}.
\end{equation}
Note that the embedding Liouville operator $\hat{L}^{\emb}$ is time-dependent.

\subsection{Connections to embedding theories}
\label{sec:embed}

We conclude this section by explicitly connecting to the formalism of real-time density matrix embedding theory (real-time DMET)~\cite{doi:10.1063/1.5012766,10.1063/5.0146973} and the closely related time-dependent complete-active-space self-consistent-field method (TD-CASSCF)~\cite{PhysRevA.88.023402, 10.1063/1.3600397}.  In particular, we focus on real-time DMET since the wavefunction ansatz in real-time DMET has the same form as the embedded wavefunction here~\cite{doi:10.1063/1.5012766},
\begin{equation}
    \ket{\Psi(t)} = \sum_{s,b} \psi_{s,b}(t) \ket{s} \otimes \ket{b} = \sum_{s,b} \psi_{s,b}(t) \ket{s,b}, \label{eq:dmetansatz}
\end{equation}
where $s$ and $b$ are configurations in impurity and bath, respectively, and $\psi_{s,b}(t)$ are the corresponding time-dependent amplitudes. The selection of a finite number of effective bath orbitals limits possible bath configurations, and the effective bath orbitals are allowed to be time-dependent. 

In the IF-MPS, the effective bath orbitals were determined by the bath 1-RDM. In contrast, in real-time DMET, the effective bath orbitals are determined by the time-dependent variational principle (TDVP), assuming the embedding wavefunction ansatz (\ref{eq:dmetansatz}). The equation of motion from TDVP is
\begin{gather}
    i \dot{\psi}_{s,b} =  \bra{s,b} \left(\hat{H} + \hat{X}\right) \ket{\Psi} , \label{eq:dmet_eom} \\
    \ket{\dot{b}} = i \hat{X} \ket{b},
\end{gather}
where $\hat{X}$ is a quadratic Hermitian operator that describes the time dependence of the effective bath orbitals, which has the same form as Eq.~\ref{eq:hatX}, but here, its elements are determined from the TDVP equations. The equation of motion in Eq.~\ref{eq:dmet_eom} defines the embedding Hamiltonian $\hat{H} + \hat{X}$ which corresponds to the embedding Liouville operator $\hat{L}^{\text{emb}}$ in Eq.~\ref{eq:embedding_liouville}. Aside from the different generators of the bath dynamics, a qualitative difference between real-time DMET and the embedding scheme that derives from the IF-MPS in this paper is the nonunitary nature of the gauge transformation $\hat{G}$. In the real-time DMET, the $\hat{X}$ operator implements unitary dynamics of the effective bath orbitals.

\section{Results}\label{sec:sec4}

\begin{figure*}[t]
    \centering
    \includegraphics[width=1.0\textwidth]{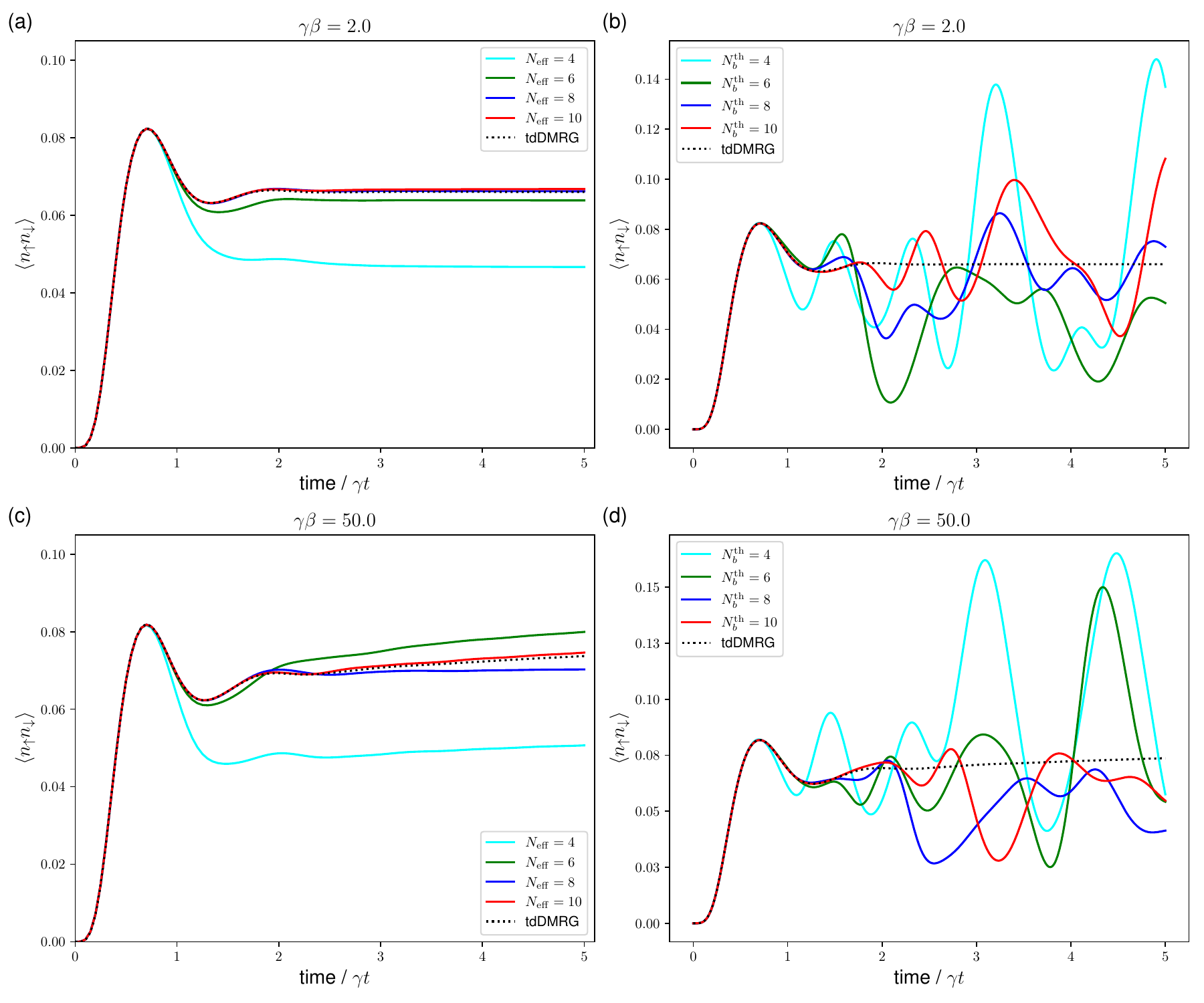}
    \caption{The time-dependence of the double occupancy of the impurity $\langle n_{\uparrow} n_{\downarrow} \rangle$ in the quench dynamics of the symmetric Anderson model with $U=2.5\pi\gamma$ and $\varepsilon_{\sigma}=-1.25\pi \gamma$. The impurity is quenched from an initial unoccupied state and a thermal bath with two different temperatures, $\gamma \beta = 2$ (a, b) and $\gamma \beta = 50$ (c, d). Reference results from time-dependent DMRG are shown (tdDMRG, black dotted). (a, c) Results from the discrete-time boundary IF-MPS using different numbers of effective bath orbitals,  $N_{\text{eff}}=$ 4 (cyan), 6 (green), 8 (blue), and 10 (red) with a time step of $\gamma \Delta t = 0.01$. (b, d) Results from the time-independent Lanczos-based bath discretization with thermofield transformation using a series of static bath discretizations, $N_b^{\text{th}} =$ 4 (cyan), 6 (green), 8 (blue), and 10 (red), where $N_b^{\text{th}}$ is the number of Lanczos vectors in the thermofield transformed bath. The finite size bath errors are clearly larger than in (a) and (c).}
    \label{fig:quench}
\end{figure*}

We now describe simulations that implement the above boundary IF-MPS formulation, including its continuous-time formulation, in the single impurity Anderson model. To generate reference results, we used
state vector propagation with a large bath discretization of the AIM. Specifically, we used
 $N_b = 40$ discrete bath orbitals ($N_b$ refers to orbitals of each spin, i.e., 40 spin $\uparrow$ and 40 spin $\downarrow$ orbitals) to approximate the bath spectral density, $J(\omega) = \sum_i |t_i|^2 \delta(\omega - E_i)$,  using a linear scheme~\cite{PhysRevB.92.155126, PhysRevB.90.235131}, as follows
\begin{equation}
    \begin{gathered}
        J(\omega) = \frac{\gamma}{\pi}\sqrt{1 - \frac{\omega^2}{W^2}}, \\
        |t_i|^2 = \int_{I_i} d \omega \: J(\omega), \\
        E_i = \frac{1}{|t_i|^2} \int_{I_i} d \omega \:
 \omega J(\omega),
    \end{gathered}
\end{equation}
where $W=10\gamma$, $\omega \in [-W,W]$, and $I_i = [-W+\frac{2W}{N_b}(i-1),-W+\frac{2W}{N_b}i]$, and the impurity Hamiltonian parameters were chosen as $U =-2\varepsilon_{\sigma}=2.5\pi\gamma$. The initial bath state was taken to be the (decoupled) thermal state with two different temperature regimes: 1) An intermediate-temperature regime with $\gamma \beta=2$ where the temperature is comparable to the other energy scales, and 2) a low-temperature regime with $\gamma \beta=50$, which is a lower temperature than the Kondo temperature, estimated to be $T_K \sim 0.07 \gamma $ \cite{Kohn_2022} ($\Gamma \beta_K = \gamma/T_K \sim 14$). The impurity was quenched from an initial unoccupied state, $\hat{\rho}_S(0)= \proj{0}$. Benchmark results were then computed using the time-dependent density matrix renormalization group (tdDMRG) with bond dimensions up to 300 using the Block2 \cite{10.1063/5.0050902, 10.1063/5.0180424} package. We carried out a thermofield transformation on the bath orbitals to reduce the bond dimension of the MPS, as described in \cite{Kohn_2022}. We then propagated the MPS using the two-site time-dependent variational principle. With the largest bond dimension 300 and a time step size of $\gamma \Delta t = 0.01$, all the local observables presented here are converged with respect to bond dimension and time step to within an estimated absolute error of $10^{-5}$.

The boundary IF-MPS calculations were carried out both in the standard discrete-time formulation as well as the continuous-time formulation of Secs.~\ref{sec:eom} and \ref{sec:ctwavefunction}. In the discrete-time case, we used the IF-MPS Slater determinant compression scheme applied to the boundary IF. The IF-MPS was constructed using both a second-order Trotter decomposition (Trotter2, Eq.~\ref{eq:Liou_evo}) and a 4th-order Trotter decomposition (Trotter4) of the dynamics. The 4th-order Trotter decomposition is based on the Forest-Ruth formula \cite{FOREST1990105, Jose-Garcia-Ripoll_2006},
\begin{gather}
    e^{-i\hat{L}\Delta t}  \approx e^{-i \hat{L}_S \theta \Delta t/2} e^{-i \hat{L}_{SB} \theta \Delta t} e^{-i \hat{L}_S(1-\theta)\Delta t/2} \nonumber \\
    e^{-i \hat{L}_{SB}(1-2 \theta)\Delta t} e^{-i \hat{L}_S(1-\theta)\Delta t/2} e^{-i \hat{L}_{SB} \theta \Delta t} e^{-i \hat{L}_S \theta \Delta t/2},
    \label{eq:trotter4}
\end{gather}
with the constant $\theta = 1/(2-2^{1/3})$ and a Trotter error of order $\mathcal{O}(\Delta t^5)$. Discrete-time boundary IF-MPS calculations were then performed for different numbers of effective bath orbitals $N_\text{eff}$ of each spin, which formally corresponds to a boundary IF-MPS with a maximal bond dimension of $4^{N_\text{eff}}$. Note, however, that the system-bath coupling in the Anderson model does not couple up and down spins. Consequently, the IF-MPS factorizes into a spin-up and spin-down IF (each of maximal bond dimension $2^{N_\text{eff}}$) and we use this factorization for a more efficient implementation of the discrete-time boundary IF-MPS. The maximum bond dimension of $2^{N_{\text{eff}}}$ can be further reduced by only choosing the available configurations from the number-conserving $U(1)$ symmetry of the embedded wavefunction. 

For the continuous-time implementation, we wrote the effective $\hat{L}^\text{emb}$ as a second-quantized fermionic operator which could then be used in state vector propagation using a higher-order integrator such as 4th-order Runge-Kutta. Note that to apply 4th-order Runge-Kutta to the state vector at time $t$, with time step $\Delta t$, requires $\hat{L}^{\emb}(t)$ at the intermediate time $t+\Delta t/2$. To obtain this, we propagated the bath 1-RDM equations of motion (Eqs.~\ref{eq:gamma_eom},~\ref{eq:gammatr_eom}) with a finer time step $\Delta t_{\text{fine}}$, also with the 4th-order Runge-Kutta integrator. Since the cost to propagate the bath 1-RDM is much lower than that for the many-body wavefunction, we used $\Delta t_{\text{fine}}$ much smaller than $\Delta t$ to minimize the time step error from the bath 1-RDM propagation. We fixed $ \gamma \Delta t_{\text{fine}} = 0.01 / 2^{6} \approx 0.00016$ in this paper.
We then used the quantum chemistry full configuration interaction (exact diagonalization) implementation in PySCF~\cite{sun2018pyscf,sun2020recent} 
to carry out the propagation of the state vector. 

One assumption for higher-order numerical propagators to be accurate is that the time-dependent embedding Liouville operator $\hat{L}^{\emb}$ is well-behaved, i.e., it does not change too sharply. However, in the initial and final periods of the time propagation, we observe that the spectrum of the Liouville operator diverges. This is because $\Gamma^{L}$ consists of nearly core and virtual orbitals at either temporal boundary and large values of $g_k$ and $g_k^{-1}$ in the gauge transformation are applied to the effective bath orbital spaces.

This issue can easily be circumvented by initially propagating the wavefunction without the gauge transformation for a short time. In this short time period, the effective bath orbitals are taken from $\Gamma^{R}$ and $\Gamma^{L}$. After the completion of this short time period, the gauge transformation is applied to the wavefunction, and subsequently, the wavefunction is propagated with the gauge-transformed equation of motion. We found that the spectrum of the Liouville operator showed instabilities up to an initial $\gamma \Delta t_i = 0.06$ and after a final $\gamma \Delta t_f = 0.1$ (see the SM). We therefore used propagation without gauges in these short periods, and between these boundary times, the wavefunction was propagated with the gauge transformation. 

\subsection{Effective bath orbitals versus bath discretization}

\begin{figure*}[t]
    \centering
    \includegraphics[width=1.0\textwidth]{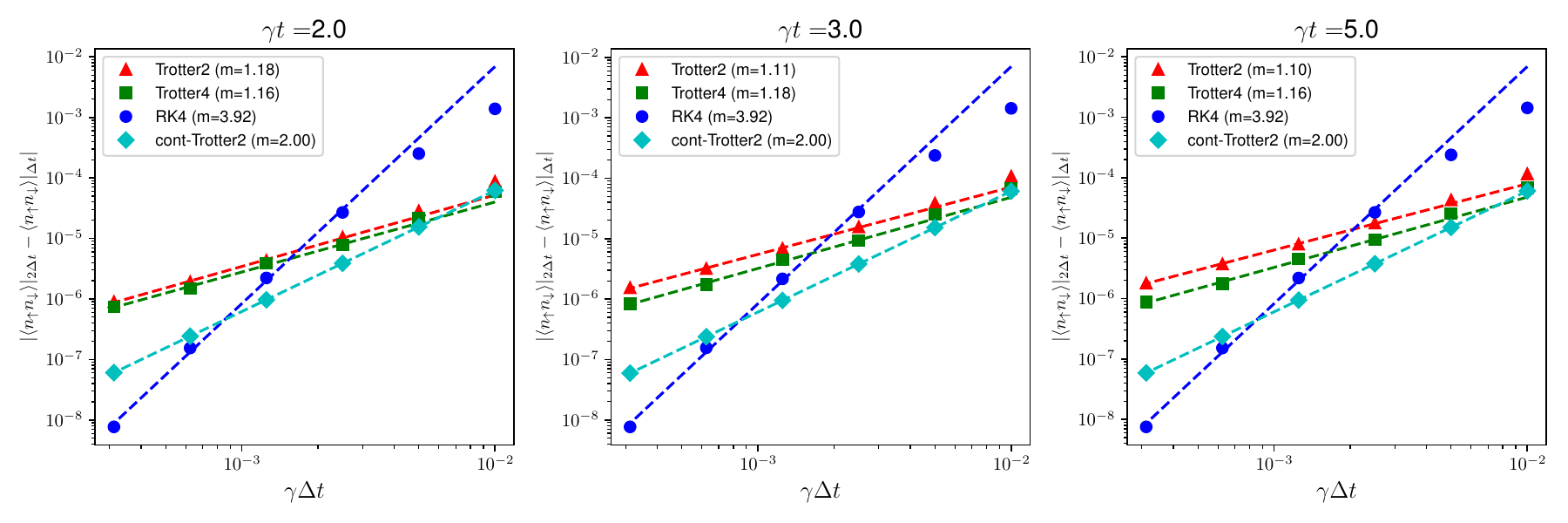}
    \caption{The difference in double occupancy $\langle n_{\uparrow} n_{\downarrow} \rangle$ between two different time steps, $\Delta t$ and $2\Delta t$ using discrete-time IF-MPS with the second- and 4th-order Trotter decomposition (labeled as Trotter2 and Trotter4 with red triangles and green squares, respectively) and continuous-time IF-MPS with the 4th order Runge Kutta method and the second-order Trotter decomposition (labeled as RK4 and cont-Trotter2 with blue circles and cyan diamonds, respectively) at propagated times, $\gamma t = 2, 3,$ and $5$, the inverse temperature, $\gamma \beta = 2$, with the number of effective bath orbitals, $N_{\eff}=8$. The results are fitted by the function, $\log |\langle n_{\uparrow} n_{\downarrow} \rangle |_{2\Delta t} - \langle n_{\uparrow} n_{\downarrow} \rangle |_{\Delta t} |= m \log \Delta t + b$, and the fits are displayed with dashed lines.}
    \label{fig:time step_errors}
\end{figure*}

As described in the analysis in Sec.~\ref{sec:ifsd}, the IF-MPS time propagation can be viewed as propagating a wavefunction in the Liouville space of the impurity and a set of effective bath orbitals. We now compare this time-dependent discretization of the (Liouville) bath with a more standard time-independent discretization.

In Fig.~\ref{fig:quench}, we show the time-dependence of the double occupancy of the impurity $\langle n_{\uparrow} n_{\downarrow} \rangle$ in the quench dynamics. We show (second-order Trotter discrete-time) IF-MPS dynamics generated using effective bath orbital numbers $N_\text{eff}=4, 6, 8, 10$ with time step $\gamma \Delta t= 0.01$ with two different temperatures, $\gamma \beta = 2$ (Fig.~\ref{fig:quench}a) and $\gamma \beta = 50$ (Fig.~\ref{fig:quench}c). We also show results from a time-independent bath discretization scheme based on a Lanczos iteration with thermofield transformation~\cite{PhysRevA.92.052116,PhysRevB.101.155134,Kohn_2022}. The thermofield transformation transforms the initial thermal state with $N_b$ bath orbitals to a Fermi sea with $N_b$ filled orbitals and $N_b$ empty orbitals and the Lanczos tridiagonalization algorithm over the filled and empty orbital space gives a truncated bath basis. We denote the number of Lanczos vectors in the truncated thermofield bath as $N_b^{\text{th}}$. We show dynamics using truncated thermofield bath discretizations with $N_b^{\text{th}}=4,6,8,10$ in Figs.~\ref{fig:quench}b and \ref{fig:quench}d. The thermofield bath dynamics was propagated using the quantum chemistry full configuration interaction method and 4th-order Runge-Kutta with time step $\Delta t = 0.005$.

As expected, the time-independent bath discretization yields substantial finite-size errors due to the limited number of Lanczos vectors supported by the bath, while the effective bath orbitals encode much more faithful dynamics. In the intermediate-temperature regime ($\gamma \beta = 2$), the dynamics is fully converged to the eye at $N_\text{eff}=8$, while the $N^\text{th}_b=10$ dynamics fails for $\gamma t > 2$. In the low-temperature regime ($\gamma \beta = 50$), the IF-MPS dynamics with $N_\text{eff}=10$ faithfully reproduces the reference result within an accuracy of $10^{-3}$.
This illustrates the compactness of the IF-MPS time-dependent bath description, which resembles the behavior seen in real-time quantum embedding studies~\cite{doi:10.1063/1.5012766} even though the bath dynamics in the current formulation is nonunitary.

\subsection{Converging to the continuous-time limit}

We now analyze the time step error incurred by varying the time step, $\gamma \Delta t \in \{0.02 / 2^n | n =0,1,2,\cdots,6\}$, at three different propagated times, $\gamma t = 2$, $3$, and $5$. For this analysis, the inverse temperature is fixed as $\gamma \beta = 2$ and the number of effective bath orbitals is fixed as $N_{\eff}=8$, which is well converged to the reference data as shown in the previous section. We examine the convergence by fitting to the function $f(\Delta t)= A \Delta t^{m} + C $. The exponent $m$ is extracted from the difference, $f(2\Delta t) - f(\Delta t) = A(2^{m}-1)\Delta t^{m} \propto \Delta t^{m}$, without the knowledge of the constant term $C$.

In Fig.~\ref{fig:time step_errors}, the differences in $\langle n_{\uparrow} n_{\downarrow} \rangle$ between two different time steps, $\Delta t$ and $2\Delta t$, $|\langle n_{\uparrow} n_{\downarrow} \rangle |_{2\Delta t} - \langle n_{\uparrow} n_{\downarrow} \rangle |_{\Delta t} |$, where $\langle \bullet \rangle |_{\Delta t} $ means that the numerical simulation was performed with a timestep $\Delta t$, are illustrated as a function of different time steps $\Delta t$. We compare four different boundary IF-MPS schemes to analyze the time step errors: 1) discrete-time boundary IF-MPS with the second-order Trotter decomposition (Trotter2, Eq.~\ref{eq:Liou_evo}), 2) discrete-time boundary IF-MPS with the 4th-order Trotter decomposition (Trotter4, Eq.~\ref{eq:trotter4}), 3) continuous-time IF-MPS propagated by the 4th-order Runge-Kutta (RK4) method, and 4) continuous-time IF-MPS with the second-order Trotter decomposition (cont-Trotter2), which will be defined further below. The exponent $m$ is extracted by fitting the differences at the four smallest time steps, and the fitted function is plotted as dashed lines.

For all three different propagated times, the exponent for the RK4 method has 4th-order time step errors, which agrees with the theoretical scaling. However, the exponent extracted from both the second-order and 4th-order Trotter decomposition in the discrete-time formulation shows that the error is first-order in time step, despite the theoretical error scaling of each Trotter decomposition without tensor network compression. We further note that the extrapolations to the continuous-time limit ($\Delta t =0$) from  Trotter2 and RK4 agree with each other up to $10^{-6}$, whereas the extrapolated results from the Trotter4 data do not (see the SM). This is because the continuous-time limit of Trotter2 (for finite bond dimension) is the same as that of RK4, but that of Trotter4 is not because it involves time evolution in both the forward and backward time directions.

The above discrepancies in the time step error scaling of the Trotter schemes imply that there is an additional first-order time step error associated with the discrete-time tensor network compression. We can show that the first-order time step error arises from the projection of the reference (i.e., $N_b=40$) bath orbital space into the embedding space. For simplicity, we demonstrate this using a time-independent projection operator $\mP$ without the gauge transformation and $\hat{L}_S = 0$, 
\begin{equation} \label{eq:proj_te}
   \mP e^{-i \hat{L} \Delta t} \mP.
\end{equation}
Expanding up to $\Delta t^2$, we obtain
\begin{equation}
    \mP -i \mP \hat{L} \mP \Delta t - \frac{1}{2} \mP \hat{L}^2 \mP (\Delta t)^2 + \mathcal{O}((\Delta t)^3).
\end{equation}
Now consider decreasing the time step $\Delta t$ to $\Delta t /2$. Then, the time-evolution operator for $\Delta t$ can be written as
\begin{equation}
    \mP e^{-i \hat{L} \Delta t/2} \mP e^{-i \hat{L} \Delta t/2} \mP,
\end{equation}
and by expanding it up to $\Delta t^2$, we find
\begin{equation}
     \mP -i \mP \hat{L} \mP \Delta t - \frac{1}{4} \left( \mP \hat{L}^2 \mP + \mP \hat{L} \mP \hat{L} \mP \right) (\Delta t)^2 + \mathcal{O}((\Delta t)^3)
\end{equation}
The two expressions agree to first-order in the time step but not to second-order in the time step due to the difference between $\mP\hat{L}^2 \mP$ and $(\mP \hat{L} \mP)^2$. Thus, after summing over $t/\Delta t$ time steps, the total propagation has first-order time step error $\mathcal{O}(\Delta t)$. The IF-MPS compression involves time-dependent projection operators $\mP(t)$ together with gauge transformations $\hat{G}(t)$. However, by assuming the smoothness of $\mP(t)$ and $\hat{G}(t)$ in time, it is easy to show that the discrete-time IF-MPS also has this first-order time step error.

Based on the above analysis, we can consider a modified Trotter decomposition method that is obtained from the continuous boundary IF-MPS. We denote this as the cont-Trotter method to distinguish it from the previous Trotter methods. 
To obtain the cont-Trotter decomposition, we make the replacement,
\begin{equation}
     \mP e^{-i \hat{L} \Delta t} \mP \to \mathcal{T} e^{-i \int \mP \hat{L} \mP dt} = \mathcal{T} e^{-i \int \hat{L}^{\emb}_{SB} dt},
\end{equation}
where the integration is over the time step interval $\Delta t$ and $\mathcal{T}$ refers to a fermionic time-ordering operator. This can also be interpreted as a generalized time-dependent Trotter decomposition \cite{Huyghebaert_1990} of the embedding Liouville operator. The time-ordered operator can be expressed using the following differential equation,
\begin{gather}
    \hat{U}(x) = \mathcal{T} e^{-i \int_{t_i}^{t_i+x} \hat{L}^{\emb}_{SB} dt}, \nonumber \\
    \frac{d}{dx} \hat{U}(x) = -i \hat{L}^{\emb}_{SB}(t_i+x) \: \hat{U}(x),
\end{gather}
and the desired operator is $\hat{U}(\Delta t)$. Since $\hat{L}^{\emb}_{SB}$ is a noninteracting operator, solving this differential equation of motion for $\hat{U}(x)$ is efficient (noting that the time-evolution operator at $x=0$ is given by the identity operator $\hat{U}(0)= \hat{I}$). We solve the differential equation using 4th-order Runge Kutta with time step $2\Delta t_{\text{fine}}$, which allows us to include all the time information of $\hat{L}^{\emb}_{SB}$ obtained from the bath 1-RDM propagation with a $\Delta t_{\text{fine}}$ time step.

In Fig.~\ref{fig:time step_errors}, time step errors from the second-order cont-Trotter decomposition of the continuous-time IF-MPS clearly show that this gives second-order time step errors. Indeed, the time step propagation error with this technique is better even than that of 4th-order Runge-Kutta except for very small time steps.

\section{Conclusions}\label{sec:sec5}

In this paper, by analyzing the tensor network influence functional (IF-MPS) for the Anderson impurity model, we derived the correct continuous-time limit via the boundary IF-MPS. We further established a correspondence to discretized bath dynamics and quantum embedding. These formal results clarify the connection between the IF-MPS and other long-standing numerical techniques. They also provide the foundation to develop improved numerical implementations, for example, through the higher-order propagators identified in this paper. 

The numerical results we obtained on the quench dynamics of the Anderson model demonstrate the advantages of the current formulation. For example, compared to discrete-time IF-MPS, by using the equations of motion we derive, we can obtain high-order convergence of the time step error. This is in contrast to increasing the order of the Trotterization in the standard discrete time approach, which does not in fact improve the time step convergence, due to the errors associated with compression. 

Our results also support the advantages of IF-MPS dynamics in a more general sense: through defining, implicitly, a time-dependent bath representation, we find we achieve a much more compact description of the influence of the bath than prior static bath discretizations.

The connections between IF-MPS and standard state vector propagation further open up the application of a wide variety of wavefunction-based simulation tools within the boundary IF-MPS framework. We plan to explore the potential of these developments in a variety of physical applications in future work including its applications to the nonequilibrium Kondo problems or nonequilibrium dynamical mean-field theory.

\section{Acknowledgements}

This material is based upon work supported by the U.S. Department of Energy, Office of Science, Office of Advanced Scientific Computing Research and Office of Basic Energy Sciences, Scientific Discovery through Advanced Computing (SciDAC) program under Award Number DE-SC0022088.

\bibliography{references} 

\begin{thebibliography}{74}%
\makeatletter
\providecommand \@ifxundefined [1]{%
 \@ifx{#1\undefined}
}%
\providecommand \@ifnum [1]{%
 \ifnum #1\expandafter \@firstoftwo
 \else \expandafter \@secondoftwo
 \fi
}%
\providecommand \@ifx [1]{%
 \ifx #1\expandafter \@firstoftwo
 \else \expandafter \@secondoftwo
 \fi
}%
\providecommand \natexlab [1]{#1}%
\providecommand \enquote  [1]{``#1''}%
\providecommand \bibnamefont  [1]{#1}%
\providecommand \bibfnamefont [1]{#1}%
\providecommand \citenamefont [1]{#1}%
\providecommand \href@noop [0]{\@secondoftwo}%
\providecommand \href [0]{\begingroup \@sanitize@url \@href}%
\providecommand \@href[1]{\@@startlink{#1}\@@href}%
\providecommand \@@href[1]{\endgroup#1\@@endlink}%
\providecommand \@sanitize@url [0]{\catcode `\\12\catcode `\$12\catcode
  `\&12\catcode `\#12\catcode `\^12\catcode `\_12\catcode `\%12\relax}%
\providecommand \@@startlink[1]{}%
\providecommand \@@endlink[0]{}%
\providecommand \url  [0]{\begingroup\@sanitize@url \@url }%
\providecommand \@url [1]{\endgroup\@href {#1}{\urlprefix }}%
\providecommand \urlprefix  [0]{URL }%
\providecommand \Eprint [0]{\href }%
\providecommand \doibase [0]{https://doi.org/}%
\providecommand \selectlanguage [0]{\@gobble}%
\providecommand \bibinfo  [0]{\@secondoftwo}%
\providecommand \bibfield  [0]{\@secondoftwo}%
\providecommand \translation [1]{[#1]}%
\providecommand \BibitemOpen [0]{}%
\providecommand \bibitemStop [0]{}%
\providecommand \bibitemNoStop [0]{.\EOS\space}%
\providecommand \EOS [0]{\spacefactor3000\relax}%
\providecommand \BibitemShut  [1]{\csname bibitem#1\endcsname}%
\let\auto@bib@innerbib\@empty
\bibitem [{\citenamefont {Hewson}(1993)}]{Hewson1993}%
  \BibitemOpen
  \bibfield  {author} {\bibinfo {author} {\bibfnamefont {A.~C.}\ \bibnamefont
  {Hewson}},\ }\href {https://doi.org/10.1017/cbo9780511470752} {\emph
  {\bibinfo {title} {The Kondo Problem to Heavy Fermions}}}\ (\bibinfo
  {publisher} {Cambridge University Press},\ \bibinfo {year}
  {1993})\BibitemShut {NoStop}%
\bibitem [{\citenamefont {Georges}\ \emph {et~al.}(1996)\citenamefont
  {Georges}, \citenamefont {Kotliar}, \citenamefont {Krauth},\ and\
  \citenamefont {Rozenberg}}]{Georges1996}%
  \BibitemOpen
  \bibfield  {author} {\bibinfo {author} {\bibfnamefont {A.}~\bibnamefont
  {Georges}}, \bibinfo {author} {\bibfnamefont {G.}~\bibnamefont {Kotliar}},
  \bibinfo {author} {\bibfnamefont {W.}~\bibnamefont {Krauth}},\ and\ \bibinfo
  {author} {\bibfnamefont {M.~J.}\ \bibnamefont {Rozenberg}},\ }\bibfield
  {title} {\bibinfo {title} {Dynamical mean-field theory of strongly correlated
  fermion systems and the limit of infinite dimensions},\ }\href
  {https://doi.org/10.1103/RevModPhys.68.13} {\bibfield  {journal} {\bibinfo
  {journal} {Rev. Mod. Phys.}\ }\textbf {\bibinfo {volume} {68}},\ \bibinfo
  {pages} {13} (\bibinfo {year} {1996})}\BibitemShut {NoStop}%
\bibitem [{\citenamefont {Knizia}\ and\ \citenamefont
  {Chan}(2012)}]{Knizia2012}%
  \BibitemOpen
  \bibfield  {author} {\bibinfo {author} {\bibfnamefont {G.}~\bibnamefont
  {Knizia}}\ and\ \bibinfo {author} {\bibfnamefont {G.~K.-L.}\ \bibnamefont
  {Chan}},\ }\bibfield  {title} {\bibinfo {title} {Density matrix embedding: A
  simple alternative to dynamical mean-field theory},\ }\href
  {https://doi.org/10.1103/PhysRevLett.109.186404} {\bibfield  {journal}
  {\bibinfo  {journal} {Phys. Rev. Lett.}\ }\textbf {\bibinfo {volume} {109}},\
  \bibinfo {pages} {186404} (\bibinfo {year} {2012})}\BibitemShut {NoStop}%
\bibitem [{\citenamefont {Wouters}\ \emph {et~al.}(2016)\citenamefont
  {Wouters}, \citenamefont {Jim\'enez-Hoyos}, \citenamefont {Sun},\ and\
  \citenamefont {Chan}}]{doi:10.1021/acs.jctc.6b00316}%
  \BibitemOpen
  \bibfield  {author} {\bibinfo {author} {\bibfnamefont {S.}~\bibnamefont
  {Wouters}}, \bibinfo {author} {\bibfnamefont {C.~A.}\ \bibnamefont
  {Jim\'enez-Hoyos}}, \bibinfo {author} {\bibfnamefont {Q.}~\bibnamefont
  {Sun}},\ and\ \bibinfo {author} {\bibfnamefont {G.~K.-L.}\ \bibnamefont
  {Chan}},\ }\bibfield  {title} {\bibinfo {title} {A practical guide to density
  matrix embedding theory in quantum chemistry},\ }\href
  {https://doi.org/10.1021/acs.jctc.6b00316} {\bibfield  {journal} {\bibinfo
  {journal} {Journal of Chemical Theory and Computation}\ }\textbf {\bibinfo
  {volume} {12}},\ \bibinfo {pages} {2706} (\bibinfo {year} {2016})},\ \bibinfo
  {note} {pMID: 27159268}\BibitemShut {NoStop}%
\bibitem [{\citenamefont {Paeckel}\ \emph {et~al.}(2019)\citenamefont
  {Paeckel}, \citenamefont {K\"ohler}, \citenamefont {Swoboda}, \citenamefont
  {Manmana}, \citenamefont {Schollw\"ock},\ and\ \citenamefont
  {Hubig}}]{PAECKEL2019167998}%
  \BibitemOpen
  \bibfield  {author} {\bibinfo {author} {\bibfnamefont {S.}~\bibnamefont
  {Paeckel}}, \bibinfo {author} {\bibfnamefont {T.}~\bibnamefont {K\"ohler}},
  \bibinfo {author} {\bibfnamefont {A.}~\bibnamefont {Swoboda}}, \bibinfo
  {author} {\bibfnamefont {S.~R.}\ \bibnamefont {Manmana}}, \bibinfo {author}
  {\bibfnamefont {U.}~\bibnamefont {Schollw\"ock}},\ and\ \bibinfo {author}
  {\bibfnamefont {C.}~\bibnamefont {Hubig}},\ }\bibfield  {title} {\bibinfo
  {title} {Time-evolution methods for matrix-product states},\ }\href
  {https://doi.org/https://doi.org/10.1016/j.aop.2019.167998} {\bibfield
  {journal} {\bibinfo  {journal} {Annals of Physics}\ }\textbf {\bibinfo
  {volume} {411}},\ \bibinfo {pages} {167998} (\bibinfo {year}
  {2019})}\BibitemShut {NoStop}%
\bibitem [{\citenamefont {Heidrich-Meisner}\ \emph {et~al.}(2009)\citenamefont
  {Heidrich-Meisner}, \citenamefont {Feiguin},\ and\ \citenamefont
  {Dagotto}}]{PhysRevB.79.235336}%
  \BibitemOpen
  \bibfield  {author} {\bibinfo {author} {\bibfnamefont {F.}~\bibnamefont
  {Heidrich-Meisner}}, \bibinfo {author} {\bibfnamefont {A.~E.}\ \bibnamefont
  {Feiguin}},\ and\ \bibinfo {author} {\bibfnamefont {E.}~\bibnamefont
  {Dagotto}},\ }\bibfield  {title} {\bibinfo {title} {Real-time simulations of
  nonequilibrium transport in the single-impurity anderson model},\ }\href
  {https://doi.org/10.1103/PhysRevB.79.235336} {\bibfield  {journal} {\bibinfo
  {journal} {Phys. Rev. B}\ }\textbf {\bibinfo {volume} {79}},\ \bibinfo
  {pages} {235336} (\bibinfo {year} {2009})}\BibitemShut {NoStop}%
\bibitem [{\citenamefont {Wolf}\ \emph {et~al.}(2014)\citenamefont {Wolf},
  \citenamefont {McCulloch},\ and\ \citenamefont
  {Schollw\"ock}}]{PhysRevB.90.235131}%
  \BibitemOpen
  \bibfield  {author} {\bibinfo {author} {\bibfnamefont {F.~A.}\ \bibnamefont
  {Wolf}}, \bibinfo {author} {\bibfnamefont {I.~P.}\ \bibnamefont
  {McCulloch}},\ and\ \bibinfo {author} {\bibfnamefont {U.}~\bibnamefont
  {Schollw\"ock}},\ }\bibfield  {title} {\bibinfo {title} {Solving
  nonequilibrium dynamical mean-field theory using matrix product states},\
  }\href {https://doi.org/10.1103/PhysRevB.90.235131} {\bibfield  {journal}
  {\bibinfo  {journal} {Phys. Rev. B}\ }\textbf {\bibinfo {volume} {90}},\
  \bibinfo {pages} {235131} (\bibinfo {year} {2014})}\BibitemShut {NoStop}%
\bibitem [{\citenamefont {Ganahl}\ \emph {et~al.}(2015)\citenamefont {Ganahl},
  \citenamefont {Aichhorn}, \citenamefont {Evertz}, \citenamefont
  {Thunstr\"om}, \citenamefont {Held},\ and\ \citenamefont
  {Verstraete}}]{Ganahl2015}%
  \BibitemOpen
  \bibfield  {author} {\bibinfo {author} {\bibfnamefont {M.}~\bibnamefont
  {Ganahl}}, \bibinfo {author} {\bibfnamefont {M.}~\bibnamefont {Aichhorn}},
  \bibinfo {author} {\bibfnamefont {H.~G.}\ \bibnamefont {Evertz}}, \bibinfo
  {author} {\bibfnamefont {P.}~\bibnamefont {Thunstr\"om}}, \bibinfo {author}
  {\bibfnamefont {K.}~\bibnamefont {Held}},\ and\ \bibinfo {author}
  {\bibfnamefont {F.}~\bibnamefont {Verstraete}},\ }\bibfield  {title}
  {\bibinfo {title} {Efficient dmft impurity solver using real-time dynamics
  with matrix product states},\ }\href
  {https://doi.org/10.1103/PhysRevB.92.155132} {\bibfield  {journal} {\bibinfo
  {journal} {Phys. Rev. B}\ }\textbf {\bibinfo {volume} {92}},\ \bibinfo
  {pages} {155132} (\bibinfo {year} {2015})}\BibitemShut {NoStop}%
\bibitem [{\citenamefont {Kohn}\ and\ \citenamefont
  {Santoro}(2022)}]{Kohn_2022}%
  \BibitemOpen
  \bibfield  {author} {\bibinfo {author} {\bibfnamefont {L.}~\bibnamefont
  {Kohn}}\ and\ \bibinfo {author} {\bibfnamefont {G.~E.}\ \bibnamefont
  {Santoro}},\ }\bibfield  {title} {\bibinfo {title} {Quench dynamics of the
  anderson impurity model at finite temperature using matrix product states:
  entanglement and bath dynamics},\ }\href
  {https://doi.org/10.1088/1742-5468/ac729b} {\bibfield  {journal} {\bibinfo
  {journal} {Journal of Statistical Mechanics: Theory and Experiment}\ }\textbf
  {\bibinfo {volume} {2022}},\ \bibinfo {pages} {063102} (\bibinfo {year}
  {2022})}\BibitemShut {NoStop}%
\bibitem [{\citenamefont {M\"{u}hlbacher}\ and\ \citenamefont
  {Rabani}(2008)}]{Muehlbacher2008}%
  \BibitemOpen
  \bibfield  {author} {\bibinfo {author} {\bibfnamefont {L.}~\bibnamefont
  {M\"{u}hlbacher}}\ and\ \bibinfo {author} {\bibfnamefont {E.}~\bibnamefont
  {Rabani}},\ }\bibfield  {title} {\bibinfo {title} {Real-time path integral
  approach to nonequilibrium many-body quantum systems},\ }\href
  {https://doi.org/10.1103/PhysRevLett.100.176403} {\bibfield  {journal}
  {\bibinfo  {journal} {Phys. Rev. Lett.}\ }\textbf {\bibinfo {volume} {100}},\
  \bibinfo {pages} {176403} (\bibinfo {year} {2008})}\BibitemShut {NoStop}%
\bibitem [{\citenamefont {Cohen}\ and\ \citenamefont
  {Rabani}(2011)}]{Cohen2011}%
  \BibitemOpen
  \bibfield  {author} {\bibinfo {author} {\bibfnamefont {G.}~\bibnamefont
  {Cohen}}\ and\ \bibinfo {author} {\bibfnamefont {E.}~\bibnamefont {Rabani}},\
  }\bibfield  {title} {\bibinfo {title} {Memory effects in nonequilibrium
  quantum impurity models},\ }\href
  {https://doi.org/10.1103/PhysRevB.84.075150} {\bibfield  {journal} {\bibinfo
  {journal} {Phys. Rev. B}\ }\textbf {\bibinfo {volume} {84}},\ \bibinfo
  {pages} {075150} (\bibinfo {year} {2011})}\BibitemShut {NoStop}%
\bibitem [{\citenamefont {Gull}\ \emph {et~al.}(2011)\citenamefont {Gull},
  \citenamefont {Millis}, \citenamefont {Lichtenstein}, \citenamefont
  {Rubtsov}, \citenamefont {Troyer},\ and\ \citenamefont {Werner}}]{Gull2011}%
  \BibitemOpen
  \bibfield  {author} {\bibinfo {author} {\bibfnamefont {E.}~\bibnamefont
  {Gull}}, \bibinfo {author} {\bibfnamefont {A.~J.}\ \bibnamefont {Millis}},
  \bibinfo {author} {\bibfnamefont {A.~I.}\ \bibnamefont {Lichtenstein}},
  \bibinfo {author} {\bibfnamefont {A.~N.}\ \bibnamefont {Rubtsov}}, \bibinfo
  {author} {\bibfnamefont {M.}~\bibnamefont {Troyer}},\ and\ \bibinfo {author}
  {\bibfnamefont {P.}~\bibnamefont {Werner}},\ }\bibfield  {title} {\bibinfo
  {title} {Continuous-time monte~carlo methods for quantum impurity models},\
  }\href {https://doi.org/10.1103/revmodphys.83.349} {\bibfield  {journal}
  {\bibinfo  {journal} {Reviews of Modern Physics}\ }\textbf {\bibinfo {volume}
  {83}},\ \bibinfo {pages} {349} (\bibinfo {year} {2011})}\BibitemShut
  {NoStop}%
\bibitem [{\citenamefont {Gull}\ \emph {et~al.}(2010)\citenamefont {Gull},
  \citenamefont {Reichman},\ and\ \citenamefont {Millis}}]{Gull2010}%
  \BibitemOpen
  \bibfield  {author} {\bibinfo {author} {\bibfnamefont {E.}~\bibnamefont
  {Gull}}, \bibinfo {author} {\bibfnamefont {D.~R.}\ \bibnamefont {Reichman}},\
  and\ \bibinfo {author} {\bibfnamefont {A.~J.}\ \bibnamefont {Millis}},\
  }\bibfield  {title} {\bibinfo {title} {Bold-line diagrammatic monte carlo
  method: General formulation and application to expansion around the
  noncrossing approximation},\ }\href
  {https://doi.org/10.1103/PhysRevB.82.075109} {\bibfield  {journal} {\bibinfo
  {journal} {Phys. Rev. B}\ }\textbf {\bibinfo {volume} {82}},\ \bibinfo
  {pages} {075109} (\bibinfo {year} {2010})}\BibitemShut {NoStop}%
\bibitem [{\citenamefont {Cohen}\ \emph {et~al.}(2013)\citenamefont {Cohen},
  \citenamefont {Gull}, \citenamefont {Reichman}, \citenamefont {Millis},\ and\
  \citenamefont {Rabani}}]{Cohen2013}%
  \BibitemOpen
  \bibfield  {author} {\bibinfo {author} {\bibfnamefont {G.}~\bibnamefont
  {Cohen}}, \bibinfo {author} {\bibfnamefont {E.}~\bibnamefont {Gull}},
  \bibinfo {author} {\bibfnamefont {D.~R.}\ \bibnamefont {Reichman}}, \bibinfo
  {author} {\bibfnamefont {A.~J.}\ \bibnamefont {Millis}},\ and\ \bibinfo
  {author} {\bibfnamefont {E.}~\bibnamefont {Rabani}},\ }\bibfield  {title}
  {\bibinfo {title} {Numerically exact long-time magnetization dynamics at the
  nonequilibrium kondo crossover of the anderson impurity model},\ }\href
  {https://doi.org/10.1103/PhysRevB.87.195108} {\bibfield  {journal} {\bibinfo
  {journal} {Phys. Rev. B}\ }\textbf {\bibinfo {volume} {87}},\ \bibinfo
  {pages} {195108} (\bibinfo {year} {2013})}\BibitemShut {NoStop}%
\bibitem [{\citenamefont {Cohen}\ \emph
  {et~al.}(2014{\natexlab{a}})\citenamefont {Cohen}, \citenamefont {Gull},
  \citenamefont {Reichman},\ and\ \citenamefont {Millis}}]{Cohen2014a}%
  \BibitemOpen
  \bibfield  {author} {\bibinfo {author} {\bibfnamefont {G.}~\bibnamefont
  {Cohen}}, \bibinfo {author} {\bibfnamefont {E.}~\bibnamefont {Gull}},
  \bibinfo {author} {\bibfnamefont {D.~R.}\ \bibnamefont {Reichman}},\ and\
  \bibinfo {author} {\bibfnamefont {A.~J.}\ \bibnamefont {Millis}},\ }\bibfield
   {title} {\bibinfo {title} {Green's functions from real-time bold-line monte
  carlo calculations: Spectral properties of the nonequilibrium anderson
  impurity model},\ }\href {https://doi.org/10.1103/PhysRevLett.112.146802}
  {\bibfield  {journal} {\bibinfo  {journal} {Phys. Rev. Lett.}\ }\textbf
  {\bibinfo {volume} {112}},\ \bibinfo {pages} {146802} (\bibinfo {year}
  {2014}{\natexlab{a}})}\BibitemShut {NoStop}%
\bibitem [{\citenamefont {Cohen}\ \emph
  {et~al.}(2014{\natexlab{b}})\citenamefont {Cohen}, \citenamefont {Reichman},
  \citenamefont {Millis},\ and\ \citenamefont {Gull}}]{Cohen2014b}%
  \BibitemOpen
  \bibfield  {author} {\bibinfo {author} {\bibfnamefont {G.}~\bibnamefont
  {Cohen}}, \bibinfo {author} {\bibfnamefont {D.~R.}\ \bibnamefont {Reichman}},
  \bibinfo {author} {\bibfnamefont {A.~J.}\ \bibnamefont {Millis}},\ and\
  \bibinfo {author} {\bibfnamefont {E.}~\bibnamefont {Gull}},\ }\bibfield
  {title} {\bibinfo {title} {Green's functions from real-time bold-line monte
  carlo},\ }\href {https://doi.org/10.1103/PhysRevB.89.115139} {\bibfield
  {journal} {\bibinfo  {journal} {Phys. Rev. B}\ }\textbf {\bibinfo {volume}
  {89}},\ \bibinfo {pages} {115139} (\bibinfo {year}
  {2014}{\natexlab{b}})}\BibitemShut {NoStop}%
\bibitem [{\citenamefont {Feynman}\ and\ \citenamefont
  {Vernon}(2000)}]{FEYNMAN2000547}%
  \BibitemOpen
  \bibfield  {author} {\bibinfo {author} {\bibfnamefont {R.}~\bibnamefont
  {Feynman}}\ and\ \bibinfo {author} {\bibfnamefont {F.}~\bibnamefont
  {Vernon}},\ }\bibfield  {title} {\bibinfo {title} {The theory of a general
  quantum system interacting with a linear dissipative system},\ }\href
  {https://doi.org/https://doi.org/10.1006/aphy.2000.6017} {\bibfield
  {journal} {\bibinfo  {journal} {Annals of Physics}\ }\textbf {\bibinfo
  {volume} {281}},\ \bibinfo {pages} {547} (\bibinfo {year}
  {2000})}\BibitemShut {NoStop}%
\bibitem [{\citenamefont {Tanimura}\ and\ \citenamefont
  {Kubo}(1989)}]{Tanimura1989}%
  \BibitemOpen
  \bibfield  {author} {\bibinfo {author} {\bibfnamefont {Y.}~\bibnamefont
  {Tanimura}}\ and\ \bibinfo {author} {\bibfnamefont {R.}~\bibnamefont
  {Kubo}},\ }\bibfield  {title} {\bibinfo {title} {Time evolution of a quantum
  system in contact with a nearly gaussian-markoffian noise bath},\ }\href
  {https://doi.org/10.1143/jpsj.58.101} {\bibfield  {journal} {\bibinfo
  {journal} {Journal of the Physical Society of Japan}\ }\textbf {\bibinfo
  {volume} {58}},\ \bibinfo {pages} {101} (\bibinfo {year} {1989})}\BibitemShut
  {NoStop}%
\bibitem [{\citenamefont {Jin}\ \emph {et~al.}(2008)\citenamefont {Jin},
  \citenamefont {Zheng},\ and\ \citenamefont {Yan}}]{10.1063/1.2938087}%
  \BibitemOpen
  \bibfield  {author} {\bibinfo {author} {\bibfnamefont {J.}~\bibnamefont
  {Jin}}, \bibinfo {author} {\bibfnamefont {X.}~\bibnamefont {Zheng}},\ and\
  \bibinfo {author} {\bibfnamefont {Y.}~\bibnamefont {Yan}},\ }\bibfield
  {title} {\bibinfo {title} {{Exact dynamics of dissipative electronic systems
  and quantum transport: Hierarchical equations of motion approach}},\
  }\bibfield  {journal} {\bibinfo  {journal} {The Journal of Chemical Physics}\
  }\textbf {\bibinfo {volume} {128}},\ \href
  {https://doi.org/10.1063/1.2938087} {10.1063/1.2938087} (\bibinfo {year}
  {2008}),\ \bibinfo {note} {234703}\BibitemShut {NoStop}%
\bibitem [{\citenamefont {Hou}\ \emph {et~al.}(2014)\citenamefont {Hou},
  \citenamefont {Wang}, \citenamefont {Zheng}, \citenamefont {Tong},
  \citenamefont {Wei},\ and\ \citenamefont {Yan}}]{Dong2014}%
  \BibitemOpen
  \bibfield  {author} {\bibinfo {author} {\bibfnamefont {D.}~\bibnamefont
  {Hou}}, \bibinfo {author} {\bibfnamefont {R.}~\bibnamefont {Wang}}, \bibinfo
  {author} {\bibfnamefont {X.}~\bibnamefont {Zheng}}, \bibinfo {author}
  {\bibfnamefont {N.}~\bibnamefont {Tong}}, \bibinfo {author} {\bibfnamefont
  {J.}~\bibnamefont {Wei}},\ and\ \bibinfo {author} {\bibfnamefont
  {Y.}~\bibnamefont {Yan}},\ }\bibfield  {title} {\bibinfo {title}
  {Hierarchical equations of motion for an impurity solver in dynamical
  mean-field theory},\ }\href {https://doi.org/10.1103/PhysRevB.90.045141}
  {\bibfield  {journal} {\bibinfo  {journal} {Phys. Rev. B}\ }\textbf {\bibinfo
  {volume} {90}},\ \bibinfo {pages} {045141} (\bibinfo {year}
  {2014})}\BibitemShut {NoStop}%
\bibitem [{\citenamefont {Arrigoni}\ \emph {et~al.}(2013)\citenamefont
  {Arrigoni}, \citenamefont {Knap},\ and\ \citenamefont {von~der
  Linden}}]{PhysRevLett.110.086403}%
  \BibitemOpen
  \bibfield  {author} {\bibinfo {author} {\bibfnamefont {E.}~\bibnamefont
  {Arrigoni}}, \bibinfo {author} {\bibfnamefont {M.}~\bibnamefont {Knap}},\
  and\ \bibinfo {author} {\bibfnamefont {W.}~\bibnamefont {von~der Linden}},\
  }\bibfield  {title} {\bibinfo {title} {Nonequilibrium dynamical mean-field
  theory: An auxiliary quantum master equation approach},\ }\href
  {https://doi.org/10.1103/PhysRevLett.110.086403} {\bibfield  {journal}
  {\bibinfo  {journal} {Phys. Rev. Lett.}\ }\textbf {\bibinfo {volume} {110}},\
  \bibinfo {pages} {086403} (\bibinfo {year} {2013})}\BibitemShut {NoStop}%
\bibitem [{\citenamefont {Dorda}\ \emph {et~al.}(2014)\citenamefont {Dorda},
  \citenamefont {Nuss}, \citenamefont {von~der Linden},\ and\ \citenamefont
  {Arrigoni}}]{PhysRevB.89.165105}%
  \BibitemOpen
  \bibfield  {author} {\bibinfo {author} {\bibfnamefont {A.}~\bibnamefont
  {Dorda}}, \bibinfo {author} {\bibfnamefont {M.}~\bibnamefont {Nuss}},
  \bibinfo {author} {\bibfnamefont {W.}~\bibnamefont {von~der Linden}},\ and\
  \bibinfo {author} {\bibfnamefont {E.}~\bibnamefont {Arrigoni}},\ }\bibfield
  {title} {\bibinfo {title} {Auxiliary master equation approach to
  nonequilibrium correlated impurities},\ }\href
  {https://doi.org/10.1103/PhysRevB.89.165105} {\bibfield  {journal} {\bibinfo
  {journal} {Phys. Rev. B}\ }\textbf {\bibinfo {volume} {89}},\ \bibinfo
  {pages} {165105} (\bibinfo {year} {2014})}\BibitemShut {NoStop}%
\bibitem [{\citenamefont {Dorda}\ \emph {et~al.}(2015)\citenamefont {Dorda},
  \citenamefont {Ganahl}, \citenamefont {Evertz}, \citenamefont {von~der
  Linden},\ and\ \citenamefont {Arrigoni}}]{PhysRevB.92.125145}%
  \BibitemOpen
  \bibfield  {author} {\bibinfo {author} {\bibfnamefont {A.}~\bibnamefont
  {Dorda}}, \bibinfo {author} {\bibfnamefont {M.}~\bibnamefont {Ganahl}},
  \bibinfo {author} {\bibfnamefont {H.~G.}\ \bibnamefont {Evertz}}, \bibinfo
  {author} {\bibfnamefont {W.}~\bibnamefont {von~der Linden}},\ and\ \bibinfo
  {author} {\bibfnamefont {E.}~\bibnamefont {Arrigoni}},\ }\bibfield  {title}
  {\bibinfo {title} {Auxiliary master equation approach within matrix product
  states: Spectral properties of the nonequilibrium anderson impurity model},\
  }\href {https://doi.org/10.1103/PhysRevB.92.125145} {\bibfield  {journal}
  {\bibinfo  {journal} {Phys. Rev. B}\ }\textbf {\bibinfo {volume} {92}},\
  \bibinfo {pages} {125145} (\bibinfo {year} {2015})}\BibitemShut {NoStop}%
\bibitem [{\citenamefont {Chen}\ \emph
  {et~al.}(2019{\natexlab{a}})\citenamefont {Chen}, \citenamefont {Cohen},\
  and\ \citenamefont {Galperin}}]{PhysRevLett.122.186803}%
  \BibitemOpen
  \bibfield  {author} {\bibinfo {author} {\bibfnamefont {F.}~\bibnamefont
  {Chen}}, \bibinfo {author} {\bibfnamefont {G.}~\bibnamefont {Cohen}},\ and\
  \bibinfo {author} {\bibfnamefont {M.}~\bibnamefont {Galperin}},\ }\bibfield
  {title} {\bibinfo {title} {Auxiliary master equation for nonequilibrium
  dual-fermion approach},\ }\href
  {https://doi.org/10.1103/PhysRevLett.122.186803} {\bibfield  {journal}
  {\bibinfo  {journal} {Phys. Rev. Lett.}\ }\textbf {\bibinfo {volume} {122}},\
  \bibinfo {pages} {186803} (\bibinfo {year} {2019}{\natexlab{a}})}\BibitemShut
  {NoStop}%
\bibitem [{\citenamefont {Chen}\ \emph
  {et~al.}(2019{\natexlab{b}})\citenamefont {Chen}, \citenamefont {Arrigoni},\
  and\ \citenamefont {Galperin}}]{Chen_2019}%
  \BibitemOpen
  \bibfield  {author} {\bibinfo {author} {\bibfnamefont {F.}~\bibnamefont
  {Chen}}, \bibinfo {author} {\bibfnamefont {E.}~\bibnamefont {Arrigoni}},\
  and\ \bibinfo {author} {\bibfnamefont {M.}~\bibnamefont {Galperin}},\
  }\bibfield  {title} {\bibinfo {title} {Markovian treatment of non-markovian
  dynamics of open fermionic systems},\ }\href
  {https://doi.org/10.1088/1367-2630/ab5ec5} {\bibfield  {journal} {\bibinfo
  {journal} {New Journal of Physics}\ }\textbf {\bibinfo {volume} {21}},\
  \bibinfo {pages} {123035} (\bibinfo {year} {2019}{\natexlab{b}})}\BibitemShut
  {NoStop}%
\bibitem [{\citenamefont {Makri}(1992)}]{Makri1992}%
  \BibitemOpen
  \bibfield  {author} {\bibinfo {author} {\bibfnamefont {N.}~\bibnamefont
  {Makri}},\ }\bibfield  {title} {\bibinfo {title} {Improved feynman
  propagators on a grid and non-adiabatic corrections within the path integral
  framework},\ }\href {https://doi.org/10.1016/0009-2614(92)85654-s} {\bibfield
   {journal} {\bibinfo  {journal} {Chemical Physics Letters}\ }\textbf
  {\bibinfo {volume} {193}},\ \bibinfo {pages} {435} (\bibinfo {year}
  {1992})}\BibitemShut {NoStop}%
\bibitem [{\citenamefont {Makri}\ and\ \citenamefont
  {Makarov}(1995)}]{Makri1995a}%
  \BibitemOpen
  \bibfield  {author} {\bibinfo {author} {\bibfnamefont {N.}~\bibnamefont
  {Makri}}\ and\ \bibinfo {author} {\bibfnamefont {D.~E.}\ \bibnamefont
  {Makarov}},\ }\bibfield  {title} {\bibinfo {title} {Tensor propagator for
  iterative quantum time evolution of reduced density matrices. i. theory},\
  }\href {https://doi.org/10.1063/1.469508} {\bibfield  {journal} {\bibinfo
  {journal} {The Journal of Chemical Physics}\ }\textbf {\bibinfo {volume}
  {102}},\ \bibinfo {pages} {4600} (\bibinfo {year} {1995})}\BibitemShut
  {NoStop}%
\bibitem [{\citenamefont {Weiss}\ \emph {et~al.}(2008)\citenamefont {Weiss},
  \citenamefont {Eckel}, \citenamefont {Thorwart},\ and\ \citenamefont
  {Egger}}]{PhysRevB.77.195316}%
  \BibitemOpen
  \bibfield  {author} {\bibinfo {author} {\bibfnamefont {S.}~\bibnamefont
  {Weiss}}, \bibinfo {author} {\bibfnamefont {J.}~\bibnamefont {Eckel}},
  \bibinfo {author} {\bibfnamefont {M.}~\bibnamefont {Thorwart}},\ and\
  \bibinfo {author} {\bibfnamefont {R.}~\bibnamefont {Egger}},\ }\bibfield
  {title} {\bibinfo {title} {Iterative real-time path integral approach to
  nonequilibrium quantum transport},\ }\href
  {https://doi.org/10.1103/PhysRevB.77.195316} {\bibfield  {journal} {\bibinfo
  {journal} {Phys. Rev. B}\ }\textbf {\bibinfo {volume} {77}},\ \bibinfo
  {pages} {195316} (\bibinfo {year} {2008})}\BibitemShut {NoStop}%
\bibitem [{\citenamefont {Segal}\ \emph {et~al.}(2010)\citenamefont {Segal},
  \citenamefont {Millis},\ and\ \citenamefont {Reichman}}]{PhysRevB.82.205323}%
  \BibitemOpen
  \bibfield  {author} {\bibinfo {author} {\bibfnamefont {D.}~\bibnamefont
  {Segal}}, \bibinfo {author} {\bibfnamefont {A.~J.}\ \bibnamefont {Millis}},\
  and\ \bibinfo {author} {\bibfnamefont {D.~R.}\ \bibnamefont {Reichman}},\
  }\bibfield  {title} {\bibinfo {title} {Numerically exact path-integral
  simulation of nonequilibrium quantum transport and dissipation},\ }\href
  {https://doi.org/10.1103/PhysRevB.82.205323} {\bibfield  {journal} {\bibinfo
  {journal} {Phys. Rev. B}\ }\textbf {\bibinfo {volume} {82}},\ \bibinfo
  {pages} {205323} (\bibinfo {year} {2010})}\BibitemShut {NoStop}%
\bibitem [{\citenamefont {Ba\~nuls}\ \emph {et~al.}(2009)\citenamefont
  {Ba\~nuls}, \citenamefont {Hastings}, \citenamefont {Verstraete},\ and\
  \citenamefont {Cirac}}]{Banuls2009}%
  \BibitemOpen
  \bibfield  {author} {\bibinfo {author} {\bibfnamefont {M.~C.}\ \bibnamefont
  {Ba\~nuls}}, \bibinfo {author} {\bibfnamefont {M.~B.}\ \bibnamefont
  {Hastings}}, \bibinfo {author} {\bibfnamefont {F.}~\bibnamefont
  {Verstraete}},\ and\ \bibinfo {author} {\bibfnamefont {J.~I.}\ \bibnamefont
  {Cirac}},\ }\bibfield  {title} {\bibinfo {title} {Matrix product states for
  dynamical simulation of infinite chains},\ }\href
  {https://doi.org/10.1103/PhysRevLett.102.240603} {\bibfield  {journal}
  {\bibinfo  {journal} {Phys. Rev. Lett.}\ }\textbf {\bibinfo {volume} {102}},\
  \bibinfo {pages} {240603} (\bibinfo {year} {2009})}\BibitemShut {NoStop}%
\bibitem [{\citenamefont {Hastings}\ and\ \citenamefont
  {Mahajan}(2015)}]{PhysRevA.91.032306}%
  \BibitemOpen
  \bibfield  {author} {\bibinfo {author} {\bibfnamefont {M.~B.}\ \bibnamefont
  {Hastings}}\ and\ \bibinfo {author} {\bibfnamefont {R.}~\bibnamefont
  {Mahajan}},\ }\bibfield  {title} {\bibinfo {title} {Connecting entanglement
  in time and space: Improving the folding algorithm},\ }\href
  {https://doi.org/10.1103/PhysRevA.91.032306} {\bibfield  {journal} {\bibinfo
  {journal} {Phys. Rev. A}\ }\textbf {\bibinfo {volume} {91}},\ \bibinfo
  {pages} {032306} (\bibinfo {year} {2015})}\BibitemShut {NoStop}%
\bibitem [{\citenamefont {Strathearn}\ \emph {et~al.}(2018)\citenamefont
  {Strathearn}, \citenamefont {Kirton}, \citenamefont {Kilda}, \citenamefont
  {Keeling},\ and\ \citenamefont {Lovett}}]{Strathearn2018}%
  \BibitemOpen
  \bibfield  {author} {\bibinfo {author} {\bibfnamefont {A.}~\bibnamefont
  {Strathearn}}, \bibinfo {author} {\bibfnamefont {P.}~\bibnamefont {Kirton}},
  \bibinfo {author} {\bibfnamefont {D.}~\bibnamefont {Kilda}}, \bibinfo
  {author} {\bibfnamefont {J.}~\bibnamefont {Keeling}},\ and\ \bibinfo {author}
  {\bibfnamefont {B.~W.}\ \bibnamefont {Lovett}},\ }\bibfield  {title}
  {\bibinfo {title} {Efficient non-markovian quantum dynamics using
  time-evolving matrix product operators},\ }\bibfield  {journal} {\bibinfo
  {journal} {Nature Communications}\ }\textbf {\bibinfo {volume} {9}},\ \href
  {https://doi.org/10.1038/s41467-018-05617-3} {10.1038/s41467-018-05617-3}
  (\bibinfo {year} {2018})\BibitemShut {NoStop}%
\bibitem [{\citenamefont {Cygorek}\ \emph {et~al.}(2022)\citenamefont
  {Cygorek}, \citenamefont {Cosacchi}, \citenamefont {Vagov}, \citenamefont
  {Axt}, \citenamefont {Lovett}, \citenamefont {Keeling},\ and\ \citenamefont
  {Gauger}}]{Cygorek2022}%
  \BibitemOpen
  \bibfield  {author} {\bibinfo {author} {\bibfnamefont {M.}~\bibnamefont
  {Cygorek}}, \bibinfo {author} {\bibfnamefont {M.}~\bibnamefont {Cosacchi}},
  \bibinfo {author} {\bibfnamefont {A.}~\bibnamefont {Vagov}}, \bibinfo
  {author} {\bibfnamefont {V.~M.}\ \bibnamefont {Axt}}, \bibinfo {author}
  {\bibfnamefont {B.~W.}\ \bibnamefont {Lovett}}, \bibinfo {author}
  {\bibfnamefont {J.}~\bibnamefont {Keeling}},\ and\ \bibinfo {author}
  {\bibfnamefont {E.~M.}\ \bibnamefont {Gauger}},\ }\bibfield  {title}
  {\bibinfo {title} {Simulation of open quantum systems by automated
  compression of arbitrary environments},\ }\href
  {https://doi.org/10.1038/s41567-022-01544-9} {\bibfield  {journal} {\bibinfo
  {journal} {Nature Physics}\ }\textbf {\bibinfo {volume} {18}},\ \bibinfo
  {pages} {662} (\bibinfo {year} {2022})}\BibitemShut {NoStop}%
\bibitem [{\citenamefont {Gribben}\ \emph {et~al.}(2022)\citenamefont
  {Gribben}, \citenamefont {Rouse}, \citenamefont {Iles-Smith}, \citenamefont
  {Strathearn}, \citenamefont {Maguire}, \citenamefont {Kirton}, \citenamefont
  {Nazir}, \citenamefont {Gauger},\ and\ \citenamefont {Lovett}}]{Gribben2022}%
  \BibitemOpen
  \bibfield  {author} {\bibinfo {author} {\bibfnamefont {D.}~\bibnamefont
  {Gribben}}, \bibinfo {author} {\bibfnamefont {D.~M.}\ \bibnamefont {Rouse}},
  \bibinfo {author} {\bibfnamefont {J.}~\bibnamefont {Iles-Smith}}, \bibinfo
  {author} {\bibfnamefont {A.}~\bibnamefont {Strathearn}}, \bibinfo {author}
  {\bibfnamefont {H.}~\bibnamefont {Maguire}}, \bibinfo {author} {\bibfnamefont
  {P.}~\bibnamefont {Kirton}}, \bibinfo {author} {\bibfnamefont
  {A.}~\bibnamefont {Nazir}}, \bibinfo {author} {\bibfnamefont {E.~M.}\
  \bibnamefont {Gauger}},\ and\ \bibinfo {author} {\bibfnamefont {B.~W.}\
  \bibnamefont {Lovett}},\ }\bibfield  {title} {\bibinfo {title} {Exact
  dynamics of nonadditive environments in non-markovian open quantum systems},\
  }\href {https://doi.org/10.1103/PRXQuantum.3.010321} {\bibfield  {journal}
  {\bibinfo  {journal} {PRX Quantum}\ }\textbf {\bibinfo {volume} {3}},\
  \bibinfo {pages} {010321} (\bibinfo {year} {2022})}\BibitemShut {NoStop}%
\bibitem [{\citenamefont {Ye}\ and\ \citenamefont
  {Chan}(2021)}]{10.1063/5.0047260}%
  \BibitemOpen
  \bibfield  {author} {\bibinfo {author} {\bibfnamefont {E.}~\bibnamefont
  {Ye}}\ and\ \bibinfo {author} {\bibfnamefont {G.~K.-L.}\ \bibnamefont
  {Chan}},\ }\bibfield  {title} {\bibinfo {title} {Constructing tensor network
  influence functionals for general quantum dynamics},\ }\href
  {https://doi.org/10.1063/5.0047260} {\bibfield  {journal} {\bibinfo
  {journal} {The Journal of Chemical Physics}\ }\textbf {\bibinfo {volume}
  {155}},\ \bibinfo {pages} {044104} (\bibinfo {year} {2021})}\BibitemShut
  {NoStop}%
\bibitem [{\citenamefont {J\o{}rgensen}\ and\ \citenamefont
  {Pollock}(2019)}]{PhysRevLett.123.240602}%
  \BibitemOpen
  \bibfield  {author} {\bibinfo {author} {\bibfnamefont {M.~R.}\ \bibnamefont
  {J\o{}rgensen}}\ and\ \bibinfo {author} {\bibfnamefont {F.~A.}\ \bibnamefont
  {Pollock}},\ }\bibfield  {title} {\bibinfo {title} {Exploiting the causal
  tensor network structure of quantum processes to efficiently simulate
  non-markovian path integrals},\ }\href
  {https://doi.org/10.1103/PhysRevLett.123.240602} {\bibfield  {journal}
  {\bibinfo  {journal} {Phys. Rev. Lett.}\ }\textbf {\bibinfo {volume} {123}},\
  \bibinfo {pages} {240602} (\bibinfo {year} {2019})}\BibitemShut {NoStop}%
\bibitem [{\citenamefont {Sonner}\ \emph {et~al.}(2021)\citenamefont {Sonner},
  \citenamefont {Lerose},\ and\ \citenamefont {Abanin}}]{SONNER2021168677}%
  \BibitemOpen
  \bibfield  {author} {\bibinfo {author} {\bibfnamefont {M.}~\bibnamefont
  {Sonner}}, \bibinfo {author} {\bibfnamefont {A.}~\bibnamefont {Lerose}},\
  and\ \bibinfo {author} {\bibfnamefont {D.~A.}\ \bibnamefont {Abanin}},\
  }\bibfield  {title} {\bibinfo {title} {Influence functional of many-body
  systems: Temporal entanglement and matrix-product state representation},\
  }\href {https://doi.org/https://doi.org/10.1016/j.aop.2021.168677} {\bibfield
   {journal} {\bibinfo  {journal} {Annals of Physics}\ }\textbf {\bibinfo
  {volume} {435}},\ \bibinfo {pages} {168677} (\bibinfo {year} {2021})},\
  \bibinfo {note} {special issue on Philip W. Anderson}\BibitemShut {NoStop}%
\bibitem [{\citenamefont {Thoenniss}\ \emph
  {et~al.}(2023{\natexlab{a}})\citenamefont {Thoenniss}, \citenamefont
  {Lerose},\ and\ \citenamefont {Abanin}}]{PhysRevB.107.195101}%
  \BibitemOpen
  \bibfield  {author} {\bibinfo {author} {\bibfnamefont {J.}~\bibnamefont
  {Thoenniss}}, \bibinfo {author} {\bibfnamefont {A.}~\bibnamefont {Lerose}},\
  and\ \bibinfo {author} {\bibfnamefont {D.~A.}\ \bibnamefont {Abanin}},\
  }\bibfield  {title} {\bibinfo {title} {Nonequilibrium quantum impurity
  problems via matrix-product states in the temporal domain},\ }\href
  {https://doi.org/10.1103/PhysRevB.107.195101} {\bibfield  {journal} {\bibinfo
   {journal} {Phys. Rev. B}\ }\textbf {\bibinfo {volume} {107}},\ \bibinfo
  {pages} {195101} (\bibinfo {year} {2023}{\natexlab{a}})}\BibitemShut
  {NoStop}%
\bibitem [{\citenamefont {Ng}\ \emph {et~al.}(2023)\citenamefont {Ng},
  \citenamefont {Park}, \citenamefont {Millis}, \citenamefont {Chan},\ and\
  \citenamefont {Reichman}}]{PhysRevB.107.125103}%
  \BibitemOpen
  \bibfield  {author} {\bibinfo {author} {\bibfnamefont {N.}~\bibnamefont
  {Ng}}, \bibinfo {author} {\bibfnamefont {G.}~\bibnamefont {Park}}, \bibinfo
  {author} {\bibfnamefont {A.~J.}\ \bibnamefont {Millis}}, \bibinfo {author}
  {\bibfnamefont {G.~K.-L.}\ \bibnamefont {Chan}},\ and\ \bibinfo {author}
  {\bibfnamefont {D.~R.}\ \bibnamefont {Reichman}},\ }\bibfield  {title}
  {\bibinfo {title} {Real-time evolution of anderson impurity models via tensor
  network influence functionals},\ }\href
  {https://doi.org/10.1103/PhysRevB.107.125103} {\bibfield  {journal} {\bibinfo
   {journal} {Phys. Rev. B}\ }\textbf {\bibinfo {volume} {107}},\ \bibinfo
  {pages} {125103} (\bibinfo {year} {2023})}\BibitemShut {NoStop}%
\bibitem [{\citenamefont {Thoenniss}\ \emph
  {et~al.}(2023{\natexlab{b}})\citenamefont {Thoenniss}, \citenamefont
  {Sonner}, \citenamefont {Lerose},\ and\ \citenamefont
  {Abanin}}]{PhysRevB.107.L201115}%
  \BibitemOpen
  \bibfield  {author} {\bibinfo {author} {\bibfnamefont {J.}~\bibnamefont
  {Thoenniss}}, \bibinfo {author} {\bibfnamefont {M.}~\bibnamefont {Sonner}},
  \bibinfo {author} {\bibfnamefont {A.}~\bibnamefont {Lerose}},\ and\ \bibinfo
  {author} {\bibfnamefont {D.~A.}\ \bibnamefont {Abanin}},\ }\bibfield  {title}
  {\bibinfo {title} {Efficient method for quantum impurity problems out of
  equilibrium},\ }\href {https://doi.org/10.1103/PhysRevB.107.L201115}
  {\bibfield  {journal} {\bibinfo  {journal} {Phys. Rev. B}\ }\textbf {\bibinfo
  {volume} {107}},\ \bibinfo {pages} {L201115} (\bibinfo {year}
  {2023}{\natexlab{b}})}\BibitemShut {NoStop}%
\bibitem [{\citenamefont {Kloss}\ \emph {et~al.}(2023)\citenamefont {Kloss},
  \citenamefont {Thoenniss}, \citenamefont {Sonner}, \citenamefont {Lerose},
  \citenamefont {Fishman}, \citenamefont {Stoudenmire}, \citenamefont
  {Parcollet}, \citenamefont {Georges},\ and\ \citenamefont
  {Abanin}}]{PhysRevB.108.205110}%
  \BibitemOpen
  \bibfield  {author} {\bibinfo {author} {\bibfnamefont {B.}~\bibnamefont
  {Kloss}}, \bibinfo {author} {\bibfnamefont {J.}~\bibnamefont {Thoenniss}},
  \bibinfo {author} {\bibfnamefont {M.}~\bibnamefont {Sonner}}, \bibinfo
  {author} {\bibfnamefont {A.}~\bibnamefont {Lerose}}, \bibinfo {author}
  {\bibfnamefont {M.~T.}\ \bibnamefont {Fishman}}, \bibinfo {author}
  {\bibfnamefont {E.~M.}\ \bibnamefont {Stoudenmire}}, \bibinfo {author}
  {\bibfnamefont {O.}~\bibnamefont {Parcollet}}, \bibinfo {author}
  {\bibfnamefont {A.}~\bibnamefont {Georges}},\ and\ \bibinfo {author}
  {\bibfnamefont {D.~A.}\ \bibnamefont {Abanin}},\ }\bibfield  {title}
  {\bibinfo {title} {Equilibrium quantum impurity problems via matrix product
  state encoding of the retarded action},\ }\href
  {https://doi.org/10.1103/PhysRevB.108.205110} {\bibfield  {journal} {\bibinfo
   {journal} {Phys. Rev. B}\ }\textbf {\bibinfo {volume} {108}},\ \bibinfo
  {pages} {205110} (\bibinfo {year} {2023})}\BibitemShut {NoStop}%
\bibitem [{\citenamefont {Chen}\ \emph
  {et~al.}(2024{\natexlab{a}})\citenamefont {Chen}, \citenamefont {Xu},\ and\
  \citenamefont {Guo}}]{PhysRevB.109.045140}%
  \BibitemOpen
  \bibfield  {author} {\bibinfo {author} {\bibfnamefont {R.}~\bibnamefont
  {Chen}}, \bibinfo {author} {\bibfnamefont {X.}~\bibnamefont {Xu}},\ and\
  \bibinfo {author} {\bibfnamefont {C.}~\bibnamefont {Guo}},\ }\bibfield
  {title} {\bibinfo {title} {Grassmann time-evolving matrix product operators
  for quantum impurity models},\ }\href
  {https://doi.org/10.1103/PhysRevB.109.045140} {\bibfield  {journal} {\bibinfo
   {journal} {Phys. Rev. B}\ }\textbf {\bibinfo {volume} {109}},\ \bibinfo
  {pages} {045140} (\bibinfo {year} {2024}{\natexlab{a}})}\BibitemShut
  {NoStop}%
\bibitem [{\citenamefont {Chen}\ \emph
  {et~al.}(2024{\natexlab{b}})\citenamefont {Chen}, \citenamefont {Xu},\ and\
  \citenamefont {Guo}}]{PhysRevB.109.165113}%
  \BibitemOpen
  \bibfield  {author} {\bibinfo {author} {\bibfnamefont {R.}~\bibnamefont
  {Chen}}, \bibinfo {author} {\bibfnamefont {X.}~\bibnamefont {Xu}},\ and\
  \bibinfo {author} {\bibfnamefont {C.}~\bibnamefont {Guo}},\ }\bibfield
  {title} {\bibinfo {title} {Real-time impurity solver using grassmann
  time-evolving matrix product operators},\ }\href
  {https://doi.org/10.1103/PhysRevB.109.165113} {\bibfield  {journal} {\bibinfo
   {journal} {Phys. Rev. B}\ }\textbf {\bibinfo {volume} {109}},\ \bibinfo
  {pages} {165113} (\bibinfo {year} {2024}{\natexlab{b}})}\BibitemShut
  {NoStop}%
\bibitem [{\citenamefont {Kretchmer}\ and\ \citenamefont
  {Chan}(2018)}]{doi:10.1063/1.5012766}%
  \BibitemOpen
  \bibfield  {author} {\bibinfo {author} {\bibfnamefont {J.~S.}\ \bibnamefont
  {Kretchmer}}\ and\ \bibinfo {author} {\bibfnamefont {G.~K.-L.}\ \bibnamefont
  {Chan}},\ }\bibfield  {title} {\bibinfo {title} {A real-time extension of
  density matrix embedding theory for non-equilibrium electron dynamics},\
  }\href {https://doi.org/10.1063/1.5012766} {\bibfield  {journal} {\bibinfo
  {journal} {The Journal of Chemical Physics}\ }\textbf {\bibinfo {volume}
  {148}},\ \bibinfo {pages} {054108} (\bibinfo {year} {2018})}\BibitemShut
  {NoStop}%
\bibitem [{Note1()}]{Note1}%
  \BibitemOpen
  \bibinfo {note} {Within the influence functional formalism, this assumption
  can be relaxed to general Gaussian system-bath initial states, including a
  quantum quench from the ground state of the model at $U=0$ to finite
  $U$.}\BibitemShut {Stop}%
\bibitem [{\citenamefont {Schmutz}(1978)}]{Schmutz1978}%
  \BibitemOpen
  \bibfield  {author} {\bibinfo {author} {\bibfnamefont {M.}~\bibnamefont
  {Schmutz}},\ }\bibfield  {title} {\bibinfo {title} {Real-time green's
  functions in many body problems},\ }\href
  {https://doi.org/10.1007/BF01323673} {\bibfield  {journal} {\bibinfo
  {journal} {Zeitschrift f{\"u}r Physik B Condensed Matter}\ }\textbf {\bibinfo
  {volume} {30}},\ \bibinfo {pages} {97} (\bibinfo {year} {1978})}\BibitemShut
  {NoStop}%
\bibitem [{\citenamefont {Dzhioev}\ and\ \citenamefont
  {Kosov}(2011)}]{10.1063/1.3548065}%
  \BibitemOpen
  \bibfield  {author} {\bibinfo {author} {\bibfnamefont {A.~A.}\ \bibnamefont
  {Dzhioev}}\ and\ \bibinfo {author} {\bibfnamefont {D.~S.}\ \bibnamefont
  {Kosov}},\ }\bibfield  {title} {\bibinfo {title} {{Super-fermion
  representation of quantum kinetic equations for the electron transport
  problem}},\ }\bibfield  {journal} {\bibinfo  {journal} {The Journal of
  Chemical Physics}\ }\textbf {\bibinfo {volume} {134}},\ \href
  {https://doi.org/10.1063/1.3548065} {10.1063/1.3548065} (\bibinfo {year}
  {2011}),\ \bibinfo {note} {044121}\BibitemShut {NoStop}%
\bibitem [{\citenamefont {Harbola}\ and\ \citenamefont
  {Mukamel}(2008)}]{HARBOLA2008191}%
  \BibitemOpen
  \bibfield  {author} {\bibinfo {author} {\bibfnamefont {U.}~\bibnamefont
  {Harbola}}\ and\ \bibinfo {author} {\bibfnamefont {S.}~\bibnamefont
  {Mukamel}},\ }\bibfield  {title} {\bibinfo {title} {Superoperator
  nonequilibrium green's function theory of many-body systems; applications to
  charge transfer and transport in open junctions},\ }\href
  {https://doi.org/https://doi.org/10.1016/j.physrep.2008.05.003} {\bibfield
  {journal} {\bibinfo  {journal} {Physics Reports}\ }\textbf {\bibinfo {volume}
  {465}},\ \bibinfo {pages} {191} (\bibinfo {year} {2008})}\BibitemShut
  {NoStop}%
\bibitem [{\citenamefont {Prosen}(2008)}]{Prosen_2008}%
  \BibitemOpen
  \bibfield  {author} {\bibinfo {author} {\bibfnamefont {T.}~\bibnamefont
  {Prosen}},\ }\bibfield  {title} {\bibinfo {title} {Third quantization: a
  general method to solve master equations for quadratic open fermi systems},\
  }\href {https://doi.org/10.1088/1367-2630/10/4/043026} {\bibfield  {journal}
  {\bibinfo  {journal} {New Journal of Physics}\ }\textbf {\bibinfo {volume}
  {10}},\ \bibinfo {pages} {043026} (\bibinfo {year} {2008})}\BibitemShut
  {NoStop}%
\bibitem [{\citenamefont {Fishman}\ and\ \citenamefont
  {White}(2015)}]{PhysRevB.92.075132}%
  \BibitemOpen
  \bibfield  {author} {\bibinfo {author} {\bibfnamefont {M.~T.}\ \bibnamefont
  {Fishman}}\ and\ \bibinfo {author} {\bibfnamefont {S.~R.}\ \bibnamefont
  {White}},\ }\bibfield  {title} {\bibinfo {title} {Compression of correlation
  matrices and an efficient method for forming matrix product states of
  fermionic gaussian states},\ }\href
  {https://doi.org/10.1103/PhysRevB.92.075132} {\bibfield  {journal} {\bibinfo
  {journal} {Phys. Rev. B}\ }\textbf {\bibinfo {volume} {92}},\ \bibinfo
  {pages} {075132} (\bibinfo {year} {2015})}\BibitemShut {NoStop}%
\bibitem [{\citenamefont {Petrica}\ \emph {et~al.}(2021)\citenamefont
  {Petrica}, \citenamefont {Zheng}, \citenamefont {Chan},\ and\ \citenamefont
  {Clark}}]{PhysRevB.103.125161}%
  \BibitemOpen
  \bibfield  {author} {\bibinfo {author} {\bibfnamefont {G.}~\bibnamefont
  {Petrica}}, \bibinfo {author} {\bibfnamefont {B.-X.}\ \bibnamefont {Zheng}},
  \bibinfo {author} {\bibfnamefont {G.~K.-L.}\ \bibnamefont {Chan}},\ and\
  \bibinfo {author} {\bibfnamefont {B.~K.}\ \bibnamefont {Clark}},\ }\bibfield
  {title} {\bibinfo {title} {Finite and infinite matrix product states for
  gutzwiller projected mean-field wave functions},\ }\href
  {https://doi.org/10.1103/PhysRevB.103.125161} {\bibfield  {journal} {\bibinfo
   {journal} {Phys. Rev. B}\ }\textbf {\bibinfo {volume} {103}},\ \bibinfo
  {pages} {125161} (\bibinfo {year} {2021})}\BibitemShut {NoStop}%
\bibitem [{\citenamefont {Schollw\"{o}ck}(2011)}]{Schollwoeck2011}%
  \BibitemOpen
  \bibfield  {author} {\bibinfo {author} {\bibfnamefont {U.}~\bibnamefont
  {Schollw\"{o}ck}},\ }\bibfield  {title} {\bibinfo {title} {The density-matrix
  renormalization group in the age of matrix product states},\ }\href
  {https://doi.org/10.1016/j.aop.2010.09.012} {\bibfield  {journal} {\bibinfo
  {journal} {Annals of Physics}\ }\textbf {\bibinfo {volume} {326}},\ \bibinfo
  {pages} {96} (\bibinfo {year} {2011})}\BibitemShut {NoStop}%
\bibitem [{\citenamefont {Yehorova}\ and\ \citenamefont
  {Kretchmer}(2023)}]{10.1063/5.0146973}%
  \BibitemOpen
  \bibfield  {author} {\bibinfo {author} {\bibfnamefont {D.}~\bibnamefont
  {Yehorova}}\ and\ \bibinfo {author} {\bibfnamefont {J.~S.}\ \bibnamefont
  {Kretchmer}},\ }\bibfield  {title} {\bibinfo {title} {{A multi-fragment
  real-time extension of projected density matrix embedding theory:
  Non-equilibrium electron dynamics in extended systems}},\ }\bibfield
  {journal} {\bibinfo  {journal} {The Journal of Chemical Physics}\ }\textbf
  {\bibinfo {volume} {158}},\ \href {https://doi.org/10.1063/5.0146973}
  {10.1063/5.0146973} (\bibinfo {year} {2023}),\ \bibinfo {note}
  {131102}\BibitemShut {NoStop}%
\bibitem [{\citenamefont {Peschel}(2012)}]{Peschel2012}%
  \BibitemOpen
  \bibfield  {author} {\bibinfo {author} {\bibfnamefont {I.}~\bibnamefont
  {Peschel}},\ }\bibfield  {title} {\bibinfo {title} {Special review:
  Entanglement in solvable many-particle models},\ }\href
  {https://doi.org/10.1007/s13538-012-0074-1} {\bibfield  {journal} {\bibinfo
  {journal} {Brazilian Journal of Physics}\ }\textbf {\bibinfo {volume} {42}},\
  \bibinfo {pages} {267} (\bibinfo {year} {2012})}\BibitemShut {NoStop}%
\bibitem [{\citenamefont {Jin}\ \emph {et~al.}(2022)\citenamefont {Jin},
  \citenamefont {Sun}, \citenamefont {Zhou},\ and\ \citenamefont
  {Tu}}]{PhysRevB.105.L081101}%
  \BibitemOpen
  \bibfield  {author} {\bibinfo {author} {\bibfnamefont {H.-K.}\ \bibnamefont
  {Jin}}, \bibinfo {author} {\bibfnamefont {R.-Y.}\ \bibnamefont {Sun}},
  \bibinfo {author} {\bibfnamefont {Y.}~\bibnamefont {Zhou}},\ and\ \bibinfo
  {author} {\bibfnamefont {H.-H.}\ \bibnamefont {Tu}},\ }\bibfield  {title}
  {\bibinfo {title} {Matrix product states for hartree-fock-bogoliubov wave
  functions},\ }\href {https://doi.org/10.1103/PhysRevB.105.L081101} {\bibfield
   {journal} {\bibinfo  {journal} {Phys. Rev. B}\ }\textbf {\bibinfo {volume}
  {105}},\ \bibinfo {pages} {L081101} (\bibinfo {year} {2022})}\BibitemShut
  {NoStop}%
\bibitem [{\citenamefont {Schuch}\ and\ \citenamefont
  {Bauer}(2019)}]{PhysRevB.100.245121}%
  \BibitemOpen
  \bibfield  {author} {\bibinfo {author} {\bibfnamefont {N.}~\bibnamefont
  {Schuch}}\ and\ \bibinfo {author} {\bibfnamefont {B.}~\bibnamefont {Bauer}},\
  }\bibfield  {title} {\bibinfo {title} {Matrix product state algorithms for
  gaussian fermionic states},\ }\href
  {https://doi.org/10.1103/PhysRevB.100.245121} {\bibfield  {journal} {\bibinfo
   {journal} {Phys. Rev. B}\ }\textbf {\bibinfo {volume} {100}},\ \bibinfo
  {pages} {245121} (\bibinfo {year} {2019})}\BibitemShut {NoStop}%
\bibitem [{\citenamefont {Manthe}\ \emph {et~al.}(1992)\citenamefont {Manthe},
  \citenamefont {Meyer},\ and\ \citenamefont {Cederbaum}}]{10.1063/1.463007}%
  \BibitemOpen
  \bibfield  {author} {\bibinfo {author} {\bibfnamefont {U.}~\bibnamefont
  {Manthe}}, \bibinfo {author} {\bibfnamefont {H.}~\bibnamefont {Meyer}},\ and\
  \bibinfo {author} {\bibfnamefont {L.~S.}\ \bibnamefont {Cederbaum}},\
  }\bibfield  {title} {\bibinfo {title} {{Wave-packet dynamics within the
  multiconfiguration Hartree framework: General aspects and application to
  NOCl}},\ }\href {https://doi.org/10.1063/1.463007} {\bibfield  {journal}
  {\bibinfo  {journal} {The Journal of Chemical Physics}\ }\textbf {\bibinfo
  {volume} {97}},\ \bibinfo {pages} {3199} (\bibinfo {year}
  {1992})}\BibitemShut {NoStop}%
\bibitem [{\citenamefont {Meyer}\ \emph {et~al.}(1990)\citenamefont {Meyer},
  \citenamefont {Manthe},\ and\ \citenamefont {Cederbaum}}]{MEYER199073}%
  \BibitemOpen
  \bibfield  {author} {\bibinfo {author} {\bibfnamefont {H.-D.}\ \bibnamefont
  {Meyer}}, \bibinfo {author} {\bibfnamefont {U.}~\bibnamefont {Manthe}},\ and\
  \bibinfo {author} {\bibfnamefont {L.}~\bibnamefont {Cederbaum}},\ }\bibfield
  {title} {\bibinfo {title} {The multi-configurational time-dependent hartree
  approach},\ }\href
  {https://doi.org/https://doi.org/10.1016/0009-2614(90)87014-I} {\bibfield
  {journal} {\bibinfo  {journal} {Chemical Physics Letters}\ }\textbf {\bibinfo
  {volume} {165}},\ \bibinfo {pages} {73} (\bibinfo {year} {1990})}\BibitemShut
  {NoStop}%
\bibitem [{\citenamefont {Verstraete}\ and\ \citenamefont
  {Cirac}(2010)}]{PhysRevLett.104.190405}%
  \BibitemOpen
  \bibfield  {author} {\bibinfo {author} {\bibfnamefont {F.}~\bibnamefont
  {Verstraete}}\ and\ \bibinfo {author} {\bibfnamefont {J.~I.}\ \bibnamefont
  {Cirac}},\ }\bibfield  {title} {\bibinfo {title} {Continuous matrix product
  states for quantum fields},\ }\href
  {https://doi.org/10.1103/PhysRevLett.104.190405} {\bibfield  {journal}
  {\bibinfo  {journal} {Phys. Rev. Lett.}\ }\textbf {\bibinfo {volume} {104}},\
  \bibinfo {pages} {190405} (\bibinfo {year} {2010})}\BibitemShut {NoStop}%
\bibitem [{\citenamefont {Ganahl}\ \emph {et~al.}(2017)\citenamefont {Ganahl},
  \citenamefont {Rinc\'on},\ and\ \citenamefont
  {Vidal}}]{PhysRevLett.118.220402}%
  \BibitemOpen
  \bibfield  {author} {\bibinfo {author} {\bibfnamefont {M.}~\bibnamefont
  {Ganahl}}, \bibinfo {author} {\bibfnamefont {J.}~\bibnamefont {Rinc\'on}},\
  and\ \bibinfo {author} {\bibfnamefont {G.}~\bibnamefont {Vidal}},\ }\bibfield
   {title} {\bibinfo {title} {Continuous matrix product states for quantum
  fields: An energy minimization algorithm},\ }\href
  {https://doi.org/10.1103/PhysRevLett.118.220402} {\bibfield  {journal}
  {\bibinfo  {journal} {Phys. Rev. Lett.}\ }\textbf {\bibinfo {volume} {118}},\
  \bibinfo {pages} {220402} (\bibinfo {year} {2017})}\BibitemShut {NoStop}%
\bibitem [{\citenamefont {Pirvu}\ \emph {et~al.}(2010)\citenamefont {Pirvu},
  \citenamefont {Murg}, \citenamefont {Cirac},\ and\ \citenamefont
  {Verstraete}}]{Pirvu_2010}%
  \BibitemOpen
  \bibfield  {author} {\bibinfo {author} {\bibfnamefont {B.}~\bibnamefont
  {Pirvu}}, \bibinfo {author} {\bibfnamefont {V.}~\bibnamefont {Murg}},
  \bibinfo {author} {\bibfnamefont {J.~I.}\ \bibnamefont {Cirac}},\ and\
  \bibinfo {author} {\bibfnamefont {F.}~\bibnamefont {Verstraete}},\ }\bibfield
   {title} {\bibinfo {title} {Matrix product operator representations},\ }\href
  {https://doi.org/10.1088/1367-2630/12/2/025012} {\bibfield  {journal}
  {\bibinfo  {journal} {New Journal of Physics}\ }\textbf {\bibinfo {volume}
  {12}},\ \bibinfo {pages} {025012} (\bibinfo {year} {2010})}\BibitemShut
  {NoStop}%
\bibitem [{Note2()}]{Note2}%
  \BibitemOpen
  \bibinfo {note} {The factor of 1/2 difference from $W^{B,2}$ comes from the
  fact that Eq.~\ref {eq:contMPS} describes double excitations of
  indistinguishable bosonic particles whereas $W^{B,2}$ takes into account
  different auxiliary fermions.}\BibitemShut {Stop}%
\bibitem [{\citenamefont {Sato}\ and\ \citenamefont
  {Ishikawa}(2013)}]{PhysRevA.88.023402}%
  \BibitemOpen
  \bibfield  {author} {\bibinfo {author} {\bibfnamefont {T.}~\bibnamefont
  {Sato}}\ and\ \bibinfo {author} {\bibfnamefont {K.~L.}\ \bibnamefont
  {Ishikawa}},\ }\bibfield  {title} {\bibinfo {title} {Time-dependent
  complete-active-space self-consistent-field method for multielectron dynamics
  in intense laser fields},\ }\href
  {https://doi.org/10.1103/PhysRevA.88.023402} {\bibfield  {journal} {\bibinfo
  {journal} {Phys. Rev. A}\ }\textbf {\bibinfo {volume} {88}},\ \bibinfo
  {pages} {023402} (\bibinfo {year} {2013})}\BibitemShut {NoStop}%
\bibitem [{\citenamefont {Miranda}\ \emph {et~al.}(2011)\citenamefont
  {Miranda}, \citenamefont {Fisher}, \citenamefont {Stella},\ and\
  \citenamefont {Horsfield}}]{10.1063/1.3600397}%
  \BibitemOpen
  \bibfield  {author} {\bibinfo {author} {\bibfnamefont {R.~P.}\ \bibnamefont
  {Miranda}}, \bibinfo {author} {\bibfnamefont {A.~J.}\ \bibnamefont {Fisher}},
  \bibinfo {author} {\bibfnamefont {L.}~\bibnamefont {Stella}},\ and\ \bibinfo
  {author} {\bibfnamefont {A.~P.}\ \bibnamefont {Horsfield}},\ }\bibfield
  {title} {\bibinfo {title} {A multiconfigurational time-dependent hartree-fock
  method for excited electronic states. i. general formalism and application to
  open-shell states},\ }\href {https://doi.org/10.1063/1.3600397} {\bibfield
  {journal} {\bibinfo  {journal} {The Journal of Chemical Physics}\ }\textbf
  {\bibinfo {volume} {134}},\ \bibinfo {pages} {244101} (\bibinfo {year}
  {2011})}\BibitemShut {NoStop}%
\bibitem [{\citenamefont {de~Vega}\ \emph {et~al.}(2015)\citenamefont
  {de~Vega}, \citenamefont {Schollw\"ock},\ and\ \citenamefont
  {Wolf}}]{PhysRevB.92.155126}%
  \BibitemOpen
  \bibfield  {author} {\bibinfo {author} {\bibfnamefont {I.}~\bibnamefont
  {de~Vega}}, \bibinfo {author} {\bibfnamefont {U.}~\bibnamefont
  {Schollw\"ock}},\ and\ \bibinfo {author} {\bibfnamefont {F.~A.}\ \bibnamefont
  {Wolf}},\ }\bibfield  {title} {\bibinfo {title} {How to discretize a quantum
  bath for real-time evolution},\ }\href
  {https://doi.org/10.1103/PhysRevB.92.155126} {\bibfield  {journal} {\bibinfo
  {journal} {Phys. Rev. B}\ }\textbf {\bibinfo {volume} {92}},\ \bibinfo
  {pages} {155126} (\bibinfo {year} {2015})}\BibitemShut {NoStop}%
\bibitem [{\citenamefont {Zhai}\ and\ \citenamefont
  {Chan}(2021)}]{10.1063/5.0050902}%
  \BibitemOpen
  \bibfield  {author} {\bibinfo {author} {\bibfnamefont {H.}~\bibnamefont
  {Zhai}}\ and\ \bibinfo {author} {\bibfnamefont {G.~K.-L.}\ \bibnamefont
  {Chan}},\ }\bibfield  {title} {\bibinfo {title} {{Low communication high
  performance ab initio density matrix renormalization group algorithms}},\
  }\bibfield  {journal} {\bibinfo  {journal} {The Journal of Chemical Physics}\
  }\textbf {\bibinfo {volume} {154}},\ \href
  {https://doi.org/10.1063/5.0050902} {10.1063/5.0050902} (\bibinfo {year}
  {2021}),\ \bibinfo {note} {224116}\BibitemShut {NoStop}%
\bibitem [{\citenamefont {Zhai}\ \emph {et~al.}(2023)\citenamefont {Zhai},
  \citenamefont {Larsson}, \citenamefont {Lee}, \citenamefont {Cui},
  \citenamefont {Zhu}, \citenamefont {Sun}, \citenamefont {Peng}, \citenamefont
  {Peng}, \citenamefont {Liao}, \citenamefont {T\"olle}, \citenamefont {Yang},
  \citenamefont {Li},\ and\ \citenamefont {Chan}}]{10.1063/5.0180424}%
  \BibitemOpen
  \bibfield  {author} {\bibinfo {author} {\bibfnamefont {H.}~\bibnamefont
  {Zhai}}, \bibinfo {author} {\bibfnamefont {H.~R.}\ \bibnamefont {Larsson}},
  \bibinfo {author} {\bibfnamefont {S.}~\bibnamefont {Lee}}, \bibinfo {author}
  {\bibfnamefont {Z.-H.}\ \bibnamefont {Cui}}, \bibinfo {author} {\bibfnamefont
  {T.}~\bibnamefont {Zhu}}, \bibinfo {author} {\bibfnamefont {C.}~\bibnamefont
  {Sun}}, \bibinfo {author} {\bibfnamefont {L.}~\bibnamefont {Peng}}, \bibinfo
  {author} {\bibfnamefont {R.}~\bibnamefont {Peng}}, \bibinfo {author}
  {\bibfnamefont {K.}~\bibnamefont {Liao}}, \bibinfo {author} {\bibfnamefont
  {J.}~\bibnamefont {T\"olle}}, \bibinfo {author} {\bibfnamefont
  {J.}~\bibnamefont {Yang}}, \bibinfo {author} {\bibfnamefont {S.}~\bibnamefont
  {Li}},\ and\ \bibinfo {author} {\bibfnamefont {G.~K.-L.}\ \bibnamefont
  {Chan}},\ }\bibfield  {title} {\bibinfo {title} {{Block2: A comprehensive
  open source framework to develop and apply state-of-the-art DMRG algorithms
  in electronic structure and beyond}},\ }\href
  {https://doi.org/10.1063/5.0180424} {\bibfield  {journal} {\bibinfo
  {journal} {The Journal of Chemical Physics}\ }\textbf {\bibinfo {volume}
  {159}},\ \bibinfo {pages} {234801} (\bibinfo {year} {2023})}\BibitemShut
  {NoStop}%
\bibitem [{\citenamefont {Forest}\ and\ \citenamefont
  {Ruth}(1990)}]{FOREST1990105}%
  \BibitemOpen
  \bibfield  {author} {\bibinfo {author} {\bibfnamefont {E.}~\bibnamefont
  {Forest}}\ and\ \bibinfo {author} {\bibfnamefont {R.~D.}\ \bibnamefont
  {Ruth}},\ }\bibfield  {title} {\bibinfo {title} {Fourth-order symplectic
  integration},\ }\href
  {https://doi.org/https://doi.org/10.1016/0167-2789(90)90019-L} {\bibfield
  {journal} {\bibinfo  {journal} {Physica D: Nonlinear Phenomena}\ }\textbf
  {\bibinfo {volume} {43}},\ \bibinfo {pages} {105} (\bibinfo {year}
  {1990})}\BibitemShut {NoStop}%
\bibitem [{\citenamefont {Garc\'ia-Ripoll}(2006)}]{Jose-Garcia-Ripoll_2006}%
  \BibitemOpen
  \bibfield  {author} {\bibinfo {author} {\bibfnamefont {J.~J.}\ \bibnamefont
  {Garc\'ia-Ripoll}},\ }\bibfield  {title} {\bibinfo {title} {Time evolution of
  matrix product states},\ }\href {https://doi.org/10.1088/1367-2630/8/12/305}
  {\bibfield  {journal} {\bibinfo  {journal} {New Journal of Physics}\ }\textbf
  {\bibinfo {volume} {8}},\ \bibinfo {pages} {305} (\bibinfo {year}
  {2006})}\BibitemShut {NoStop}%
\bibitem [{\citenamefont {Sun}\ \emph {et~al.}(2018)\citenamefont {Sun},
  \citenamefont {Berkelbach}, \citenamefont {Blunt}, \citenamefont {Booth},
  \citenamefont {Guo}, \citenamefont {Li}, \citenamefont {Liu}, \citenamefont
  {McClain}, \citenamefont {Sayfutyarova}, \citenamefont {Sharma} \emph
  {et~al.}}]{sun2018pyscf}%
  \BibitemOpen
  \bibfield  {author} {\bibinfo {author} {\bibfnamefont {Q.}~\bibnamefont
  {Sun}}, \bibinfo {author} {\bibfnamefont {T.~C.}\ \bibnamefont {Berkelbach}},
  \bibinfo {author} {\bibfnamefont {N.~S.}\ \bibnamefont {Blunt}}, \bibinfo
  {author} {\bibfnamefont {G.~H.}\ \bibnamefont {Booth}}, \bibinfo {author}
  {\bibfnamefont {S.}~\bibnamefont {Guo}}, \bibinfo {author} {\bibfnamefont
  {Z.}~\bibnamefont {Li}}, \bibinfo {author} {\bibfnamefont {J.}~\bibnamefont
  {Liu}}, \bibinfo {author} {\bibfnamefont {J.~D.}\ \bibnamefont {McClain}},
  \bibinfo {author} {\bibfnamefont {E.~R.}\ \bibnamefont {Sayfutyarova}},
  \bibinfo {author} {\bibfnamefont {S.}~\bibnamefont {Sharma}}, \emph
  {et~al.},\ }\bibfield  {title} {\bibinfo {title} {Pyscf: the python-based
  simulations of chemistry framework},\ }\href
  {https://doi.org/10.1002/wcms.1340} {\bibfield  {journal} {\bibinfo
  {journal} {Wiley Interdisciplinary Reviews: Computational Molecular Science}\
  }\textbf {\bibinfo {volume} {8}},\ \bibinfo {pages} {e1340} (\bibinfo {year}
  {2018})}\BibitemShut {NoStop}%
\bibitem [{\citenamefont {Sun}\ \emph {et~al.}(2020)\citenamefont {Sun},
  \citenamefont {Zhang}, \citenamefont {Banerjee}, \citenamefont {Bao},
  \citenamefont {Barbry}, \citenamefont {Blunt}, \citenamefont {Bogdanov},
  \citenamefont {Booth}, \citenamefont {Chen}, \citenamefont {Cui} \emph
  {et~al.}}]{sun2020recent}%
  \BibitemOpen
  \bibfield  {author} {\bibinfo {author} {\bibfnamefont {Q.}~\bibnamefont
  {Sun}}, \bibinfo {author} {\bibfnamefont {X.}~\bibnamefont {Zhang}}, \bibinfo
  {author} {\bibfnamefont {S.}~\bibnamefont {Banerjee}}, \bibinfo {author}
  {\bibfnamefont {P.}~\bibnamefont {Bao}}, \bibinfo {author} {\bibfnamefont
  {M.}~\bibnamefont {Barbry}}, \bibinfo {author} {\bibfnamefont {N.~S.}\
  \bibnamefont {Blunt}}, \bibinfo {author} {\bibfnamefont {N.~A.}\ \bibnamefont
  {Bogdanov}}, \bibinfo {author} {\bibfnamefont {G.~H.}\ \bibnamefont {Booth}},
  \bibinfo {author} {\bibfnamefont {J.}~\bibnamefont {Chen}}, \bibinfo {author}
  {\bibfnamefont {Z.-H.}\ \bibnamefont {Cui}}, \emph {et~al.},\ }\bibfield
  {title} {\bibinfo {title} {Recent developments in the pyscf program
  package},\ }\href {https://doi.org/10.1063/5.0006074} {\bibfield  {journal}
  {\bibinfo  {journal} {The Journal of chemical physics}\ }\textbf {\bibinfo
  {volume} {153}} (\bibinfo {year} {2020})}\BibitemShut {NoStop}%
\bibitem [{\citenamefont {de~Vega}\ and\ \citenamefont
  {Ba\~nuls}(2015)}]{PhysRevA.92.052116}%
  \BibitemOpen
  \bibfield  {author} {\bibinfo {author} {\bibfnamefont {I.}~\bibnamefont
  {de~Vega}}\ and\ \bibinfo {author} {\bibfnamefont {M.-C.}\ \bibnamefont
  {Ba\~nuls}},\ }\bibfield  {title} {\bibinfo {title} {Thermofield-based
  chain-mapping approach for open quantum systems},\ }\href
  {https://doi.org/10.1103/PhysRevA.92.052116} {\bibfield  {journal} {\bibinfo
  {journal} {Phys. Rev. A}\ }\textbf {\bibinfo {volume} {92}},\ \bibinfo
  {pages} {052116} (\bibinfo {year} {2015})}\BibitemShut {NoStop}%
\bibitem [{\citenamefont {N\"u\ss{}eler}\ \emph {et~al.}(2020)\citenamefont
  {N\"u\ss{}eler}, \citenamefont {Dhand}, \citenamefont {Huelga},\ and\
  \citenamefont {Plenio}}]{PhysRevB.101.155134}%
  \BibitemOpen
  \bibfield  {author} {\bibinfo {author} {\bibfnamefont {A.}~\bibnamefont
  {N\"u\ss{}eler}}, \bibinfo {author} {\bibfnamefont {I.}~\bibnamefont
  {Dhand}}, \bibinfo {author} {\bibfnamefont {S.~F.}\ \bibnamefont {Huelga}},\
  and\ \bibinfo {author} {\bibfnamefont {M.~B.}\ \bibnamefont {Plenio}},\
  }\bibfield  {title} {\bibinfo {title} {Efficient simulation of open quantum
  systems coupled to a fermionic bath},\ }\href
  {https://doi.org/10.1103/PhysRevB.101.155134} {\bibfield  {journal} {\bibinfo
   {journal} {Phys. Rev. B}\ }\textbf {\bibinfo {volume} {101}},\ \bibinfo
  {pages} {155134} (\bibinfo {year} {2020})}\BibitemShut {NoStop}%
\bibitem [{\citenamefont {Huyghebaert}\ and\ \citenamefont
  {Raedt}(1990)}]{Huyghebaert_1990}%
  \BibitemOpen
  \bibfield  {author} {\bibinfo {author} {\bibfnamefont {J.}~\bibnamefont
  {Huyghebaert}}\ and\ \bibinfo {author} {\bibfnamefont {H.~D.}\ \bibnamefont
  {Raedt}},\ }\bibfield  {title} {\bibinfo {title} {Product formula methods for
  time-dependent schrodinger problems},\ }\href
  {https://doi.org/10.1088/0305-4470/23/24/019} {\bibfield  {journal} {\bibinfo
   {journal} {Journal of Physics A: Mathematical and General}\ }\textbf
  {\bibinfo {volume} {23}},\ \bibinfo {pages} {5777} (\bibinfo {year}
  {1990})}\BibitemShut {NoStop}%
\end{thebibliography}%

\pagebreak
\setcounter{equation}{0}
\setcounter{figure}{0}
\setcounter{table}{0}
\setcounter{page}{1}
\setcounter{section}{0}
\makeatletter
\renewcommand{\thesection}{SM-\Roman{section}}
\renewcommand{\theequation}{S\arabic{equation}}
\renewcommand{\thefigure}{S\arabic{figure}}
\renewcommand{\bibnumfmt}[1]{[S#1]}
\renewcommand{\citenumfont}[1]{S#1}


\title{Supplemental Materials for ``Tensor network influence functionals in the continuous-time limit: connections to quantum embedding, bath discretization, and higher-order time propagation''}

\begin{CJK*}{UTF8}{mj}
\author{Gunhee Park (박건희)}
\affiliation{Division of Engineering and Applied Science, California Institute of Technology, Pasadena, CA 91125, USA}
\author{Nathan Ng}
\affiliation{Department of Chemistry, Columbia University, New York, New York 10027, United States}
\author{David R.\ Reichman}
\affiliation{Department of Chemistry, Columbia University, New York, New York 10027, United States}
\author{Garnet Kin-Lic Chan}
\affiliation{Division of Chemistry and Chemical Engineering, California Institute of Technology, Pasadena, California 91125, USA}

\maketitle
\end{CJK*}
\onecolumngrid

\section{Additional data}

\subsection{Diagonal elements of impurity reduced density operator}

\begin{figure}[h]
    \centering
    \includegraphics[width=0.9\columnwidth]{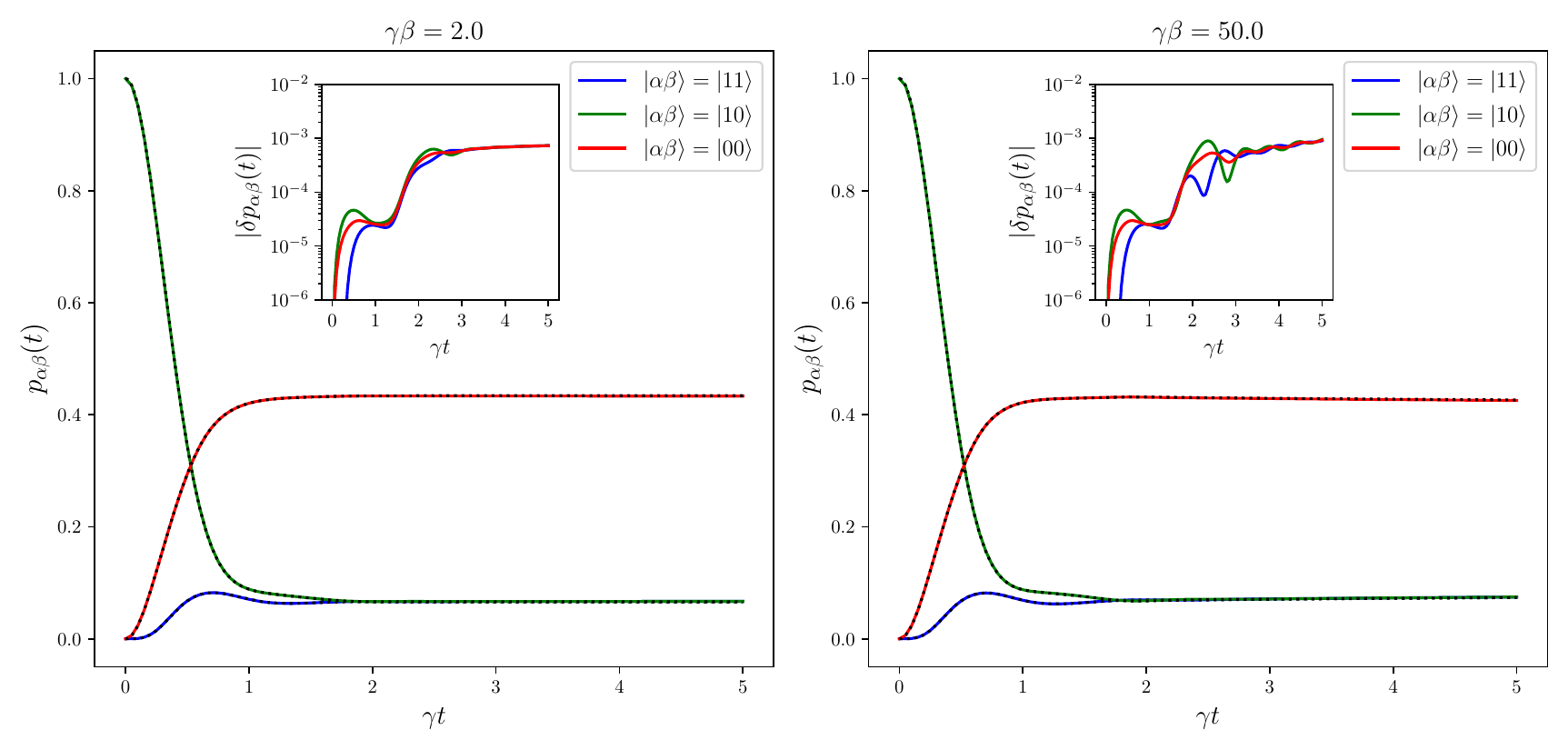}
    \caption{Diagonal elements of the impurity reduced density operator $p_{\alpha \beta}(t) = \bra{\alpha \beta}\hat{\rho}_S(t) \ket{\alpha \beta}$ of the symmetric Anderson model with $U=2.5\pi\gamma$ and $\varepsilon_{\sigma}=-1.25\pi \gamma$. The impurity is quenched from an unoccupied initial state with a thermal bath at temperature $\gamma \beta = 2$ (left) and $50$ (right). Dotted lines are results from time-dependent DMRG (tdDMRG) with a bond-dimension 300 and solid lines are results from the discrete-time boundary IF-MPS with the number of effective bath orbitals $N_{\eff}=10$ and a timestep of $\gamma \Delta t = 0.01$. \textbf{(inset)} Absolute deviations in $p_{\alpha \beta}(t)$ compared to the results of tdDMRG.}
    \label{fig:SM_quench}
\end{figure}

In the main text, we showed the time-dependence of the double occupancy, $\langle n_{\uparrow} n_{\downarrow} \rangle$, following a quench dynamics of the symmetric Anderson model with $U=2.5\pi\gamma$ and $\varepsilon_{\sigma}=-1.25\pi \gamma$ from the initially unoccupied impurity state and a thermal bath of temperature $\gamma \beta = 2$ and $50$. In this section, we present additional data for diagonal elements of the impurity reduced density operator, $p_{\alpha \beta}(t) = \bra{\alpha \beta}\hat{\rho}_S(t) \ket{\alpha \beta}$, where $\alpha, \beta \in \{0,1\}$ is the configuration of spin $\uparrow$ and $\downarrow$. For example, the double occupancy can be expressed as $\langle n_{\uparrow} n_{\downarrow} \rangle = p_{11}$. In Fig.~\ref{fig:SM_quench}, we compare results from the discrete-time formulation of the boundary IF with $N_{\eff}=10$ effective bath orbitals to the benchmark result from time-dependent DMRG (tdDMRG) with bond-dimension 300, as explained in the main text. The density operator elements are normalized so the trace is $1$, $\tr \hat{\rho}_S(t) = 1$. The maximum errors are of the order of $10^{-3}$ compared to the results from tdDMRG for both temperatures.

\subsection{Spectrum of embedding Liouville operator}

\begin{figure}[h]
    \centering
    \includegraphics[width=0.6\columnwidth]{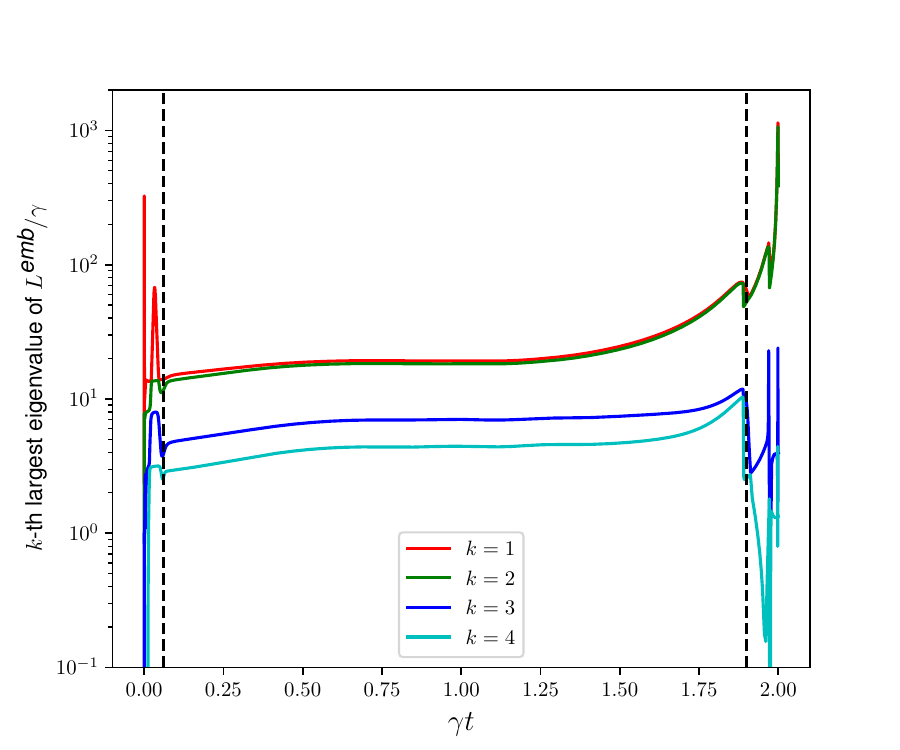}
    \caption{The 4 largest eigenvalues of the embedding Liouville operator, $\hat{L}^{\emb}$, constructed from the IF with the total propagated time, $\gamma t =2$. The spectrum shows the instabilities at the initial and final boundary times, which are denoted by dashed lines at $\gamma \Delta t_i = 0.06$ and $\gamma \Delta t_f = 0.1$.}
    \label{fig:SM_spectrum}
\end{figure}

In this section, we show the instabilities of the embedding Liouville operator, $\hat{L}^{\emb}$, close to the boundary times. Fig.~\ref{fig:SM_spectrum} shows the 4 largest eigenvalues of $\hat{L}^{\emb}$ constructed from the boundary IF with total propagated time, $\gamma t =2$. The spectrum shows the instabilities at the initial and final boundary times, which are denoted by dashed lines at $\gamma \Delta t_i = 0.06$ and $\gamma \Delta t_f = 0.1$. We observe similar instabilities for other total propagation times up to $\gamma t = 5$.

\subsection{Temporal Entanglement Entropy of Boundary IF}
In Ref.~\cite{SONNER2021168677}, the authors reported the vanishing temporal entanglement entropy scaling in the continuous-time limit from the Hamiltonian dynamics of the one-dimensional spin chain, and \cite{PhysRevB.107.125103} shows that this scaling also holds for SIAM. We showed that the equation of motion constructed in the continuous-time limit leads to unitary dynamics in the super-Fock bath orbital space, which preserves the spectrum of the 1-RDM, and hence the spectrum consists of only 0 and 1 with zero entanglement entropy of the IF.

\begin{figure}[h]
    \centering
    \includegraphics[width=0.6\columnwidth]{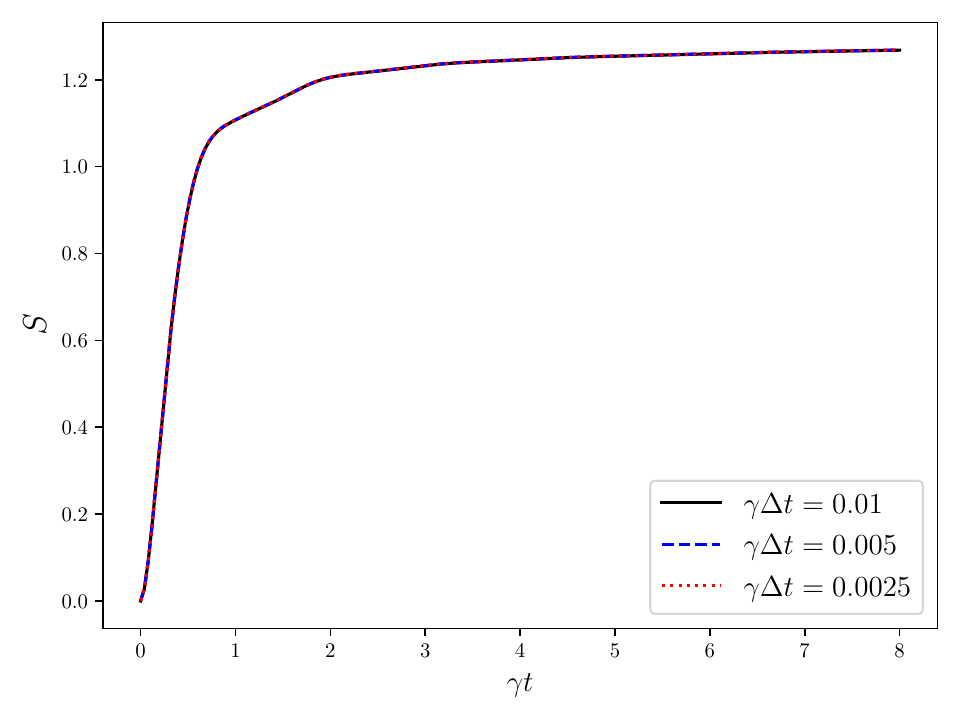}
    \caption{Temporal entanglement entropy of the boundary influence functional at the half-cut with three different timesteps, $\gamma \Delta t = 0.01, 0.005$, and $0.0025$, as a function of total propagated time. The other parameters are set to be the same as the main text. The figure clearly shows that the values of entanglement entropy from three different timesteps overlap each other. }
    \label{fig:EEplot}
\end{figure}

The boundary IF was proposed in the main text to modify the behavior of the IF to yield a meaningful continuous-time limit. The tensor elements satisfy the form of a continuous-matrix product state and hence its physical properties including the entanglement entropy are well-defined as the discretization is changed.  In Fig.~\ref{fig:EEplot}, we numerically demonstrate that the temporal entanglement entropy is almost invariant to the time discretization. The entanglement entropy of the boundary IF is increasing very slowly over the propagated time, which allows us to achieve an efficient classical simulation in the long-time limit.

\begin{figure}[t]
    \centering
    \includegraphics[width=0.65\columnwidth]{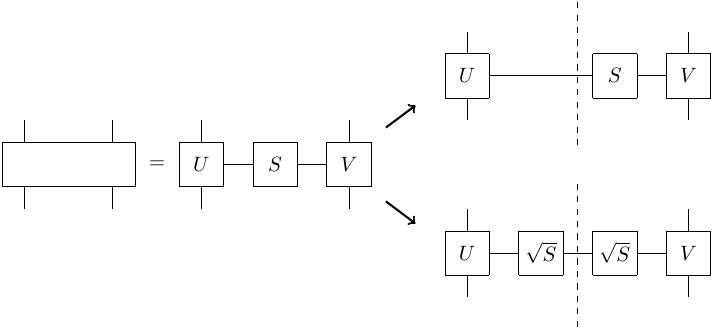}
    \caption{Influence functional tensors can be constructed from gates after splitting the gate with a singular value decomposition. The singular values can be either absorbed to the influence functional tensors (right upper side, we denote this the `absorb' gauge) or split to both sides (right lower side, we denote this the `split' gauge).}
    \label{fig:BoundaryIF}
\end{figure}

\subsection{Temporal Entanglement Entropy of one-dimensional Ising model}

As shown in the above section, the entanglement entropy of the IF depends on the boundary splitting. In this section, we show that the above boundary splitting also allows for a  temporal entanglement entropy that converges to a nontrivial value as a function of timestep size in the one-dimensional Ising model (which represents IF dynamics with a non-quadratic interacting bath). Fig.~\ref{fig:BoundaryIF} shows data from two different splitting schemes of the time-evolution operators, the first one absorbs singular values to the boundary, which corresponds to the ordinary influence functional in \cite{SONNER2021168677}, and the second one splits the singular values, which corresponds to the boundary influence functional. 

In this section, we denote this difference as a boundary `gauge', or simply gauge for short, where we call the first gauge the  `absorb' gauge and the second, the `split' gauge. However, we stress that the gauge term here is different from the `gauge' in the main text, where the gauge referred to the gauge structure along the temporal direction. We demonstrate the effect of the boundary gauge on the temporal entanglement entropy for a one-dimensional quantum Ising chain with both transverse and longitudinal magnetic fields,
\begin{equation}
    H = - \sum_i (J \sigma_i^z \sigma_{i+1}^z + g \sigma_i^x + h  \sigma_i^z),
\end{equation}
initialized in a polarized product state, $\ket{Z+}= \lim_{N \to \infty} \ket{0}^{\otimes N}$. We fix $J=1$ and tune the other two parameters to test both integrable ($g=0.5, \: h=0.0$) and nonintegrable ($g=-1.05, \: h=0.5$) systems with three different timesteps, $\Delta t =$ 0.01, 0.02, and 0.04. We use second-order Trotter decomposition to represent the time-evolution operators.

\begin{figure}[t]
    \centering
    \includegraphics[width=0.95\columnwidth]{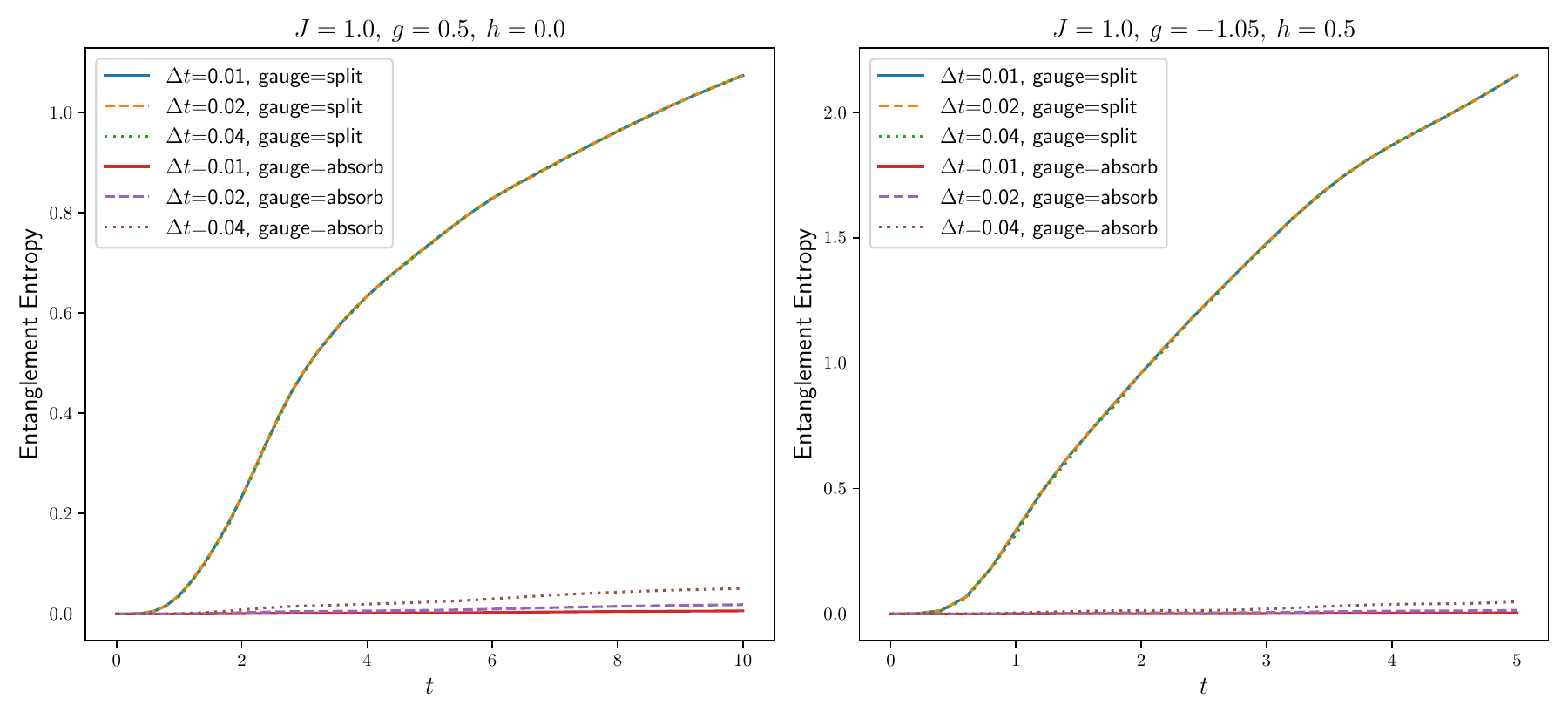}
    \caption{Temporal Entanglement entropy from the quench dynamics of the quantum Ising model with both transverse and longitudinal magnetic field, with a polarized initial state, $\ket{Z+}= \lim_{N \to \infty} \ket{0}^{\otimes N}$. We consider two sets of parameters for the Hamiltonian, one corresponds to an integrable system, $J=1.0, \: g=0.5, \: h=0.0$ (left), and the other to a nonintegrable system, $J=1.0, \: g=-1.05, \: h=0.5$ (right) with three different timesteps, $\Delta t = $ 0.01, 0.02, and 0.04 with two different boundary gauges. It clearly shows that the absorb gauge leads to zero temporal entanglement entropy as the timestep goes to zero whereas the split boundary gauge allows us to get an almost timestep-invariant temporal entanglement entropy with a non-trivial temporal entanglement entropy in the continuous-time limit for both integrable and nonintegrable models. }
    \label{fig:EE_IF_Ising1d}
\end{figure}

Fig.~\ref{fig:EE_IF_Ising1d} shows the temporal entanglement entropy as a function of propagated time in integrable and nonintegrable models for three different timesteps with two different boundary gauges. The values of the temporal entanglement entropy are converged with respect to the bond dimensions and the singular value cutoffs. It clearly shows that the absorb gauge leads to zero temporal entanglement entropy as the timestep goes to zero, which agrees with \cite{SONNER2021168677}, whereas the split boundary gauge allows us to get an almost timestep-invariant temporal entanglement entropy that avoids the zero temporal entanglement entropy behavior for both integrable and nonintegrable models. This shows that the boundary influence functional with the split gauge contains the correct continuous-time limit and that its entanglement entropy properly reflects the complexity of the classical simulation without timestep dependence. Additionally, it avoids the numerical instabilities associated with small singular values that appear in the absorb gauge due to the zero entanglement entropy when using small timesteps.

\begin{figure}[t]
    \centering
    \includegraphics[width=0.55\columnwidth]{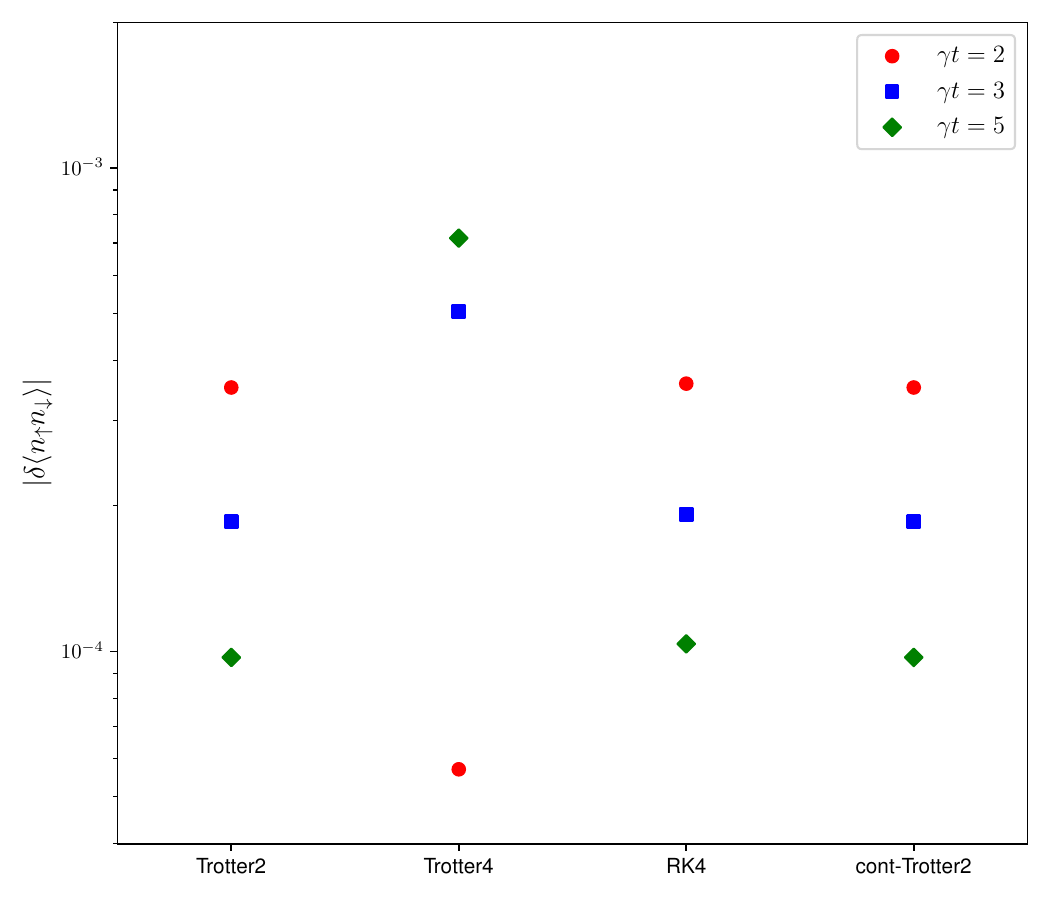}
    \caption{The values of the double occupancy compared to the results of tdDMRG after the extrapolation of the timestep to zero for the four different methods described in the main text with the three different propagated times. The difference between the results from Trotter2, RK4, and cont-Trotter2 is within $10^{-5}$ whereas the results from Trotter4 extrapolate to a different limit compared to the other three methods.}
    \label{fig:extrapolated_timestep}
\end{figure}

\subsection{Timestep extrapolated data}

In this section, we report the values of double occupancy compared to the results of tdDMRG after the zero-timestep extrapolation for four different methods described in the main text with three different propagated times (Fig.~\ref{fig:extrapolated_timestep}). These results show that the extrapolated dynamics from the discrete-time IF-MPS and the continuous-time IF-MPS are consistent, as the difference between the results from Trotter2, RK4, and cont-Trotter2 lies within $10^{-5}$. 

However, the results from Trotter4 have different errors in comparison to the other three methods. This is because the discrete-time IF-MPS from Trotter2 and the continuous-time IF-MPS converge to the same object in the continuous-time limit but the discrete-time IF-MPS from Trotter4 does not since it involves time evolution both in the forward and backward time direction.

\subsection{Computational time}
We here analyze the computational time for the IF-MPS simulation to support claims of efficiency. There are two different sources of computational cost - one is from the IF-MPS construction with Gaussian operations and the other is from many-body state-vector propagation. Given the number of time-steps, $N_t$, the former computation scales as $\mathcal{O}(N_b^3 N_t)$ and the latter computation scales as $\mathcal{O}(2^{N_{\eff}+N_{\text{imp}}} N_t)$ in the discrete-time formulation of IF-MPS. The exponential factor $2^{N_{\eff}+N_{\text{imp}}}$ originates from the full set of configurations of the Hilbert space with $N_{\eff}+N_{\text{imp}}$ orbitals. In the continuous-time IF-MPS, it has a finer time-step, $\Delta t_{\text{fine}}$, and $N_{t,\text{fine}} = t/\Delta t_{\text{fine}}$, so its Gaussian computation scales as $\mathcal{O}(N_b^3 N_{t, \text{fine}})$.

Based on computations using a local desktop computer with an Intel Core i7-12700 processor, in the discrete-time IF-MPS calculation for the single-time observable with $\gamma t = 5$, $N_b = 40$, and $N_t = 500$, the Gaussian IF-MPS construction took about 10 seconds. The many-body state-vector propagation took about 2 seconds, 10 seconds, 1 minute, and 10 minutes for $N_{\eff}=$ 4, 6, 8, 10, respectively. In contrast, the reference tdDMRG calculation took on the order of a few hours. 

For the continuous-time IF-MPS calculation, the computational costs differ due to the use of a finer time-step $N_{t,\text{fine}}$ in the Gaussian IF-MPS construction. For the calculations in the main text, we chose an extremely fine time grid with $N_{t,\text{fine}}=32000$ ($\Delta t_{\text{fine}} \sim 0.00016$) as a proof-of-principle validation, such that the Gaussian IF-MPS construction took about 20 minutes. In practice, this time can be reduced by choosing a less fine time grid. 

The computational time for the Gaussian IF-MPS construction increases as $\mathcal{O}(N_b^3)$ as one approaches the continuum limit of a discretized bath. The convergence to the continuum limit of a discretized bath can be improved by adopting other discretization schemes, such as logarithmic discretization, orthogonal polynomial-based discretization, or numerically optimized discretization.

\section{Super-fermion representation}

In this section, we describe the super-fermion representation in more detail. The super-fermion representation vectorizes the density operator, $\hat{\rho}$, giving a wavefunction in a Liouville space, $\kett{\rho}$. For brevity, spin indices will be omitted unless necessary. The Liouville space forms a super-Fock space with twice the number of orbitals. Formally, it can be found by applying the density operator to a `left vacuum' state, $\kett{I}$, $\kett{\rho} = \hat{\rho} \kett{I}$,
\begin{equation}
    \kett{I}= \prod_i (c_i^{\dag} + \Tilde{c}_i^{\dag})\kett{0} =  \exp \left( \sum_i c_i^{\dag} \Tilde{c}_i \right) \ket{\bm{0}} \otimes \ket{\bm{1}},
\end{equation}
where the empty orbitals, $\ket{\bm{0}}$, are from the original Fock space, and the fully occupied orbitals, $\ket{\bm{1}}$, are from the extended Fock space. The application of $\hat{\rho}$ into $\kett{I}$ can be interpreted as a particle-hole transformation of the bra of the density operator, and $\kett{I}$ can be interpreted as a set of fermionic Bell pairs. It is easy to see the following relation,
\begin{equation}
    c_i \kett{I} = \Tilde{c}_i \kett{I}, \quad c_i^{\dag} \kett{I} = - \Tilde{c}_i^{\dag} \kett{I}.
\end{equation}\label{eq:SM_superfer_rel}
For example, for the thermal initial bath, $\hat{\rho}_B$, with an inverse temperature, $\beta$,
\begin{equation}
    \hat{\rho}_B =  \left( \frac{1}{1+e^{-\beta E_i}} \proj{0} + \frac{e^{-\beta E_i}}{1+e^{-\beta E_i}} \proj{1} \right)^{\otimes N_b} =  \left( f_{-}(E_i,\beta) \proj{0}+ f_{+}(E_i,\beta) \proj{1} \right)^{\otimes N_b} ,
\end{equation}
the super-fermion representation, $\kett{\rho_B}$, is
\begin{equation}
    \kett{\rho_B} = \prod_{i} \left( f_{+}(E_{i},\beta) \hat{c}^{\dag}_{i} + 
f_{-}(E_{i},\beta) \hat{\Tilde{c}}^{\dag}_{i}  \right)\kett{0}.
\end{equation}
The expectation value for an observable $\hat{O}$ is given by,
\begin{equation}
    \tr \left[ \hat{O} \hat{\rho} \right] = \braa{I} \hat{O} \kett{\rho}.
\end{equation}
Therefore, $\kett{\tr_B}$ in the main text has the same expression as $\kett{I}$. From the von Neumann equation,
\begin{equation}
    i \frac{d}{dt} \hat{\rho} = [\hat{H}, \hat{\rho}] = \hat{H} \hat{\rho} - \hat{\rho} \hat{H},
\end{equation}
by applying the left vacuum $\kett{I}$ on both sides, we can find the time evolution equation for $\kett{\rho}$,
\begin{equation}
    i \frac{d}{dt} \hat{\rho} \kett{I}=  i \frac{d}{dt} \kett{\rho} = \hat{L} \kett{\rho},
\end{equation}
with the Liouville operator, $\hat{L}$, which can be found from $[\hat{H},\hat{\rho}]\kett{I}$. Using the relationships, Eq.~\ref{eq:SM_superfer_rel}, and the Hamiltonian for SIAM, 
\begin{equation}
    \begin{gathered}
    \hat{H} = \hat{H}_S + \hat{H}_{SB}+\hat{H}_{B},\\
    \hat{H}_S = U \hat{n}_{\uparrow} \hat{n}_{\downarrow} + \sum_{\sigma} \varepsilon_{\sigma} \hat{n}_{\sigma}, \\
    \hat{H}_{SB} = \sum_{i, \sigma} \left( t_i \hat{c}^{\dag}_{i, \sigma} \hat{d}_{\sigma} + \hc \right),\\
    \hat{H}_{B} = \sum_{i, \sigma} E_{i} \hat{c}^{\dag}_{i, \sigma} \hat{c}_{i,\sigma},
    \end{gathered}
\end{equation}
The corresponding Liouville operator can be found as,
\begin{equation}
    \begin{gathered}
    \hat{L} = \hat{L}_S + \hat{L}_{SB}+\hat{L}_{B},\\
    \hat{L}_S = U \hat{n}_{\uparrow} \hat{n}_{\downarrow} -U (1-\hat{\Tilde{n}}_{\uparrow})(1-\hat{\Tilde{n}}_{\downarrow}) 
    + \sum_{\sigma}  \varepsilon_{\sigma} (\hat{n}_{\sigma} + \hat{\Tilde{n}}_{\sigma}),  \\
    \hat{L}_{SB} = \sum_{i, \sigma} \left( t_i \hat{c}^{\dag}_{i, \sigma} \hat{d}_{\sigma} +t_i \hat{\Tilde{c}}^{\dag}_{i, \sigma} \hat{\Tilde{d}}_{\sigma} + \hc \right),\\
    \hat{L}_{B} = \sum_{i, \sigma} E_{i} (\hat{c}^{\dag}_{i, \sigma} \hat{c}_{i,\sigma} + \hat{\Tilde{c}}^{\dag}_{i,\sigma} \hat{\Tilde{c}}_{i,\sigma} ),
    \end{gathered}
\end{equation}
with constant terms omitted. For example, $\hat{\rho}c_i^{\dag}c_j \kett{I}$ ($i \neq j$) can be expressed as,
\begin{equation}
    \hat{\rho}c_i^{\dag}c_j \kett{I} = \hat{\rho}c_i^{\dag} \Tilde{c}_j \kett{I} = - \hat{\rho} \Tilde{c}_j c_i^{\dag} \kett{I} =  \hat{\rho} \Tilde{c}_j \Tilde{c}_i^{\dag} \kett{I} = - \Tilde{c}_i^{\dag} \Tilde{c}_j \kett{\rho}.
\end{equation}
Note that the dynamics preserves its trace,
\begin{equation}
    0 = i \frac{d}{dt} \tr [\hat{\rho}] = \braa{I} \hat{L} \kett{\rho},
\end{equation}
for any $\rho$, so $\braa{I}\hat{L} = 0$. This is the reason why $\kett{I}$ is called the left vacuum state.

\section{Evaluation of 1-RDM}

\subsection{Boundary IF-MPS}\label{sec:SM-bd-MPS}

We first express $\hat{U}^{SB}$ in a single particle basis.

\begin{equation}
    U^{SB} = \begin{pmatrix}  & & \vline & & \\ & K^{SS}_{ss'} & \vline & K^{SB}_{si'} & \\  & & \vline & & \\ \hline & & \vline & & \\ & K^{BS}_{is'} & \vline & K^{BB}_{ii'} & \\ & & \vline & & \end{pmatrix}
\end{equation}

Representing it in Grassmann variables, ${\eta}$ and ${\xi}$,

\begin{align}
    \bra{{\Bar{\eta}},{\Bar{\xi}}} \hat{U}^{SB} \ket{{\eta},{\xi}} &= \exp \Bigl[  \Bar{\eta}_s K^{SS}_{ss'} \eta_{s'} + \Bar{\eta}_s K^{SB}_{si'}\xi_{i'} + \Bar{\xi}_i K^{BS}_{is'} \eta_{s'} + \Bar{\xi}_i K^{BB}_{ii'} \xi_{i'} \Bigr] \nonumber \\
    &= \exp \Bigl[\Bar{\eta}_s K^{SS}_{ss'} \eta_{s'} + \Bar{\xi}_i K^{BB}_{ii'} \xi_{i'}  \Bigr] \exp \Bigl[ \Bar{\eta}_s K^{SB}_{si'}\xi_{i'} + \Bar{\xi}_i K^{BS}_{is'} \eta_{s'}\Bigr]
\end{align}

The terms, $K^{SS}$ and $K^{BB}$, are factored out and the terms, $K^{SB}$ and $K^{BS}$, couple $S$ and $B$. $K^{SB}$ and $K^{BS}$ can be split with SVD, $K^{SB}=U^a \Sigma^a V^{a }$, $K^{BS}=V^{b}\Sigma^b U^{b}$. Using the resolution of identity,
\begin{equation}
    \begin{gathered}
        \exp \Bigl[ \Bar{\eta}_s K^{SB}_{si'} \xi_{i'} \Bigr] = \int \prod_p d\Bar{\phi}^{a}_{p} d\phi^{a}_{p} \exp (-\Bar{\phi}^{a}_p \phi^a_p) \exp \Bigl[ \Bar{\eta}_s U^{a}_{sp} (\Sigma^a)^{1/2}_p \phi^{a}_p + \Bar{\phi}^{a}_p  (\Sigma^a)^{1/2}_p V^{a}_{pi'} \xi_{i'} \Bigr], \\
        \exp \Bigl[ \Bar{\xi}_i K^{BS}_{is'} \eta_{s'} \Bigr] = \int \prod_p d \Bar{\phi}^{b}_{p} d\phi^{b}_{p} \exp (-\Bar{\phi}^{b}_p \phi^b_p) \exp \Bigl[ \Bar{\xi}_i V^{b}_{ip} (\Sigma^b)^{1/2}_p \phi^{b}_p +\Bar{\phi}^{b}_p  (\Sigma^b)^{1/2}_p U^{b}_{ps'} \eta_{s'} \Bigr],
    \end{gathered}
\end{equation}
$\hat{U}^{SB}$ can be expressed as,
\begin{equation}
    \begin{gathered}
        \bra{{\Bar{\eta}},{\Bar{\xi}}} \hat{U}^{SB} \ket{{\eta},{\xi}} = \exp \Bigl[\Bar{\eta}_s K^{SS}_{ss'} \eta_{s'} +  \Bar{\eta}_s U^{a}_{sp} (\Sigma^a)^{1/2}_p \phi^{a}_p + \Bar{\phi}^{b}_p  (\Sigma^b)^{1/2}_p U^{b}_{ps'} \eta_{s'} \Bigr] \\
        \times \int \mathcal{D}\Bar{\phi} \mathcal{D}\phi \exp (-\Bar{\phi}^{a}_p \phi^a_p-\Bar{\phi}^{b}_p \phi^b_p) \exp \Bigl[ \Bar{\xi}_i K^{BB}_{ii'} \xi_{i'} + \Bar{\phi}^{a}_p  (\Sigma^a)^{1/2}_p V^{a}_{pi'} \xi_{i'} + \Bar{\xi}_i V^{b}_{ip} (\Sigma^b)^{1/2}_p \phi^{b}_p \Bigr],
    \end{gathered}
\end{equation}
where $\mathcal{D}\Bar{\phi} \mathcal{D}\phi = \prod_p d\Bar{\phi}^{a}_{p} d\phi^{a}_{p} \Bar{\phi}^{b}_{p} d\phi^{b}_{p} $. It is equivalent to the contraction of two noninteracting gates, $\hat{U}^{SB} = \hat{W}^S \cdot \hat{W}^B$. $\hat{W}^S$ and $\hat{W}^B$ can be written in a single particle basis as follows,
\begin{equation}
    \begin{gathered}
        W^{S} = \begin{pmatrix}  & & \vline & & \\ & K^{SS}_{ss'} & \vline &U^{a}_{sp} (\Sigma^a)^{1/2}_p &\\  & & \vline & & \\ \hline & & \vline & & \\ &(\Sigma^b)^{1/2}_p U^{b}_{ps'} & \vline & 0 & \\ & & \vline & & \end{pmatrix}, \\
        W^{B} = \begin{pmatrix}  & & \vline & & \\ & 0 & \vline & (\Sigma^a)^{1/2}_p V^{a}_{pi'} & \\  & & \vline & & \\ \hline & & \vline & & \\ & V^{b}_{ip} (\Sigma^b)^{1/2}_p & \vline &  K^{BB}_{ii'} & \\ & & \vline & & \end{pmatrix}.
    \end{gathered}
\end{equation}

\subsection{Iterative update of 1-RDM}
The initial bath 1-RDM of the right IF state, $\Gamma^{R}$ can be represented by a Slater determinant, $C$, after a purification,
\begin{equation}\label{eq:SM_C_1RDM}
    C = \left[ \begin{array}{c}
    \\
       \sqrt{\Gamma^{R}}  \\
       \\
         \hline \\
         \sqrt{I-\Gamma^{R}}
         \\
         \\
         
    \end{array} \right],
\end{equation}
which represents $N_b$ occupied electrons in $2N_b$ orbitals and where the time evolution operator is applied to the first $N_b$ orbitals. The time evolution operator can be either the Trotterized time evolution operator, $\hat{U}^{SB}$, or the boundary IF, $\hat{W}^B$. Here, we use the notation, $\hat{W}^B$, in a general sense, including both unitary and nonunitary time evolution ($\hat{W}^B$ in the boundary IF is nonunitary). We denote the single-particle representation of $\hat{W}^B$ as follows,
\begin{equation}
    W^B = \begin{pmatrix}  & & \vline & & \\ & K^{SS}_{ss'} & \vline & K^{SB}_{si'} & \\  & & \vline & & \\ \hline & & \vline & & \\ & K^{BS}_{is'} & \vline & K^{BB}_{ii'} & \\ & & \vline & & \end{pmatrix}.
\end{equation}
The column, $[K^{SB}, K^{BB}]^T$, is applied to the Slater determinant, and the other column, $[K^{SS},K^{BS}]$, introduces additional occupied electrons. A transformed Slater determinant, $C'$, is,
\begin{equation}
    C' = \begin{pmatrix}
         & & \vline & & \\ & K^{BB} \sqrt{\Gamma^R} & \vline & K^{BS} & \\  & & \vline & & \\ \hline & & \vline & & \\ & K^{SB} \sqrt{\Gamma^R} & \vline & K^{SS} & \\ & & \vline & & \\  \hline & & \vline & & \\ & 0 & \vline & I &\\  & & \vline & &  \\ \hline & & \vline & & \\ & \sqrt{I-\Gamma^R} & \vline & 0 & \\  & & \vline & & 
    \end{pmatrix}.
\end{equation}
Note that each column is not normalized. To normalize the coefficient matrix, we compute the overlap matrix, $S = C^{\dag}C$, and then the normalized Slater determinant can be expressed as, $C' = C S^{-1/2}$. The updated 1-RDM, $\Gamma'^{R}$ is,
\begin{equation}
    \Gamma'^{R} = \begin{pmatrix}
         & & \vline & & \\ & K^{BB} \sqrt{\Gamma^R} & \vline & K^{BS} & \\  & & \vline & &
    \end{pmatrix} \cdot S^{-1} \cdot \begin{pmatrix}
        & & \\
        & \sqrt{\Gamma^R} K^{BB \dag} & \\
        & &\\
        \hline & & \\
        & K^{BS \dag} & \\
        & &
    \end{pmatrix}
\end{equation}

\subsection{Gauge transformation of 1-RDM}

Given the Slater determinant representation of the 1-RDM, as in Eq.~\ref{eq:SM_C_1RDM}, a gauge transformation, $\hat{G}$, is applied to the first $N_b$ orbitals. The unnormalized gauge transformed Slater determinant, $C_G$, can be written as,
\begin{equation}
    C_G = \left[ \begin{array}{c}
    \\
       G\sqrt{\Gamma^{R}}  \\
       \\
         \hline \\
         \sqrt{I-\Gamma^{R}}
         \\
         \\
         
    \end{array} \right]
\end{equation}
and hence the overlap matrix is $C_G^{\dag}C_G = \sqrt{\Gamma^R}G^{\dag}G\sqrt{\Gamma^R}+I-\Gamma^R$. Therefore, the gauge transformed 1-RDM, ${\Gamma}^G$, is,
\begin{equation}
    {\Gamma}^G = [C_G (C_G^{\dag}C_G)^{-1}C_G^{\dag}]_B = G \sqrt{\Gamma^R} (\sqrt{\Gamma^R}G^{\dag}G\sqrt{\Gamma^R}+I-\Gamma^R)^{-1} \sqrt{\Gamma^R}G^{\dag}
\end{equation}

In practice, we can construct the gauge transformed coefficient matrix and 1-RDM without the core and virtual basis in $\Gamma^G$ with the eigenvalues $\lambda < \epsilon$ or $\lambda > 1 - \epsilon$. We first construct $C_G$ with the chosen basis and project out the gauge transformed core basis. The 1-RDM ${\Gamma}^G$ is constructed afterward.

\subsection{Equation of motion of 1-RDM in the continuous-time limit}

In this section, we derive the equation of motion for the 1-RDM, $\Gamma$, in the continuous-time limit of the iterative 1-RDM update. In the continuous-time limit, $\Delta t \to 0$, and taking terms up to first-order in $\Delta t$, $\hat{W}^B$ has the following form,
\begin{equation}
    W^B = \begin{pmatrix}  & & \vline & & \\ & 0 & \vline & 
 -i \sqrt{\Delta t} K^{SB}_{si'} & \\  & & \vline & & \\ \hline & & \vline & & \\ & -i \sqrt{\Delta t} K^{BS}_{is'} & \vline & I-i \Delta t K^{BB}_{ii'} & \\ & & \vline & & \end{pmatrix}.
\end{equation}
The updated Slater determinant is,
\begin{equation}
    C' = \begin{pmatrix}
         & & \vline & & \\ & (I-i\Delta t K^{BB}) \sqrt{\Gamma^R} & \vline &  -i \sqrt{\Delta t} K^{BS} & \\  & & \vline & & \\ \hline & & \vline & & \\ & -i \sqrt{\Delta t} K^{SB} \sqrt{\Gamma^R} & \vline & 0 & \\ & & \vline & & \\  \hline & & \vline & & \\ & 0 & \vline & I &\\  & & \vline & &  \\ \hline & & \vline & & \\ & \sqrt{I-\Gamma^R} & \vline & 0 & \\  & & \vline & & 
    \end{pmatrix}.
\end{equation}
The overlap matrix from $C'$ is
\begin{equation}
    S = I + \sqrt{\Delta t} \begin{pmatrix}
    0&  -i \sqrt{\Gamma^B} K^{BS} \\ iK^{BS \dag} & 0 \end{pmatrix}  + \Delta t  \begin{pmatrix}
    \sqrt{\Gamma^R}(K^{SB \dag}K^{SB} +iK^{BB \dag}-iK^{BB})\sqrt{\Gamma^R} &  0 \\ 0 & K^{BS \dag} K^{BS} 
\end{pmatrix} 
\end{equation}
The updated 1-RDM, $\Gamma(t+\Delta t)=[C' S^{-1} C'^{\dag}]_B$, is,
\begin{equation}
    \begin{gathered}
         \Gamma(t+\Delta t) = \Gamma -i \Delta t K^{BB} \Gamma +i \Delta t \Gamma K^{BB \dag} + \Delta t K^{BS} K^{BS \dag} \\
         - \Delta t K^{BS} K^{BS \dag} \Gamma - \Delta t \Gamma K^{BS} K^{BS \dag} + \Delta t \Gamma [ K^{BS}K^{BS \dag} - K^{SB \dag} K^{SB} +i(K^{BB}-K^{BB \dag})] \Gamma
         \end{gathered}
\end{equation}
In the continuous-time limit, the equation of motion is given by,
\begin{equation}
    \frac{d\Gamma}{dt} = K^{BS}K^{BS\dag} - i (K^{BB} \Gamma - \Gamma K^{BB\dag}) - \{K^{BS}K^{BS \dag},\Gamma\} + \Gamma [K^{BS}K^{BS \dag} - K^{SB \dag} K^{SB} +i(K^{BB}-K^{BB \dag})]\Gamma
\end{equation}
Given $\Gamma = \Gamma^{R}$, $K^{SB} = t^B$, $K^{BS} = t^{B\dag}$, $K^{BB}=L_{B}$, with $L_B = L_B^{\dag}$, and the equation of motion is given by,
\begin{equation}
     \frac{d \Gamma^R}{dt} = t^{B \dag}t^B -i[{L}_B,\Gamma^R] - \{t^{B \dag}t^B,\Gamma^R\}.
\end{equation}
Given $\Gamma = \Gamma^{L}$, $K^{SB} = t^B$, $K^{BS} = t^{B\dag}$, $K^{BB}=-L_{B}$, and the equation of motion is given by,
\begin{equation}\label{eq:SM_gammatr_eom}
    \frac{d \Gamma^{L}}{dt} = t^{B \dag}t^B +i[{L}_B,\Gamma^{L}] - \{t^{B \dag}t^B,\Gamma^{L}\}.
\end{equation}
Given $\Gamma = {\Gamma}^G$, $K^{SB}=\kappa^1$, $K^{BS}=\kappa^2$, $K^{BB}={L}_{GB}$, and the equation of motion is given by,
\begin{equation}\label{eq:SM_gammatilde_eom}
    \frac{d{\Gamma}^G}{dt} = \kappa^2 \kappa^{2 \dag} -i ({L}_{GB} {\Gamma}^G - {\Gamma}^G {L}_{GB}^{\dag})-\{\kappa^2 \kappa^{2 \dag},{\Gamma}^G\} +{\Gamma}^G \left[ \kappa^2 \kappa^{2 \dag} - \kappa^{1 \dag} \kappa^{1} +i({L}_{GB} - {L}_{GB}^{\dag})  \right]{\Gamma}^G
\end{equation}

\subsection{Equation of motion of gauge transformation and effective orbital basis}
The gauge transformation is expressed with the eigenvalues, $\nu_k$, and eigenvectors, $R^{L}_{ik}$, of the left 1-RDM, $\Gamma^{L}$, $G_{ki}=g_k R^{L*}_{ik}$, where $g_k = \sqrt{\nu_k / (1-\nu_k)}$. Therefore, the equation of motion of the gauge transformation can be written in terms of $\Dot{\nu}_k$ and $\Dot{R}^{L}_{ik}$.
\begin{equation}
    \left[\frac{d\hat{G}}{dt}\hat{G}^{-1}\right]_{kl} = \dot{g}_k g^{-1}_k \delta_{kl} -  \sum_i g_k R^{L *}_{ik} \dot{R}^{L }_{il} g^{-1}_l, \quad \text{with} \quad \dot{g}_k g^{-1}_k  = \frac{1}{2\nu_k(1-\nu_k)} \dot{\nu}_k.
\end{equation}
The time dependence of the eigenvalues and eigenvectors can be computed from the perturbation of $\Gamma^{L}$ to $\Gamma^{L}- dt  \dot{\Gamma}^{L}$ (the minus sign is from the fact that the left 1-RDM is propagated in the inverse time). Perturbation theory gives the following expressions,
\begin{equation}
    \begin{gathered}
        \dot{\nu}_k  = - \left[R^{L \dag} \dot{\Gamma}^{L} R^{L} \right]_{kk} = -\sum_a |\sum_i t^{B}_{ai}R^{L}_{ik}|^2(1-2\nu_k),\\
        \left[R^{L\dag}\dot{R}^{L}\right]_{kl} = - \frac{\left[R^{L \dag} \dot{\Gamma}^{L} R^{L} \right]_{kl}}{\nu_l - \nu_k} \: (k \neq l) = \frac{1-\nu_k-\nu_l}{\nu_k-\nu_l}\left[R^{L\dag}t^{B\dag}t^{B}R^{L}\right]_{kl} +i\left[R^{L\dag}\hat{L}_B R^{L}\right]_{kl},
    \end{gathered}
\end{equation}
where $\dot{\Gamma}^{L}$ is given by Eq.~\ref{eq:SM_gammatr_eom}.

The equation of motion of the effective orbital basis can be derived similarly, which contributes to the Liouville operator, $\hat{X}=-i \sum_{m,n} \left[R^{\eff \dag} \dot{R}^{\eff} \right]_{mn}\hat{a}_{m}^{\dag} \hat{a}_{n}$.
\begin{equation}
    \left[R^{\eff \dag} \dot{R}^{\eff} \right]_{mn} = \frac{\left[R^{\eff \dag} \dot{{\Gamma}}^G R^{\eff} \right]_{mn}}{\Tilde{\nu}_n - \Tilde{\nu}_m},
\end{equation}
where $\Tilde{\nu}_m$ is the eigenvalue of ${\Gamma}^G$ for the $m$-th effective orbital eigenvector.

\subsection{Spectrum of $\Gamma$ and ${\Gamma}^G$}\label{sec:spectrum_gamma}

Here, we prove certain properties of eigenvalues of $\Gamma$ and ${\Gamma}^G$, namely that if $\nu$ is an eigenvalue of $\Gamma$ or ${\Gamma}^G$, $1-\nu$ is also an eigenvalue of $\Gamma$ or ${\Gamma}^G$. It is useful to start from the original density operator picture before the particle-hole transformation on one of the Fock spaces. 

\begin{figure}[h]
    \centering
    \includegraphics[width=0.8\textwidth]{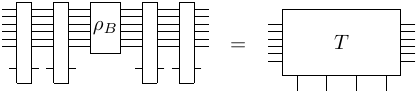}
    \caption{Diagrammatic representation of tensor $T$, which refers to the tensor from a bipartite cut of the whole influence functional tensor before the $k$-th timestep ($k=2$ in the figure).}
    \label{fig:SMfig1}
\end{figure}

First, to prove this property for $\Gamma$, we take the tensor, $T$, originating from a bipartite cut at the $k$-th timestep (Fig.~\ref{fig:SMfig1}), and represent it using Grassmann variables, $\eta$ and $\xi$, which represent the impurity and bath, respectively.
\begin{equation}
    T = \exp \Bigl[ \Bar{\eta}_s g^{SS}_{ss'} \eta_{s'} +\Bar{\eta}_s g^{SB}_{si'} \xi_{i'} + \Bar{\xi}_i g^{BS}_{is'} \eta_{s'}  + \Bar{\xi}_i g^{BB}_{ii'} \xi_{i'}  \Bigr]
\end{equation}
We represent the phase in the exponential with the matrix $G$,
\begin{equation}
    G = \begin{pmatrix}  & & \vline & & \\ & g^{SS}_{ss'} & \vline & g^{SB}_{si'} & \\  & & \vline & & \\ \hline & & \vline & & \\ & g^{BS}_{is'} & \vline & g^{BB}_{ii'} & \\ & & \vline & & \end{pmatrix}.
\end{equation}
where Grassmann variables with bars (without bars) indicate fermions from the left (right) Hilbert space. From the nature of unitary time evolution, $G$ is a Hermitian matrix, $G = G^{\dag}$, and therefore diagonalizable, $G = U\Sigma U^{\dag}$. After particle-hole transformation to the right Hilbert space, the tensor $T$ can be represented as a Slater determinant,
\begin{equation}
    \ket{T} = \prod_p \Bigl( \frac{\Sigma_p}{\sqrt{1+\Sigma_p^2}} f_p^{\dag} + \frac{1}{\sqrt{1+\Sigma_p^2}} \Tilde{f}_p^{\dag} \Bigr)\ket{0},
\end{equation}
where $f_p^{\dag}$ and $\Tilde{f}_p^{\dag}$ are defined as follows,
\begin{equation}
    \begin{gathered}
        f_p^{\dag} = U_{ip} c^{\dag}_{i} + U_{sp} d^{\dag}_s, \\
        \Tilde{f}_p^{\dag} = U_{ip} \Tilde{c}^{\dag}_i + U_{sp} \Tilde{d}^{\dag}_s.
    \end{gathered}
\end{equation}
The 1-RDM of the bath, $\Gamma$, corresponds to the 1-RDM expectation values of $c_i^{\dag}$ and $\Tilde{c}_i^{\dag}$. $\Gamma$ can be represented as,
\begin{equation}
    \Gamma_{ij} = \begin{pmatrix} & & \vline & & \\ &U_{ip} \frac{\Sigma_p^2}{1+\Sigma_p^2} U^{*}_{jp} & \vline & U_{ip} \frac{\Sigma_p}{1+\Sigma_p^2} U^{*}_{jp} &\\ & & \vline & & \\ \hline & & \vline & & \\&U_{ip} \frac{\Sigma_p}{1+\Sigma_p^2} U^{*}_{jp} & \vline & U_{ip} \frac{1}{1+\Sigma_p^2} U^{*}_{jp} & \\ & & \vline & & \end{pmatrix} = \begin{pmatrix}& & \vline & &\\ & M_1 & \vline & M_2 & \\ & & \vline & & \\ \hline & & \vline & & \\&M_2 & \vline & M_3 & \\ & & \vline & & \end{pmatrix},
\end{equation}
where $M_3 = I-M_1$. Given this form, $\Gamma$ satisfies $R \Gamma R^{\dag} = I - \Gamma$, where $R = \begin{pmatrix}
    0&  -I \\ I & 0 
\end{pmatrix}$. Therefore, if $v$ is an eigenvector of $\Gamma$ with eigenvalue $\nu$, $Rv$ is also an eigenvector of $\Gamma$ with eigenvalue $1-\nu$.

The spectrum of ${\Gamma}^G$ can be similarly established by taking the IF tensor after tracing out the bath. The phase matrix $G$ of the tensor is still a Hermitian matrix and hence the 1-RDM for the orbitals in one partition satisfies the same properties. The spectrum of ${\Gamma}^G$ is the same as that of the 1-RDM of the partition from the IF tensor, so it also has eigenvalue pairs, $\nu$ and $1-\nu$.

\end{document}